\documentclass[a4paper,11pt]{article}
\pdfoutput=1
\usepackage{./auxi/jheppub}
\usepackage[utf8]{inputenc}

\usepackage{slashed} % for Feynman slashed notation
\usepackage{mathtools} % for subequation numbering
\usepackage{tikz}
\usepackage{dsfont} % for displaying the blackboard 1 properly
\usetikzlibrary{positioning,arrows,calc}
\usetikzlibrary{arrows.meta}
\usetikzlibrary{decorations.pathmorphing}
\usetikzlibrary{decorations.markings}
\usetikzlibrary{shapes.geometric}
\usetikzlibrary{patterns}
\usepackage{xparse}
\usepackage{diagbox} % for category cells in tables
\usepackage{tensor} % for typesetting mixed tensor indices
\usepackage{subcaption} % to use subfigure environment (pics side by side)
\DeclareMathOperator{\sgn}{sgn}
\newcommand{\Li}{{\normalfont\text{Li}}}

\newcommand{\eqn}[1]{eq.~(\ref{#1})}
\newcommand{\vvev}[1]{\langle\!\langle\, #1 \, \rangle\!\rangle}
\newcommand{\vev}[1]{\langle\, #1 \, \rangle}

\newif\ifstartcompletesineup
\newif\ifendcompletesineup
\pgfkeys{
    /pgf/decoration/.cd,
    start up/.is if=startcompletesineup,
    start up=true,
    start up/.default=true,
    start down/.style={/pgf/decoration/start up=false},
    end up/.is if=endcompletesineup,
    end up=true,
    end up/.default=true,
    end down/.style={/pgf/decoration/end up=false}
}
\pgfdeclaredecoration{complete sines}{initial}
{
    \state{initial}[
        width=+0pt,
        next state=upsine,
        persistent precomputation={
            \ifstartcompletesineup
                \pgfkeys{/pgf/decoration automaton/next state=upsine}
                \ifendcompletesineup
                    \pgfmathsetmacro\matchinglength{
                        0.5*\pgfdecoratedinputsegmentlength / (ceil(0.5* \pgfdecoratedinputsegmentlength / \pgfdecorationsegmentlength) )
                    }
                \else
                    \pgfmathsetmacro\matchinglength{
                        0.5 * \pgfdecoratedinputsegmentlength / (ceil(0.5 * \pgfdecoratedinputsegmentlength / \pgfdecorationsegmentlength ) - 0.499)
                    }
                \fi
            \else
                \pgfkeys{/pgf/decoration automaton/next state=downsine}
                \ifendcompletesineup
                    \pgfmathsetmacro\matchinglength{
                        0.5* \pgfdecoratedinputsegmentlength / (ceil(0.5 * \pgfdecoratedinputsegmentlength / \pgfdecorationsegmentlength ) - 0.4999)
                    }
                \else
                    \pgfmathsetmacro\matchinglength{
                        0.5 * \pgfdecoratedinputsegmentlength / (ceil(0.5 * \pgfdecoratedinputsegmentlength / \pgfdecorationsegmentlength ) )
                    }
                \fi
            \fi
            \setlength{\pgfdecorationsegmentlength}{\matchinglength pt}
        }] {}
    \state{downsine}[width=\pgfdecorationsegmentlength,next state=upsine]{
        \pgfpathsine{\pgfpoint{0.5\pgfdecorationsegmentlength}{0.5\pgfdecorationsegmentamplitude}}
        \pgfpathcosine{\pgfpoint{0.5\pgfdecorationsegmentlength}{-0.5\pgfdecorationsegmentamplitude}}
    }
    \state{upsine}[width=\pgfdecorationsegmentlength,next state=downsine]{
        \pgfpathsine{\pgfpoint{0.5\pgfdecorationsegmentlength}{-0.5\pgfdecorationsegmentamplitude}}
        \pgfpathcosine{\pgfpoint{0.5\pgfdecorationsegmentlength}{0.5\pgfdecorationsegmentamplitude}}
}
    \state{final}{}
}

\definecolor{bgbox}{RGB}{255,254,230}
\definecolor{setupplane}{RGB}{230,230,230}
\definecolor{gluoncolor}{RGB}{207,54,108}
\definecolor{vertexcolor}{RGB}{53,152,219}
\definecolor{SEcolor}{RGB}{176,156,255}
\definecolor{blobcolor}{RGB}{190,180,230}
\tikzset{
corner/.style={line width=1pt,dashed,draw=black,dash pattern=on 6pt off 4pt},
scalar/.style={line width=1pt,draw=black},
gluon/.style={line width=1pt,decorate, draw=gluoncolor,
    decoration={complete sines,aspect=0,amplitude=1.25mm,segment length=1.5mm,start up,end up}},
ghost/.style={line width=1pt,loosely dotted,draw=black},
wilson/.style={line width=2pt,draw=black},
 }
\NewDocumentCommand\semiloop{O{black}mmmO{}O{above}}
{%
\draw[#1] let \p1 = ($(#3)-(#2)$) in (#3) arc (#4:({#4+180}):({0.5*veclen(\x1,\y1)})node[midway, #6] {#5};)
}
\newcommand\x{.6}

\newcommand{\setup}{\begin{tikzpicture}[scale = 3, cross/.style={path picture={
\draw[red]
(path picture bounding box.south east) -- (path picture bounding box.north west) (path picture bounding box.south west) -- (path picture bounding box.north east);
}}]	
\node[draw, red, circle, cross, scale = .75, thick, label={[text width=1.3cm, align=center]left:defect \\ $z,\bar{z}=0$}] (defect) at (-1,0) {};
\node[draw, circle, fill, scale = .5, thick, label={right:$\mathcal{O}(1,1)$}] (phi1) at (1,0) {};
\node[draw, circle, fill, scale = .5, thick, label={above:$\mathcal{O}(z,\bar{z})$}] (phi2) at (.3,-.3) {};
\draw[dashed] (defect) -- node[above, sloped] {$\bar{z}=0$} (0,1) -- node[above, sloped] {$z=1$} (phi1) -- node[below, sloped] {$\bar{z}=1$} (0,-1) -- node[below, sloped] {$z=0$} (defect);
\end{tikzpicture}}
\newcommand{\crossingsymmetry}{\begin{tikzpicture}[baseline={([yshift=-.55ex]current bounding box.center)}]
\draw[color=gluoncolor,fill=gluoncolor,line width=1pt] (1.25,1.8) circle (.1);
\draw[color=gluoncolor,fill=gluoncolor,line width=1pt] (.5,.7) circle (.1);
\draw[wilson] (2,0) -- (2,2.5);
\draw[ghost] (1.5,2) .. controls (.4,2.5) and (-.35,.5) .. (.4,.4) .. controls (1.25,.1) and (2.1,1.65) .. (1.5,2);
\node[] at (2.75,1.25) {$=$};
\draw[color=gluoncolor,fill=gluoncolor,line width=1pt] (4.25,1.8) circle (.1);
\draw[color=gluoncolor,fill=gluoncolor,line width=1pt] (3.5,.7) circle (.1);
\draw[wilson] (5,0) -- (5,2.5);
\draw[ghost] (3.9,1.8) .. controls (3.9,1.2) and (5.4,1.1) .. (5.4,1.8) .. controls (5.4,2.4) and (3.9,2.5) .. (3.9,1.8);
\end{tikzpicture}}

\newcommand{\propagatorS}{\begin{tikzpicture}[baseline={([yshift=-.35ex]current bounding box.center)}]
\draw[scalar] (0,.5) -- (1,.5);
\draw[fill=white] (0,.5) circle (.075);
\draw[fill=white] (1,.5) circle (.075);
\node[] at (0,.1) {\footnotesize \makebox[\widthof{$\smash{\mu,a}$}][c]{$\smash{i,a}$}};
\node[] at (0,.9) {\footnotesize $1$};
\node[] at (1,.1) {\footnotesize \makebox[\widthof{$\smash{\mu,a}$}][c]{$\smash{j,b}$}};
\node[] at (1,.9) {\footnotesize $2$};
\end{tikzpicture}}
\newcommand{\propagatorG}{\begin{tikzpicture}[baseline={([yshift=-.35ex]current bounding box.center)}]
\draw[gluon] (0,.5) -- (1,.5);
\draw[fill=white] (0,.5) circle (.075);
\draw[fill=white] (1,.5) circle (.075);
\node[] at (0,.1) {\footnotesize \makebox[\widthof{$\mu,a$}][c]{$\smash{\mu,a}$}};
\node[] at (0,.9) {\footnotesize $1$};
\node[] at (1,.1) {\footnotesize \makebox[\widthof{$\nu,b$}][c]{$\smash{\nu,b}$}};
\node[] at (1,.9) {\footnotesize $2$};
\end{tikzpicture}}
\newcommand{\propagatorF}{\begin{tikzpicture}[baseline={([yshift=-.35ex]current bounding box.center)}]
\draw[scalar] (0,.5) -- (1,.5);
\draw[fill=white] (0,.5) circle (.075);
\draw[fill=white] (1,.5) circle (.075);
\draw[-Latex] (.62,.5)--(.63,.5);
\node[] at (0,.1) {\footnotesize \makebox[\widthof{$\smash{\mu,a}$}][c]{$\smash{a}$}};
\node[] at (0,.9) {\footnotesize $1$};
\node[] at (1,.1) {\footnotesize \makebox[\widthof{$\smash{\nu,b}$}][c]{$\smash{b}$}};
\node[] at (1,.9) {\footnotesize $2$};
\end{tikzpicture}}
\newcommand{\propagatorGh}{\begin{tikzpicture}[baseline={([yshift=-.35ex]current bounding box.center)}]
\draw[ghost] (0,.5) -- (1,.5);
\draw[fill=white] (0,.5) circle (.075);
\draw[fill=white] (1,.5) circle (.075);
\node[] at (0,.1) {\footnotesize \makebox[\widthof{$\smash{\mu,a}$}][c]{$\smash{a}$}};
\node[] at (0,.9) {\footnotesize $1$};
\node[] at (1,.1) {\footnotesize \makebox[\widthof{$\smash{\nu,b}$}][c]{$\smash{b}$}};
\node[] at (1,.9) {\footnotesize $2$};
\end{tikzpicture}}
\newcommand{\vanishtwoptSS}{\begin{tikzpicture}[baseline={([yshift=-.55ex]current bounding box.center)}]
\draw[wilson] (0,0) -- (0,1);
\draw[scalar] (0,.2) -- (.5,.5);
\draw[scalar] (0,.8) -- (.5,.5);
\draw[gluon] (.5,.5) -- (1,.5);
\draw[fill=white] (0,.2) circle (.075);
\draw[fill=white] (0,.8) circle (.075);
\draw[fill=blobcolor] (1.25,.5) circle (.25);
\draw[fill=vertexcolor,vertexcolor] (.5,.5) circle (.075);
\end{tikzpicture}}
\newcommand{\vanishtwoptSG}{\begin{tikzpicture}[baseline={([yshift=-.55ex]current bounding box.center)}]
\draw[wilson] (0,0) -- (0,1);
\draw[gluon] (0,.2) -- (.5,.5);
\draw[scalar] (0,.8) -- (.5,.5);
\draw[scalar] (.5,.5) -- (1,.5);
\draw[fill=white] (0,.2) circle (.075);
\draw[fill=white] (0,.8) circle (.075);
\draw[fill=blobcolor] (1.25,.5) circle (.25);
\draw[fill=vertexcolor,vertexcolor] (.5,.5) circle (.075);
\end{tikzpicture}}
\newcommand{\vanishtwoptGG}{\begin{tikzpicture}[baseline={([yshift=-.55ex]current bounding box.center)}]
\draw[wilson] (0,0) -- (0,1);
\draw[gluon] (0,.2) -- (.5,.5);
\draw[gluon] (0,.8) -- (.5,.5);
\draw[gluon] (.5,.5) -- (1,.5);
\draw[fill=white] (0,.2) circle (.075);
\draw[fill=white] (0,.8) circle (.075);
\draw[fill=blobcolor] (1.25,.5) circle (.25);
\draw[fill=vertexcolor,vertexcolor] (.5,.5) circle (.075);
\end{tikzpicture}}
\newcommand{\vanishBPS}{\begin{tikzpicture}[baseline={([yshift=-.55ex]current bounding box.center)}]
\draw[scalar] (.9,.5) circle (.4);
\draw[fill=white,scalar] (1.5,.5) circle (.4);
\draw[gluon] (0,.5) -- (.5,.5);
\draw[fill=white] (1.2,.775) circle (.075);
\draw[fill=white] (1.2,.225) circle (.075);
\draw[fill=vertexcolor,vertexcolor] (.5,.5) circle (.075);
\end{tikzpicture}}
\newcommand{\diagTLzero}{\scalebox{\x}{\begin{tikzpicture}[baseline={([yshift=-.55ex]current bounding box.center)}]
\draw[scalar] (.5,1.1375) circle (.6);
\draw[fill=white,line width=1pt] (.5,.5) circle (.25);
\draw[fill=white,line width=1pt] (.5,1.775) circle (.25);
\draw[fill=white] (.27,.61) circle (.075);
\draw[fill=white] (.73,.61) circle (.075);
\draw[fill=white] (.27,1.665) circle (.075);
\draw[fill=white] (.73,1.665) circle (.075);
\end{tikzpicture}}}
\newcommand{\diagTLone}{\scalebox{\x}{\begin{tikzpicture}[baseline={([yshift=-.55ex]current bounding box.center)}]
\draw[wilson] (0,.15) -- (0,2.1);
\draw[scalar] (0,.5) -- (2.25,.5);
\draw[scalar] (0,1.775) -- (2.25,1.775);
\draw[scalar] (2.5,.75) -- (2.5,1.55);
\draw[fill=white,line width=1pt] (2.5,.5) circle (.25);
\draw[fill=white,line width=1pt] (2.5,1.775) circle (.25);
\draw[fill=white] (0,.5) circle (.075);
\draw[fill=white] (0,1.775) circle (.075);
\draw[fill=white] (2.25,.5) circle (.075);
\draw[fill=white] (2.5,.75) circle (.075);
\draw[fill=white] (2.25,1.775) circle (.075);
\draw[fill=white] (2.5,1.55) circle (.075);
\end{tikzpicture}}}
\newcommand{\diagSEone}{\scalebox{\x}{\begin{tikzpicture}[baseline={([yshift=-.55ex]current bounding box.center)}]
\draw[wilson] (0,.15) -- (0,2.1);
\draw[scalar] (0,.5) -- (2.25,.5);
\draw[scalar] (0,1.775) -- (2.25,1.775);
\draw[scalar] (2.5,.75) -- (2.5,1.55);
\draw[fill=white,line width=1pt] (2.5,.5) circle (.25);
\draw[fill=white,line width=1pt] (2.5,1.775) circle (.25);
\draw[fill=white] (0,.5) circle (.075);
\draw[fill=white] (0,1.775) circle (.075);
\draw[fill=white] (2.25,.5) circle (.075);
\draw[fill=white] (2.5,.75) circle (.075);
\draw[fill=white] (2.25,1.775) circle (.075);
\draw[fill=white] (2.5,1.55) circle (.075);
\draw[fill=SEcolor,SEcolor] (2.5,1.1375) circle (.15);
\end{tikzpicture}}}
\newcommand{\diagSEthree}{\scalebox{\x}{\begin{tikzpicture}[baseline={([yshift=-.55ex]current bounding box.center)}]
\draw[wilson] (0,.15) -- (0,2.1);
\draw[scalar] (0,.5) -- (2.25,.5);
\draw[scalar] (0,1.775) -- (2.25,1.775);
\draw[scalar] (2.5,.75) -- (2.5,1.55);
\draw[fill=white,line width=1pt] (2.5,.5) circle (.25);
\draw[fill=white,line width=1pt] (2.5,1.775) circle (.25);
\draw[fill=white] (0,.5) circle (.075);
\draw[fill=white] (0,1.775) circle (.075);
\draw[fill=white] (2.25,.5) circle (.075);
\draw[fill=white] (2.5,.75) circle (.075);
\draw[fill=white] (2.25,1.775) circle (.075);
\draw[fill=white] (2.5,1.55) circle (.075);
\draw[fill=SEcolor,SEcolor] (1.25,.5) circle (.15);
\end{tikzpicture}}}
\newcommand{\diagSEtwo}{\scalebox{\x}{\begin{tikzpicture}[baseline={([yshift=-.55ex]current bounding box.center)}]
\draw[wilson] (0,.15) -- (0,2.1);
\draw[scalar] (0,.5) -- (2.25,.5);
\draw[scalar] (0,1.775) -- (2.25,1.775);
\draw[scalar] (2.5,.75) -- (2.5,1.55);
\draw[fill=white,line width=1pt] (2.5,.5) circle (.25);
\draw[fill=white,line width=1pt] (2.5,1.775) circle (.25);
\draw[fill=white] (0,.5) circle (.075);
\draw[fill=white] (0,1.775) circle (.075);
\draw[fill=white] (2.25,.5) circle (.075);
\draw[fill=white] (2.5,.75) circle (.075);
\draw[fill=white] (2.25,1.775) circle (.075);
\draw[fill=white] (2.5,1.55) circle (.075);
\draw[fill=SEcolor,SEcolor] (1.25,1.775) circle (.15);
\end{tikzpicture}}}
\newcommand{\diagHhcone}{\scalebox{\x}{\begin{tikzpicture}[baseline={([yshift=-.55ex]current bounding box.center)}]
\draw[wilson] (0,.15) -- (0,2.1);
\draw[gluon] (1.25,.5) -- (1.25,1.775);
\draw[scalar] (0,.5) -- (2.25,.5);
\draw[scalar] (0,1.775) -- (2.25,1.775);
\draw[scalar] (2.5,.75) -- (2.5,1.55);
\draw[fill=white,line width=1pt] (2.5,.5) circle (.25);
\draw[fill=white,line width=1pt] (2.5,1.775) circle (.25);
\draw[fill=white] (0,.5) circle (.075);
\draw[fill=white] (0,1.775) circle (.075);
\draw[fill=white] (2.25,.5) circle (.075);
\draw[fill=white] (2.5,.75) circle (.075);
\draw[fill=white] (2.25,1.775) circle (.075);
\draw[fill=white] (2.5,1.55) circle (.075);
\draw[fill=vertexcolor,vertexcolor] (1.25,.5) circle (.075);
\draw[fill=vertexcolor,vertexcolor] (1.25,1.775) circle (.075);
\end{tikzpicture}}}
\newcommand{\diagHhcthree}{\scalebox{\x}{\begin{tikzpicture}[baseline={([yshift=-.55ex]current bounding box.center)}]
\draw[wilson] (0,.15) -- (0,2.1);
\draw[gluon] (1.25,.5) arc (-180:-325:.73);
\draw[scalar] (0,.5) -- (2.25,.5);
\draw[scalar] (0,1.775) -- (2.25,1.775);
\draw[scalar] (2.5,.75) -- (2.5,1.55);
\draw[fill=white,line width=1pt] (2.5,.5) circle (.25);
\draw[fill=white,line width=1pt] (2.5,1.775) circle (.25);
\draw[fill=white] (0,.5) circle (.075);
\draw[fill=white] (0,1.775) circle (.075);
\draw[fill=white] (2.25,.5) circle (.075);
\draw[fill=white] (2.5,.75) circle (.075);
\draw[fill=white] (2.25,1.775) circle (.075);
\draw[fill=white] (2.5,1.55) circle (.075);
\draw[fill=vertexcolor,vertexcolor] (1.25,.5) circle (.075);
\draw[fill=vertexcolor,vertexcolor] (2.5,1.1375) circle (.075);
\end{tikzpicture}}}
\newcommand{\diagHhctwo}{\scalebox{\x}{\begin{tikzpicture}[baseline={([yshift=-.55ex]current bounding box.center)}]
\draw[wilson] (0,.15) -- (0,2.1);
\draw[gluon] (1.25,1.775) arc (180:325:.73);
\draw[scalar] (0,.5) -- (2.25,.5);
\draw[scalar] (0,1.775) -- (2.25,1.775);
\draw[scalar] (2.5,.75) -- (2.5,1.55);
\draw[fill=white,line width=1pt] (2.5,.5) circle (.25);
\draw[fill=white,line width=1pt] (2.5,1.775) circle (.25);
\draw[fill=white] (0,.5) circle (.075);
\draw[fill=white] (0,1.775) circle (.075);
\draw[fill=white] (2.25,.5) circle (.075);
\draw[fill=white] (2.5,.75) circle (.075);
\draw[fill=white] (2.25,1.775) circle (.075);
\draw[fill=white] (2.5,1.55) circle (.075);
\draw[fill=vertexcolor,vertexcolor] (1.25,1.775) circle (.075);
\draw[fill=vertexcolor,vertexcolor] (2.5,1.1375) circle (.075);
\end{tikzpicture}}}
\newcommand{\diagXone}{\scalebox{\x}{\begin{tikzpicture}[baseline={([yshift=-.55ex]current bounding box.center)}]
\draw[wilson] (0,.15) -- (0,2.1);
\draw[scalar] (0,.5) -- (2.25,1.775);
\draw[scalar] (0,1.775) -- (2.25,.5);
\draw[scalar] (2.5,.75) -- (2.5,1.55);
\draw[fill=white,line width=1pt] (2.5,.5) circle (.25);
\draw[fill=white,line width=1pt] (2.5,1.775) circle (.25);
\draw[fill=white] (0,.5) circle (.075);
\draw[fill=white] (0,1.775) circle (.075);
\draw[fill=white] (2.25,.5) circle (.075);
\draw[fill=white] (2.5,.75) circle (.075);
\draw[fill=white] (2.25,1.775) circle (.075);
\draw[fill=white] (2.5,1.55) circle (.075);
\draw[fill=vertexcolor,vertexcolor] (1.125,1.1375) circle (.075);
\end{tikzpicture}}}
\newcommand{\diagXhctwo}{\scalebox{\x}{\begin{tikzpicture}[baseline={([yshift=-.55ex]current bounding box.center)}]
\draw[wilson] (0,.15) -- (0,2.1);
\draw[scalar] (0,.5) -- (2.25,.5);
\draw[scalar] (0,1.775) -- (2.5,.75);
\draw[scalar] (2.25,1.775) arc (70:170:.75);
\draw[scalar] (2.5,1.55) arc (-35:-120:.95);
\draw[fill=white,line width=1pt] (2.5,.5) circle (.25);
\draw[fill=white,line width=1pt] (2.5,1.775) circle (.25);
\draw[fill=white] (0,.5) circle (.075);
\draw[fill=white] (0,1.775) circle (.075);
\draw[fill=white] (2.25,.5) circle (.075);
\draw[fill=white] (2.5,.75) circle (.075);
\draw[fill=white] (2.25,1.775) circle (.075);
\draw[fill=white] (2.5,1.55) circle (.075);
\draw[fill=vertexcolor,vertexcolor] (1.3,1.25) circle (.075);
\end{tikzpicture}}}
\newcommand{\diagXhcthree}{\scalebox{\x}{\begin{tikzpicture}[baseline={([yshift=-.55ex]current bounding box.center)}]
\draw[wilson] (0,.15) -- (0,2.1);
\draw[scalar] (0,1.775) -- (2.25,1.775);
\draw[scalar] (0,.5) -- (2.5,1.55);
\draw[scalar] (2.25,.5) arc (-70:-170:.75);
\draw[scalar] (2.5,.75) arc (35:120:.95);
\draw[fill=white,line width=1pt] (2.5,.5) circle (.25);
\draw[fill=white,line width=1pt] (2.5,1.775) circle (.25);
\draw[fill=white] (0,.5) circle (.075);
\draw[fill=white] (0,1.775) circle (.075);
\draw[fill=white] (2.25,.5) circle (.075);
\draw[fill=white] (2.5,.75) circle (.075);
\draw[fill=white] (2.25,1.775) circle (.075);
\draw[fill=white] (2.5,1.55) circle (.075);
\draw[fill=vertexcolor,vertexcolor] (1.3,1.05) circle (.075);
\end{tikzpicture}}}
\newcommand{\diagCtwo}{\scalebox{\x}{\begin{tikzpicture}[baseline={([yshift=-.55ex]current bounding box.center)}]
\draw[wilson] (0,.15) -- (0,2.1);
\draw[corner] (1.25,.5) arc (-180:-310:.8);
\draw[scalar] (0,.5) -- (2.25,.5);
\draw[scalar] (0,1.775) -- (2.25,1.775);
\draw[scalar] (2.5,.75) -- (2.5,1.55);
\draw[fill=white,line width=1pt] (2.5,.5) circle (.25);
\draw[fill=white,line width=1pt] (2.5,1.775) circle (.25);
\draw[fill=white] (0,.5) circle (.075);
\draw[fill=white] (0,1.775) circle (.075);
\draw[fill=white] (2.25,.5) circle (.075);
\draw[fill=white] (2.5,.75) circle (.075);
\draw[fill=white] (2.25,1.775) circle (.075);
\draw[fill=white] (2.5,1.55) circle (.075);
\draw[fill=vertexcolor,vertexcolor] (1.25,.5) circle (.075);
\draw[fill=vertexcolor,vertexcolor] (2.5,1.1375) circle (.075);
\end{tikzpicture}}}
\newcommand{\diagCone}{\scalebox{\x}{\begin{tikzpicture}[baseline={([yshift=-.55ex]current bounding box.center)}]
\draw[wilson] (0,.15) -- (0,2.1);
\draw[corner] (1.25,1.775) arc (180:310:.8);
\draw[scalar] (0,.5) -- (2.25,.5);
\draw[scalar] (0,1.775) -- (2.25,1.775);
\draw[scalar] (2.5,.75) -- (2.5,1.55);
\draw[fill=white,line width=1pt] (2.5,.5) circle (.25);
\draw[fill=white,line width=1pt] (2.5,1.775) circle (.25);
\draw[fill=white] (0,.5) circle (.075);
\draw[fill=white] (0,1.775) circle (.075);
\draw[fill=white] (2.25,.5) circle (.075);
\draw[fill=white] (2.5,.75) circle (.075);
\draw[fill=white] (2.25,1.775) circle (.075);
\draw[fill=white] (2.5,1.55) circle (.075);
\draw[fill=vertexcolor,vertexcolor] (1.25,1.775) circle (.075);
\draw[fill=vertexcolor,vertexcolor] (2.5,1.1375) circle (.075);
\end{tikzpicture}}}
\newcommand{\diagCthree}{\scalebox{\x}{\begin{tikzpicture}[baseline={([yshift=-.55ex]current bounding box.center)}]
\draw[wilson] (0,.15) -- (0,2.1);
\draw[corner] (0,1.125) arc (240:350:.83);
\draw[scalar] (0,.5) -- (2.25,.5);
\draw[scalar] (0,1.775) -- (2.25,1.775);
\draw[scalar] (2.5,.75) -- (2.5,1.55);
\draw[fill=white,line width=1pt] (2.5,.5) circle (.25);
\draw[fill=white,line width=1pt] (2.5,1.775) circle (.25);
\draw[fill=white] (0,.5) circle (.075);
\draw[fill=white] (0,1.125) circle (.075);
\draw[fill=white] (0,1.775) circle (.075);
\draw[fill=white] (2.25,.5) circle (.075);
\draw[fill=white] (2.5,.75) circle (.075);
\draw[fill=white] (2.25,1.775) circle (.075);
\draw[fill=white] (2.5,1.55) circle (.075);
\draw[fill=vertexcolor,vertexcolor] (1.25,1.775) circle (.075);
\end{tikzpicture}}}
\newcommand{\diagCfour}{\scalebox{\x}{\begin{tikzpicture}[baseline={([yshift=-.55ex]current bounding box.center)}]
\draw[wilson] (0,.15) -- (0,2.1);
\draw[corner] (0,1.125) arc (-240:-350:.83);
\draw[scalar] (0,.5) -- (2.25,.5);
\draw[scalar] (0,1.775) -- (2.25,1.775);
\draw[scalar] (2.5,.75) -- (2.5,1.55);
\draw[fill=white,line width=1pt] (2.5,.5) circle (.25);
\draw[fill=white,line width=1pt] (2.5,1.775) circle (.25);
\draw[fill=white] (0,.5) circle (.075);
\draw[fill=white] (0,1.125) circle (.075);
\draw[fill=white] (0,1.775) circle (.075);
\draw[fill=white] (2.25,.5) circle (.075);
\draw[fill=white] (2.5,.75) circle (.075);
\draw[fill=white] (2.25,1.775) circle (.075);
\draw[fill=white] (2.5,1.55) circle (.075);
\draw[fill=vertexcolor,vertexcolor] (1.25,.5) circle (.075);
\end{tikzpicture}}}
\newcommand{\diagIYIone}{\scalebox{\x}{\begin{tikzpicture}[baseline={([yshift=-.55ex]current bounding box.center)}]
\draw[wilson] (0,.15) -- (0,2.1);
\draw[gluon] (0,1.125) -- (2.5,1.125);
\draw[scalar] (0,.5) -- (2.25,.5);
\draw[scalar] (0,1.775) -- (2.25,1.775);
\draw[scalar] (2.5,.75) -- (2.5,1.55);
\draw[fill=white,line width=1pt] (2.5,.5) circle (.25);
\draw[fill=white,line width=1pt] (2.5,1.775) circle (.25);
\draw[fill=white] (0,.5) circle (.075);
\draw[fill=white] (0,1.125) circle (.075);
\draw[fill=white] (0,1.775) circle (.075);
\draw[fill=white] (2.25,.5) circle (.075);
\draw[fill=white] (2.5,.75) circle (.075);
\draw[fill=white] (2.25,1.775) circle (.075);
\draw[fill=white] (2.5,1.55) circle (.075);
\draw[fill=vertexcolor,vertexcolor] (2.5,1.1375) circle (.075);
\end{tikzpicture}}}
\newcommand{\diagIYIthree}{\scalebox{\x}{\begin{tikzpicture}[baseline={([yshift=-.55ex]current bounding box.center)}]
\draw[wilson] (0,.15) -- (0,2.1);
\draw[gluon] (0,1.125) arc (-240:-370:.8);
\draw[scalar] (0,.5) -- (2.25,.5);
\draw[scalar] (0,1.775) -- (2.25,1.775);
\draw[scalar] (2.5,.75) -- (2.5,1.55);
\draw[fill=white,line width=1pt] (2.5,.5) circle (.25);
\draw[fill=white,line width=1pt] (2.5,1.775) circle (.25);
\draw[fill=white] (0,.5) circle (.075);
\draw[fill=white] (0,1.125) circle (.075);
\draw[fill=white] (0,1.775) circle (.075);
\draw[fill=white] (2.25,.5) circle (.075);
\draw[fill=white] (2.5,.75) circle (.075);
\draw[fill=white] (2.25,1.775) circle (.075);
\draw[fill=white] (2.5,1.55) circle (.075);
\draw[fill=vertexcolor,vertexcolor] (1.25,.5) circle (.075);
\end{tikzpicture}}}
\newcommand{\diagIYItwo}{\scalebox{\x}{\begin{tikzpicture}[baseline={([yshift=-.55ex]current bounding box.center)}]
\draw[wilson] (0,.15) -- (0,2.1);
\draw[gluon] (0,1.125) arc (240:370:.8);
\draw[scalar] (0,.5) -- (2.25,.5);
\draw[scalar] (0,1.775) -- (2.25,1.775);
\draw[scalar] (2.5,.75) -- (2.5,1.55);
\draw[fill=white,line width=1pt] (2.5,.5) circle (.25);
\draw[fill=white,line width=1pt] (2.5,1.775) circle (.25);
\draw[fill=white] (0,.5) circle (.075);
\draw[fill=white] (0,1.125) circle (.075);
\draw[fill=white] (0,1.775) circle (.075);
\draw[fill=white] (2.25,.5) circle (.075);
\draw[fill=white] (2.5,.75) circle (.075);
\draw[fill=white] (2.25,1.775) circle (.075);
\draw[fill=white] (2.5,1.55) circle (.075);
\draw[fill=vertexcolor,vertexcolor] (1.25,1.775) circle (.075);
\end{tikzpicture}}}
\newcommand{\diagHd}{\scalebox{\x}{\begin{tikzpicture}[baseline={([yshift=-.55ex]current bounding box.center)}]
\draw[wilson] (0,.15) -- (0,2.1);
\draw[scalar] (1.25,.5) -- (1.25,1.775);
\draw[gluon] (0,.5) -- (1.25,.5);
\draw[scalar] (1.25,.5) -- (2.25,.5);
\draw[gluon] (0,1.775) -- (1.25,1.775);
\draw[scalar] (1.25,1.775) -- (2.25,1.775);
\draw[scalar] (2.5,.75) -- (2.5,1.55);
\draw[fill=white,line width=1pt] (2.5,.5) circle (.25);
\draw[fill=white,line width=1pt] (2.5,1.775) circle (.25);
\draw[fill=white] (0,.5) circle (.075);
\draw[fill=white] (0,1.775) circle (.075);
\draw[fill=white] (2.25,.5) circle (.075);
\draw[fill=white] (2.5,.75) circle (.075);
\draw[fill=white] (2.25,1.775) circle (.075);
\draw[fill=white] (2.5,1.55) circle (.075);
\draw[fill=vertexcolor,vertexcolor] (1.25,.5) circle (.075);
\draw[fill=vertexcolor,vertexcolor] (1.25,1.775) circle (.075);
\end{tikzpicture}}}
\newcommand{\diagXdtwo}{\scalebox{\x}{\begin{tikzpicture}[baseline={([yshift=-.55ex]current bounding box.center)}]
\draw[wilson] (0,.15) -- (0,2.1);
\draw[gluon] (0,.5) -- (1.125,1.1375);
\draw[gluon] (0,1.775) -- (1.125,1.1375);
\draw[scalar] (1.125,1.1375) -- (2.25,1.775);
\draw[scalar] (1.125,1.1375) -- (2.25,.5);
\draw[scalar] (2.5,.75) -- (2.5,1.55);
\draw[fill=white,line width=1pt] (2.5,.5) circle (.25);
\draw[fill=white,line width=1pt] (2.5,1.775) circle (.25);
\draw[fill=white] (0,.5) circle (.075);
\draw[fill=white] (0,1.775) circle (.075);
\draw[fill=white] (2.25,.5) circle (.075);
\draw[fill=white] (2.5,.75) circle (.075);
\draw[fill=white] (2.25,1.775) circle (.075);
\draw[fill=white] (2.5,1.55) circle (.075);
\draw[fill=vertexcolor,vertexcolor] (1.125,1.1375) circle (.075);
\end{tikzpicture}}}
\newcommand{\diagYY}{\scalebox{\x}{\begin{tikzpicture}[baseline={([yshift=+.45ex]current bounding box.center)}]
\draw[wilson] (0,.15) -- (0,2.1);
\draw[gluon] (0,.5) .. controls (1,.9) and (2,-.25) .. (2.5,0) .. controls (3.5,-.2) and (3.8,.9) .. (2.95,1.1375);
\draw[gluon] (0,1.775) .. controls (.75,1.75) and (1.5,1.25) .. (1.9,1.1375);
\draw[scalar] (2.5,1.1375) circle (.6);
\draw[fill=white,line width=1pt] (2.5,.5) circle (.25);
\draw[fill=white,line width=1pt] (2.5,1.775) circle (.25);
\draw[fill=white] (0,.5) circle (.075);
\draw[fill=white] (0,1.775) circle (.075);
\draw[fill=white] (2.27,.61) circle (.075);
\draw[fill=white] (2.73,.61) circle (.075);
\draw[fill=white] (2.27,1.665) circle (.075);
\draw[fill=white] (2.73,1.665) circle (.075);
\draw[fill=vertexcolor,vertexcolor] (1.9,1.1375) circle (.075);
\draw[fill=vertexcolor,vertexcolor] (3.1,1.1375) circle (.075);
\end{tikzpicture}}}
\newcommand{\planarone}{\scalebox{\x}{\begin{tikzpicture}[baseline={([yshift=-.55ex]current bounding box.center)}]
\draw[wilson] (0,.15) -- (0,2.1);
\draw[scalar] (0,.54) -- (2.375,.54);
\draw[scalar] (0,.93) -- (2.375,.93);
\draw[scalar] (0,1.32) -- (2.375,1.32);
\draw[scalar] (0,1.71) -- (2.375,1.71);
\draw[fill=white,line width=1pt] (0,1.71) circle (.075);
\draw[fill=white,line width=1pt] (0,1.32) circle (.075);
\draw[fill=white,line width=1pt] (0,.93) circle (.075);
\draw[fill=white,line width=1pt] (0,.54) circle (.075);
\draw[fill=white,line width=1pt] (2.5,.735) circle (.25);
\draw[fill=white,line width=1pt] (2.5,1.515) circle (.25);
\draw[fill=white] (2.35,.93) circle (.075);
\draw[fill=white] (2.35,.54) circle (.075);
\draw[fill=white] (2.35,1.71) circle (.075);
\draw[fill=white] (2.35,1.32) circle (.075);
\end{tikzpicture}}}

\newcommand{\diagoneptfunction}{\scalebox{\x}{\begin{tikzpicture}[baseline={([yshift=-.55ex]current bounding box.center)}]
\draw[wilson] (0,.15) -- (0,2.1);
\draw[scalar] (0,1.35) -- (2.375,1.35);
\draw[scalar] (0,.93) -- (2.375,.93);
\draw[fill=white,line width=1pt] (2.5,1.1375) circle (.25);
\draw[fill=white] (0,1.35) circle (.075);
\draw[fill=white] (0,.93) circle (.075);
\draw[fill=white] (2.375,1.35) circle (.075);
\draw[fill=white] (2.375,.93) circle (.075);
\end{tikzpicture}}}

\newcommand{\wilsonrulezero}{\begin{tikzpicture}[baseline={([yshift=-.55ex]current bounding box.center)}]
\draw[wilson] (0,.2) -- (0,1.3);
\end{tikzpicture}}
\newcommand{\wilsonruleS}{\begin{tikzpicture}[baseline={([yshift=-.55ex]current bounding box.center)}]
\draw[wilson] (0,.2) -- (0,1.3);
\draw[scalar] (0,.75) -- (.75,.75);
\draw[fill=white] (0,.75) circle (.075);
\end{tikzpicture}}
\newcommand{\wilsonruleG}{\begin{tikzpicture}[baseline={([yshift=-.55ex]current bounding box.center)}]
\draw[wilson] (0,.2) -- (0,1.3);
\draw[gluon] (0,.75) -- (.75,.75);
\draw[fill=white] (0,.75) circle (.075);
\end{tikzpicture}}
\newcommand{\wilsonruleSS}{\begin{tikzpicture}[baseline={([yshift=-.55ex]current bounding box.center)}]
\draw[wilson] (0,.2) -- (0,1.3);
\draw[scalar] (0,.5) -- (.75,.5);
\draw[scalar] (0,1) -- (.75,1);
\draw[fill=white] (0,.5) circle (.075);
\draw[fill=white] (0,1) circle (.075);
\end{tikzpicture}}
\newcommand{\wilsonruleGG}{\begin{tikzpicture}[baseline={([yshift=-.55ex]current bounding box.center)}]
\draw[wilson] (0,.2) -- (0,1.3);
\draw[gluon] (0,.5) -- (.75,.5);
\draw[gluon] (0,1) -- (.75,1);
\draw[fill=white] (0,.5) circle (.075);
\draw[fill=white] (0,1) circle (.075);
\end{tikzpicture}}
\newcommand{\wilsonruleSSG}{\begin{tikzpicture}[baseline={([yshift=-.55ex]current bounding box.center)}]
\draw[wilson] (0,.1) -- (0,1.4);
\draw[scalar] (0,.375) -- (.75,.375);
\draw[gluon] (0,.75) -- (.75,.75);
\draw[scalar] (0,1.125) -- (.75,1.125);
\draw[fill=white] (0,.375) circle (.075);
\draw[fill=white] (0,.75) circle (.075);
\draw[fill=white] (0,1.125) circle (.075);
\end{tikzpicture}}
\newcommand{\wilsonruleSSSS}{\begin{tikzpicture}[baseline={([yshift=-.55ex]current bounding box.center)}]
\draw[wilson] (0,0) -- (0,1.5);
\draw[scalar] (0,.3) -- (.75,.3);
\draw[scalar] (0,.6) -- (.75,.6);
\draw[scalar] (0,.9) -- (.75,.9);
\draw[scalar] (0,1.2) -- (.75,1.2);
\draw[fill=white] (0,.3) circle (.075);
\draw[fill=white] (0,.6) circle (.075);
\draw[fill=white] (0,.9) circle (.075);
\draw[fill=white] (0,1.2) circle (.075);
\end{tikzpicture}}
\newcommand{\propagatorSSEnotext}{\begin{tikzpicture}[baseline={([yshift=-.35ex]current bounding box.center)}]
\draw[scalar] (0,.5) -- (1,.5);
\draw[fill=white] (0,.5) circle (.075);
\draw[fill=white] (1,.5) circle (.075);
\draw[fill=SEcolor,SEcolor] (.5,.5) circle (.15);
\end{tikzpicture}}
\newcommand{\SSEone}{\begin{tikzpicture}[baseline={([yshift=-.4ex]current bounding box.center)}]
\draw[scalar] (0,.5) -- (2,.5);
\draw[gluon] (.5,.5) arc (180:-20:.5);
\draw[fill=white] (0,.5) circle (.075);
\draw[fill=white] (2,.5) circle (.075);
\draw[fill=vertexcolor,vertexcolor] (.5,.5) circle (.075);
\draw[fill=vertexcolor,vertexcolor] (1.5,.5) circle (.075);
\end{tikzpicture}}
\newcommand{\SSEtwo}{\begin{tikzpicture}[baseline={([yshift=-.5ex]current bounding box.center)}]
\draw[scalar] (0,.5) -- (.5,.5);
\draw[scalar] (1.5,.5) -- (2,.5);
\draw[scalar,postaction={decorate}] (.5,.5) arc (180:0:.5);
\draw[-Latex] (1.1,1)--(1.11,1);
\draw[scalar,postaction={decorate}] (.5,.5) arc (-180:0:.5);
\draw[-Latex] (.9,0)--(.89,0);
\draw[fill=white] (0,.5) circle (.075);
\draw[fill=white] (2,.5) circle (.075);
\draw[fill=vertexcolor,vertexcolor] (.5,.5) circle (.075);
\draw[fill=vertexcolor,vertexcolor] (1.5,.5) circle (.075);
\end{tikzpicture}}
\newcommand{\SSEthree}{\begin{tikzpicture}[baseline={([yshift=-.75ex]current bounding box.center)}]
\draw[scalar] (0,.5) -- (2,.5);
\draw[scalar] (1,1) circle (.5);
\draw[fill=white] (0,.5) circle (.075);
\draw[fill=white] (2,.5) circle (.075);
\draw[fill=vertexcolor,vertexcolor] (1,.5) circle (.075);
\end{tikzpicture}}
\newcommand{\SSEfour}{\begin{tikzpicture}[baseline={([yshift=-1.05ex]current bounding box.center)}]
\draw[scalar] (0,.5) -- (2,.5);
\draw[gluon] (1,1) circle (.5);
\draw[fill=white] (0,.5) circle (.075);
\draw[fill=white] (2,.5) circle (.075);
\draw[fill=vertexcolor,vertexcolor] (1,.5) circle (.075);
\end{tikzpicture}}
\newcommand{\vertexSSG}{\begin{tikzpicture}[baseline={([yshift=-.35ex]current bounding box.center)}]
\draw[gluon] (0,.5) -- (.75,.5);
\draw[scalar] (.75,.5) -- (1.25,1);
\draw[scalar] (.75,.5) -- (1.25,0);
\draw[fill=vertexcolor,vertexcolor] (.75,.5) circle (.075);
\draw[fill=white] (0,.5) circle (.075);
\draw[fill=white] (1.25,1) circle (.075);
\draw[fill=white] (1.25,0) circle (.075);
\node[anchor=west] at (1.3,1) {\footnotesize $i,a$};
\node[] at (1.25,1.3) {\footnotesize $1$};
\node[anchor=west] at (1.3,0) {\footnotesize $j,b$};
\node[] at (1.25,-.3) {\footnotesize $2$};
\node[] at (0,.1) {\footnotesize $\mu,c$};
\node[] at (0,.9) {\footnotesize $3$};
\end{tikzpicture}}
\newcommand{\vertexSSSS}{\begin{tikzpicture}[baseline={([yshift=-.35ex]current bounding box.center)}]
\draw[scalar] (0,0) -- (1.25,1);
\draw[scalar] (0,1) -- (1.25,0);
\draw[fill=vertexcolor,vertexcolor] (.625,.5) circle (.075);
\draw[fill=white] (0,0) circle (.075);
\draw[fill=white] (0,1) circle (.075);
\draw[fill=white] (1.25,1) circle (.075);
\draw[fill=white] (1.25,0) circle (.075);
\node[anchor=west] at (1.3,1) {\footnotesize $i,a$};
\node[] at (1.25,1.3) {\footnotesize $1$};
\node[anchor=west] at (1.3,0) {\footnotesize $j,b$};
\node[] at (1.25,-.3) {\footnotesize $2$};
\node[anchor=east] at (-0.05,1) {\footnotesize $k,c$};
\node[] at (0,1.3) {\footnotesize $3$};
\node[anchor=east] at (-0.05,0) {\footnotesize $l,d$};
\node[] at (0,-.3) {\footnotesize $4$};
\end{tikzpicture}}
\newcommand{\vertexSSGG}{\begin{tikzpicture}[baseline={([yshift=-.35ex]current bounding box.center)}]
\draw[scalar] (.625,.5) -- (1.25,1);
\draw[scalar] (.625,.5) -- (1.25,0);
\draw[gluon] (0,0) -- (.625,.5);
\draw[gluon] (0,1) -- (.625,.5);
\draw[fill=vertexcolor,vertexcolor] (.625,.5) circle (.075);
\draw[fill=white] (0,0) circle (.075);
\draw[fill=white] (0,1) circle (.075);
\draw[fill=white] (1.25,1) circle (.075);
\draw[fill=white] (1.25,0) circle (.075);
\node[anchor=west] at (1.3,1) {\footnotesize $i,a$};
\node[] at (1.25,1.3) {\footnotesize $1$};
\node[anchor=west] at (1.3,0) {\footnotesize $j,b$};
\node[] at (1.25,-.3) {\footnotesize $2$};
\node[anchor=east] at (-0.05,1) {\footnotesize $\mu,c$};
\node[] at (0,1.3) {\footnotesize $3$};
\node[anchor=east] at (-0.05,0) {\footnotesize $\nu,d$};
\node[] at (0,-.3) {\footnotesize $4$};
\end{tikzpicture}}
\newcommand{\Hinsertone}{\begin{tikzpicture}[baseline={([yshift=-.55ex]current bounding box.center)}]
\draw[scalar] (0,0) -- (1.25,0);
\draw[scalar] (0,1) -- (1.25,1);
\draw[gluon] (.625,0) -- (.625,1);
\draw[fill=vertexcolor,vertexcolor] (.625,0) circle (.075);
\draw[fill=vertexcolor,vertexcolor] (.625,1) circle (.075);
\draw[fill=white] (0,0) circle (.075);
\draw[fill=white] (0,1) circle (.075);
\draw[fill=white] (1.25,1) circle (.075);
\draw[fill=white] (1.25,0) circle (.075);
\node[anchor=west] at (1.3,1) {\footnotesize $i,a$};
\node[] at (1.25,1.3) {\footnotesize $1$};
\node[anchor=west] at (1.3,0) {\footnotesize $j,b$};
\node[] at (1.25,-.3) {\footnotesize $2$};
\node[anchor=east] at (-0.05,1) {\footnotesize $k,c$};
\node[] at (0,1.3) {\footnotesize $3$};
\node[anchor=east] at (-0.05,0) {\footnotesize $l,d$};
\node[] at (0,-.3) {\footnotesize $4$};
\end{tikzpicture}}
\newcommand{\Hinserttwo}{\begin{tikzpicture}[baseline={([yshift=-.55ex]current bounding box.center)}]
\draw[gluon] (0,0) -- (.625,0);
\draw[gluon] (0,1) -- (.625,1);
\draw[scalar] (1.25,0) -- (.625,0);
\draw[scalar] (1.25,1) -- (.625,1);
\draw[scalar] (.625,0) -- (.625,1);
\draw[fill=vertexcolor,vertexcolor] (.625,0) circle (.075);
\draw[fill=vertexcolor,vertexcolor] (.625,1) circle (.075);
\draw[fill=white] (0,0) circle (.075);
\draw[fill=white] (0,1) circle (.075);
\draw[fill=white] (1.25,1) circle (.075);
\draw[fill=white] (1.25,0) circle (.075);
\node[anchor=west] at (1.3,1) {\footnotesize $i,a$};
\node[] at (1.25,1.3) {\footnotesize $1$};
\node[anchor=west] at (1.3,0) {\footnotesize $j,b$};
\node[] at (1.25,-.3) {\footnotesize $2$};
\node[anchor=east] at (-0.05,1) {\footnotesize $\mu,c$};
\node[] at (0,1.3) {\footnotesize $3$};
\node[anchor=east] at (-0.05,0) {\footnotesize $\nu,d$};
\node[] at (0,-.3) {\footnotesize $4$};
\end{tikzpicture}}

\newcommand{\diagYonetwotwo}{\scalebox{\x}{\begin{tikzpicture}[baseline={([yshift=-.55ex]current bounding box.center)}]
\draw[scalar] (0,.5) -- (1,.5);
\draw[scalar] (1.5,.5) circle (.5);
\draw[fill=white] (0,.5) circle (.075);
\draw[fill=white] (2,.5) circle (.075);
\draw[fill=vertexcolor,vertexcolor] (1,.5) circle (.075);
\end{tikzpicture}}}
\newcommand{\diagXoneonetwothree}{\scalebox{\x}{\begin{tikzpicture}[baseline={([yshift=-.55ex]current bounding box.center)}]
\draw[scalar] (.5,1) -- (1,.5);
\draw[scalar] (.5,0) -- (1,.5);
\draw[scalar] (1.5,.5) circle (.5);
\draw[fill=white] (.5,1) circle (.075);
\draw[fill=white] (.5,0) circle (.075);
\draw[fill=white] (2,.5) circle (.075);
\draw[fill=vertexcolor,vertexcolor] (1,.5) circle (.075);
\end{tikzpicture}}}
\newcommand{\diagFonethreeonefour}{\scalebox{\x}{\begin{tikzpicture}[baseline={([yshift=-.55ex]current bounding box.center)}]
\draw[scalar] (0,0) -- (.625,0);
\draw[scalar] (0,1) -- (.625,1);
\draw[scalar] (.625,0) arc (270:450:.5);
\draw[scalar] (.625,0) -- (.625,1);
\draw[fill=vertexcolor,vertexcolor] (.625,0) circle (.075);
\draw[fill=vertexcolor,vertexcolor] (.625,1) circle (.075);
\draw[fill=white] (0,0) circle (.075);
\draw[fill=white] (0,1) circle (.075);
\draw[fill=white] (1.125,.5) circle (.075);
\end{tikzpicture}}}
\usepackage{simplewick}
\usepackage{dsfont}
\usepackage{float}
\usepackage{pgfplots}
\pgfplotsset{compat=1.14}

\let\oldbfseries=\bfseries
\let\oldmdseries=\mdseries
\let\oldnormalfont=\normalfont
\renewcommand{\bfseries}{\oldbfseries\boldmath}
\renewcommand{\mdseries}{\oldmdseries\unboldmath}
\renewcommand{\normalfont}{\oldnormalfont\unboldmath}

\makeatletter
\newlength{\apb@width}
\newcommand{\autoparbox}[2][c]{\settowidth{\apb@width}{#2}\parbox[#1]{\apb@width}{#2}}

\makeatother

\def\Nm{{\mathcal{N}}}

\newcommand{\beq}{\begin{equation}}
\newcommand{\eeq}{\end{equation}}

\definecolor{nicegreen}{rgb}{0.1,0.6,0.1}

\newmuskip\pFqskip
\mathchardef\pFcomma=\mathcode`,

\makeatletter
\renewcommand*\env@matrix[1][\arraystretch]{%
  \edef\arraystretch{#1}%
  \hskip -\arraycolsep
  \let\@ifnextchar\new@ifnextchar
  \array{*\c@MaxMatrixCols c}}
\makeatother

\graphicspath{{./figures/}}

\title{\center{Two-point correlator of chiral primary operators with a Wilson line defect
in $\mathcal{N}=4$ SYM }}

\author[1]{Julien Barrat,}
\author[2]{Pedro Liendo,}
\author[1]{Jan Plefka.}

\affiliation[1]{Institut f\"ur Physik und IRIS Adlershof, Humboldt-Universit{\"a}t zu Berlin, Zum Gro{\ss}en Windkanal 6, 12489 Berlin, Germany}
\affiliation[2]{DESY Hamburg, Theory Group, Notkestra{\ss}e 85, D-22607 Hamburg, Germany}

\emailAdd{julien.barrat@hu-berlin.de, pedro.liendo@desy.de, jan.plefka@hu-berlin.de}

\preprint{HU-EP-20/33-RTG, DESY 20-191}

\abstract{We study the two-point function of the stress-tensor multiplet of $\mathcal{N}=4$ SYM in the presence of a line defect. To be more precise, we focus on the single-trace operator of conformal dimension two that sits in the $20'$ irrep of the $\mathfrak{so}(6)_\text{R}$ R-symmetry, and add a Maldacena-Wilson line to the configuration which makes the two-point function non-trivial. We use a combination of perturbation theory and defect CFT techniques to obtain results up to next-to-leading order in the coupling constant. Being a defect CFT correlator, there exist two (super)conformal block expansions which capture defect and bulk data respectively. We present a closed-form formula for the defect CFT data, which allows to write an efficient Taylor series for the correlator in the limit when one of the operators is close to the line. The bulk channel is technically harder and closed-form formulae are particularly challenging to obtain, nevertheless we use our analysis to check against well-known data of $\mathcal{N}=4$ SYM. In particular, we recover the correct anomalous dimensions of a famous tower of twist-two operators (which includes the Konishi multiplet), and successfully compare the one-point function of the stress-tensor multiplet with results obtained using matrix-model techniques.
}

\begin{document}

\setcounter{tocdepth}{2}
\maketitle
\setcounter{page}{1}

%!TEX root = ../2pt_function_wline.tex
%%%%%%%%%%%%%%%%%%%%%%%%%%%%%%%%%%%%%%%%%%%%%

\section{Introduction}
\label{sec:intro}

The maximally extended supersymmetric Yang-Mills theory in four dimensional space-time
($\mathcal{N}=4$ SYM) \cite{Brink:1976bc} is a prime theoretical laboratory in quantum field theory.
It stands out in particular due to its
quantum exact conformal symmetry, a hidden integrability in the large $N$ color limit 
\cite{Beisert:2010jr} of the SU($N$) model, as well 
as its relevance within the AdS/CFT correspondence \cite{Maldacena:1997re,Gubser:1998bc,Witten:1998qj}. As a consequence an enormous wealth
of results has been obtained to date -- including exact non-perturbative ones.
Three classes of (pre)--observables that are prototypical for gauge theories can be identified:
scattering amplitudes, correlation functions of local gauge invariant operators and 
expectation values of non-local Wilson loop operators. They may be severely constrained by
the manifest superconformal as well as hidden symmetries of the $\mathcal{N}=4$ SYM theory.
Setting aside the sector of scattering amplitudes\footnote{Here the tree-level all multiplicity
as well as four and five point to all-orders in the coupling are known exactly.}, the
canonical representatives of the local and non-local classes are: 
\begin{enumerate}
\item The length-two local operator built from the scalar field $\phi^{i}(x)$:
\begin{equation}
\label{eq:O20def}
\mathcal{O}_{20'}(x):= \text{Tr}[\phi^i (x) \phi^j (x)]
-\frac{\delta^{ij}}{6}\text{Tr}[\phi^k (x) \phi_k (x)]\,,
\end{equation}
which transforms in the $20'$ of the $\mathfrak{so}(6)_\text{R}$ R-symmetry group. It is a chiral
primary ($1/2$-BPS) operator of the conformal symmetry group, as a consequence
of which its two-point and three-point correlation functions are protected from radiative corrections
and given exactly by their free field theory approximation
\cite{Lee:1998bxa}. The $\mathcal{O}_{20'}(x)$ operators 
sit in the so-called stress tensor multiplet of the underlying superconformal symmetry group
that includes the stress tensor as well as the Lagrangian of $\mathcal{N}=4$ SYM itself as 
superconformal descendants. Non-trivial results emerge for the four-point correlation 
function of the $\mathcal{O}_{20'}(x)$ \cite{DHoker:1999kzh,Dolan:2000ut,Eden:2000mv,Arutyunov:2003ad} and beyond 
\cite{Drukker:2008pi,Drukker:2009sf}.
\item Supersymmetric Wilson or Maldacena-Wilson loops \cite{Maldacena:1998im}:
\begin{equation}
\label{eq:WLdef}
\mathcal{W}_{C} := \frac{1}{N} \text{Tr}\ \mathcal{P} \exp \oint_C d\tau \bigl( i \tensor{\dot{x}}{_{\smash{\mu}}} \tensor{A}{^{\smash{\mu}}} + | \dot{x} | \tensor{\theta}{_i} \tensor{\phi}{^i} \bigr),
\end{equation}
defined by a specific curve $C$ in space-time parametrized via $x^{\mu}(\tau)$ and $\theta_{i}\theta^{i}=1$. Its vacuum expectation value is finite for
smooth curves, and trivial, i.e.~$\langle \mathcal{W}_{l}\rangle =1$, for straight line geometries \cite{Erickson:2000af}. For the circular contours the non-trivial result is known exactly  \cite{Erickson:2000af,Drukker:2000rr,Pestun:2007rz}\footnote{ It takes the radius-independent form
$\langle \mathcal{W}_{c}\rangle = \frac{2}{\sqrt{\lambda}}I_{1}(\sqrt{\lambda})$ with
the coupling $\lambda=g^{2}N$ in the large $N$ 't Hooft limit.}
 and related to the straight line configuration
through an (anomalous) conformal transformation.
\end{enumerate}
Given the prominence of these simplest operators within the $\mathcal{N}=4$ SYM theory, the
study of correlation functions involving \emph{both} operators suggests itself. 
The correlation function of a single chiral primary operator sitting outside of 
a straight line or circular Maldacena-Wilson loop  
$\langle \mathcal{W}_{C}\mathcal{O}_{20'}(x)\rangle$
is fixed by conformal symmetry up to a function of the coupling constant \cite{Berenstein:1998ij,Zarembo:2002ph} which may be computed exactly \cite{Semenoff:2001xp,Giombi:2009ds}. Correlators involving cusped/straight lines with a defect operator were also studied in integrability, and many results are known non-perturbatively \cite{Beccaria:2017rbe,Kim:2017sju,Cooke:2017qgm}. However, a lot less is known for the situation of placing \emph{two} local operators outside of a Maldacena-Wilson line at generic positions $\langle \mathcal{W}_{c/l}\mathcal{O}_{20'}(x_1) \mathcal{O}_{20'}(x_2)\rangle$. Previous work \cite{Buchbinder:2012vr} considered this correlator at weak coupling at leading order (LO) and at strong coupling using the AdS/CFT duality to the minimal string surfaces in $AdS_{5}\times S^{5}$, see also the very recent \cite{Beccaria:2020ykg}. 
Moreover, configurations with operators \textit{along} the line have received some attention recently: correlators of 1/2-BPS operators in the resulting 1d CFT have been studied using holography \cite{Giombi:2017cqn}, the conformal bootstrap \cite{Liendo:2018ukf}, and by looking at protected subsectors \cite{Bonini:2015fng,Giombi:2018qox,Giombi:2018hsx}, while integrability has access to some non-perturbative CFT data, even away from the protected sectors \cite{Grabner:2020nis}.

\bigskip

In this work we consider the perturbative study of a Maldacena-Wilson line with two generic bulk chiral primaries 
at next-to-leading order (NLO) in the coupling constant. Unlike one-point functions, the coordinate dependence of two-point correlators in the presence of a defect is not fixed by symmetry. Two-point functions depend on two conformal invariants $z$ and $\bar{z}$ similar to the two cross-ratios that characterize four-point correlators in standard CFTs.  Motivated by the modern conformal bootstrap \cite{Poland:2018epd}, recent years have seen significant progress in  understanding the structure of defect CFT correlators. 

There exist two representations for two-point functions in the presence of a line defect \cite{Billo:2016cpy}. On the one hand, one can fuse the two local operators using the usual operator product expansion (OPE), and then evaluate the resulting one-point functions in the presence of the defect. The second representation is to consider the defect operator expansion, in which one of the local operators is written as an infinite sum of local excitations on the defect. Both these expansions are characterized by corresponding conformal blocks and the associated \textit{CFT data} (i.e. the spectrum of operators and the OPE coefficients multiplying the blocks), and they are usually referred to as the \textit{bulk channel} and \textit{defect channel} expansions.

Equality of these two representations is the starting point for the bootstrap program of defect CFTs \cite{Gaiotto:2013nva,Billo:2016cpy,Lauria:2017wav,Lauria:2018klo,Lemos:2017vnx,Soderberg:2017oaa,Liendo:2019jpu,Lauria:2020emq}. In this work we will not use the bootstrap approach but rather perform an explicit perturbative computation up to next-to-leading order (NLO). However, as it will be clear in the main text, the use of modern defect CFT results will be crucial for our success, in particular, correlators of 1/2-BPS operators are severely constrained by Ward identities which were carefully studied in \cite{Liendo:2016ymz}. These powerful constraints together with a novel combination of perturbation theory, defect CFT techniques and numerical integration will allow us to obtain an exact formula for the defect channel data, which is basically a solution of the problem and can be used to produce an efficient Taylor series around the defect OPE limit $(z,\bar{z}) \sim 0$. We perform a similar analysis for the bulk channel, however this is technically more complicated and closed-form expressions are hard to obtain. We nevertheless extract a large amount of CFT data and compare it to known results, finding a perfect match for an infinite set of anomalous dimensions, as well as for the one-point function of the $\mathcal{O}_{20'}(x)$ operator. 

\bigskip

The outline of the paper is as follows. In section \ref{sec:preliminaries} we review the kinematics of the two-point function of the $\mathcal{O}_{20'}$ chiral primary and its superconformal block expansion in the defect and bulk channels. We then present our perturbative computation in section \ref{sec:perturbative}; we find analytic formulae for some of the integrals, and when this is not possible we perform some numerical consistency checks using the Ward identities. In section \ref{sec:cft_data} we extract the CFT data in the defect and bulk channels by combining our perturbative calculation with the block expansion. We present an exact formula for the defect data, and perform various non-trivial checks of our results. We conclude in section \ref{sec:conclusions}, and collect technical details in the appendices.

%!TEX root = ../2pt_function_wline.tex
%%%%%%%%%%%%%%%%%%%%%%%%%%%%%%%%%%%%%%%%%%%%%

\section{Preliminaries}
\label{sec:preliminaries}

We start by reviewing the analysis of \cite{Liendo:2016ymz}, where the basic kinematics of two-point functions in the presence of a 1/2-BPS line defect were studied. We focus on the correlator of two $\mathcal{O}_{20'}(x)$ operators in the presence of a Maldacena-Wilson line, however we should point out that the kinematics reviewed here are more general, and apply to any line defect that preserves half of the supersymmetry.

\subsection{Two-Point Function of Single-Trace Operators with Wilson-Line Defect}
\label{subsec:twopointfunction}

\begin{figure}[t]
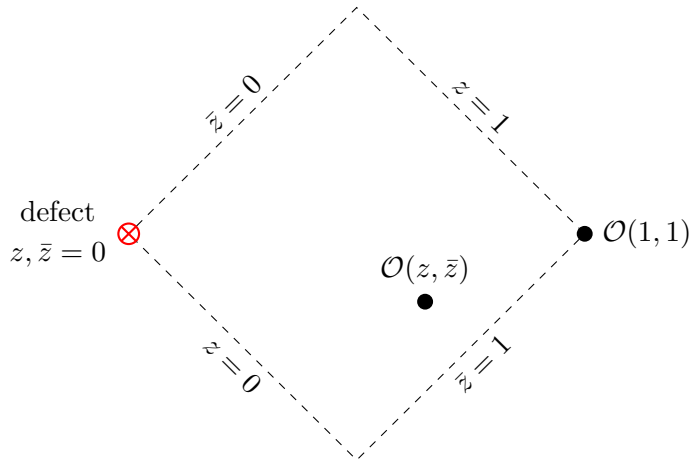

\centering
\setup
\caption{Setup for the two-point function of single-trace operators $\mathcal{O}_{20'}$ in the presence of a Wilson-line defect $\mathcal{W}_l$ in Euclidean $4$d space. Using scaling symmetry, the coordinate system can be chosen such that $x_1 := (1,0,0,0)$, while the second operator can be described with the help of complex coordinates $z, \bar{z}$ as indicated in the text. The line extends out of the page (in the $\tau$-direction) and is orthogonal to the complex plane.}
\label{fig1}
\end{figure}

\noindent First, it is convenient to rewrite the single-trace chiral primary operator 
$\mathcal{O}_{20'}(x)$ of \eqn{eq:O20def} 
as
\begin{equation}
\mathcal{O}(x) := u_i u_j\ \text{Tr}[ \phi^i (x) \phi^j (x)] \,,
\label{eq:singletraceoperators}
\end{equation}
where the $\mathfrak{so}(6)_\text{R}$ vectors $u_{i}$'s obey $u^{2}=0$ and are necessarily complex.
This is a convenient bookkeeping device in order to ensure the tracelessness condition of the
chiral primaries $\mathcal{O}_{20'}(x)$. These operators are $1/2$-BPS and as reviewed in the introduction their two- and three-point functions are protected when there is no defect present. If one inserts a straight Maldacena-Wilson line, the  kinematics changes drastically. Consider
\begin{equation}
\mathcal{W}_{l} := \frac{1}{N} \text{Tr}\ \mathcal{P} \exp \int_{-\infty}^{\infty} d\tau \bigl( i \tensor{\dot{x}}{^{\smash{\mu}}} \tensor{A}{_{\smash{\mu}}} + | \dot{x} | \tensor{\theta}{_i} \tensor{\phi}{^i} \bigr) \,,
\label{eq:MaldacenaWilsonLoop}
\end{equation}
with $\tensor{A}{_{\smash{\mu}}} := \tensor{T}{_a} \tensor*{A}{^a_{\smash{\mu}}}$, $\tensor{\phi}{_i} := \tensor{T}{_a} \tensor*{\phi}{_i^a}$, where $T_a$ is a generator of the gauge group of the SYM theory and $i=1,\ldots, 6$ an $\mathfrak{so}(6)_\text{R}$ index. We work in Euclidean space after Wick rotation and parameterize the line by $x^{\mu}(\tau)=(0,0,0,\tau)$, which implies $\tensor{\dot{x}}{^\mu}=(0,0,0,1)$.

\bigskip

Our main goal is to compute the following two-point function in the presence of the defect:
\[
\vvev{\mathcal{O}(x_{1})\, \mathcal{O}(x_{2})}:=
\vev{\mathcal{W}_{l}\, \mathcal{O}(x_{1})\, \mathcal{O}(x_{2})} \,.
\]
Bosonic constraints on this type of correlators were originally studied in \cite{Billo:2016cpy} (see also \cite{Lauria:2017wav, Lauria:2018klo, Liendo:2019jpu}). A defect with such a geometry preserves the symmetry $\mathfrak{so}(1,2) \times \mathfrak{so}(3)$, where the first group corresponds to the $1$d conformal group on the line and the second one to rotations orthogonal to the defect. The quantum number associated to the $\mathfrak{so}(1,2)$ symmetry is commonly referred to as the \textit{transverse spin}, while the one corresponding to $\mathfrak{so}(3)$ is called the \textit{parallel spin}. Because this is a supersymmetric setup,  we must also consider the $\mathfrak{sp}(4)_\text{R}$ R-symmetry. The bosonic subalgebras plus the fermionic generators form the $\mathfrak{osp}(4|4)$ \textit{defect} superalgebra. Representations of $\mathfrak{osp}(4|4)$ are labeled by their conformal dimension $\hat{\Delta}$, transverse spin $s$ as well as by the $\mathfrak{sp}(4)_\text{R}$ Dynkin labels $[a,b]$. Following \cite{Liendo:2016ymz} we denote 1/2-BPS multiplets of $\mathfrak{osp}(4|4)$ by $(B,\pm)_{\hat{k}}$, with $\hat{k}$ labeling the $[0,\hat{k}]$ irreducible representations of the superconformal primary.\bigskip

The geometry of the system becomes more transparent if we choose a suitable frame. We can use scaling symmetry such that the operator $\mathcal{O}(x_1)$ is located at $x_1 = (1,0,0,0)$, while the second operator may be put in the 
$x$-$y$-plane, and hence its most general coordinates are $x_2 = (x, y , 0, 0)$. Since our setup contains two degrees of freedom, it is also convenient to define complex coordinates of the following form:
\begin{equation*}
z := x + i y \,, \quad \bar{z} := x - i y \,.
\end{equation*}
These coordinates are conformal invariants similar to the two cross-ratios that characterize four-point functions in CFTs without a defect. The configuration is depicted in fig.~\ref{fig1}.\footnote{In Lorentzian signature $z$ and $\bar{z}$ are real and independent.}

In addition to the spacetime invariants $(z,\bar{z})$ we can also define the variable $\omega$:
\begin{equation}
\frac{4 \omega}{(1-\omega)^2} := - \frac{(u_1 \cdot \theta) (u_2 \cdot \theta)}{(u_1 \cdot u_2)}\,,
\end{equation}
which plays the role of an R-symmetry invariant. This awkward-looking definition is justified by the appearance of the R-symmetry blocks given in \eqref{eq:bulkRsymmetry} and \eqref{eq:defectRsymmetry}. We recall that the $u$'s and $\theta$ couple respectively to the single-trace operators and to the line defect.

It will be convenient for later to define the following combination of bosonic invariants:

\begin{equation}
\Omega := \frac{\sqrt{z \bar{z}}}{(1-z)(1-\bar{z})}\, \frac{(1-\omega)^2}{4 \omega}\,.
\label{eq:invariant}
\end{equation}
Using $\Omega$ we may rewrite the correlator as
\begin{equation}
\vvev{ \mathcal{O}(x_1) \mathcal{O}(x_2) } = 
\frac{(u_1 \cdot \theta)^2 (u_2 \cdot \theta)^2}{x_1^2 x_2^2} \mathcal{F}(z , \bar{z} , \omega)\,,
\label{eq:superconformalinvariance}
\end{equation}
with
\begin{equation}
\label{eq:bigOmega_expansion}
\mathcal{F} (z , \bar{z} , \omega) = \sum_{c=0}^2 \Omega^{2-c} F_c (z , \bar{z})\,.
\end{equation}

The sum reflects the fact that there are three R-symmetry channels, which correspond to zero, one and two contractions of the $u_1, u_2$ of the single-trace operators with the $\theta$ vector of the Wilson-line defect. The interpretation of each channel depends on which OPE decomposition we are considering. In the bulk channel expansion they correspond to  $[0,k,0]$ irreps of $\mathfrak{so}(6)_\text{R}$ for $k=0,1,2$, while in the defect channel they label $[0,\hat{k}]$ irreps of $\mathfrak{sp}(4)_\text{R}$ for $\hat{k}=0,1,2$. All the $\omega$ dependence is now contained in $\Omega$, and the only unknowns left are the spacetime functions $F_c$. The latter have the nice property of being constant at the first two orders in perturbation theory, as we will show explicitly in section \ref{subsec:identityandleadingorders}.

\subsection{Superconformal Ward Identities}
\label{subsec:superconformalwardidentities}

Conformal and R-symmetry invariance imply that the two-point function can be written in terms of $(z,\bar{z})$ and $\omega$. These constraints however do not capture the full power of superconformal symmetry. In particular, invariance under the full superconformal algebra imposes the following analyticity property \cite{Dolan:2004mu,Liendo:2016ymz}:
\begin{subequations}
\begin{gather}
\left. \left( \partial_z + \frac{1}{2} \partial_\omega \right) \mathcal{F}(z,\bar{z},\omega) \right|_{z = \omega} = 0\,, \label{subeq:firstWI} \\
\left. \left( \partial_{\bar{z}} + \frac{1}{2} \partial_\omega \right) \mathcal{F}(z,\bar{z},\omega) \right|_{\bar{z} = \omega} = 0\,, \label{subeq:secondWI}
\end{gather}
\label{eq:WI}
\end{subequations}
which guarantees the absence of spurious poles when $z=\omega$ and $\bar{z}=\omega$.

It is easy to check that any power of the invariant $\Omega$ fulfills the Ward identities, i.e.
\begin{equation}
\left. \left( \partial_z + \frac{1}{2} \partial_\omega \right) \Omega^k \right|_{z = \omega} = 0 = \left. \left( \partial_{\bar{z}} + \frac{1}{2} \partial_\omega \right) \Omega^k \right|_{\bar{z} = \omega} \,.
\label{eq:invariantWI}
\end{equation}

Using the representation $\eqref{eq:bigOmega_expansion}$, we see that the three R-symmetry channels are not all independent. They satisfy differential constraints that we refer to as \textit{reduced Ward identities}:

\begin{equation}
\sum_{c=0}^2 \left( \frac{1}{4} \frac{1 - z}{1 - \bar{z}} \frac{\sqrt{z \bar{z}}}{z} \right)^{2-c} \partial_z F_c (z , \bar{z}) = 0 \,.
\label{eq:reducedWI}
\end{equation}
A second analogous identity follows straightforwardly from eq. (\ref{subeq:secondWI}).

In favorable situations, it is possible to solve the differential constraints in terms of a fewer number of independent functions,  like in $4d$ $\Nm=2$ and $\Nm=4$ SCFTs with no defects \cite{Dolan:2000ut,Dolan:2004mu}. For our system of equations however this is not possible. As pointed out in \cite{Liendo:2016ymz}, equations \eqref{subeq:firstWI} and \eqref{subeq:secondWI} are closely related to the Ward identities of $3d$ SCFTs in which the differential constraints are harder to solve.\footnote{See \cite{Dolan:2004mu} for an extensive discussion on how to solve Ward identities in different dimensions.} 

These identities provide a powerful check of the perturbative results that we will derive in section \ref{sec:perturbative}. In particular, the reduced Ward identities simplify considerably in the collinear limit $z = \bar{z}$, in which eq. (\ref{eq:reducedWI}) becomes

\begin{subequations}
\begin{gather}
F_0 (x,x) + 4 F_1 (x,x) + 16 F_2(x,x) = c_1, \quad \text{for } x>0\,, \\
F_0 (x,x) - 4 F_1 (x,x) + 16 F_2(x,x) = c_2, \quad \text{for } x<0\,.
\end{gather}
\label{eq:reducedWIcollinear}%
\end{subequations}
Here $c_1$ and $c_2$ are unknown numerical constants that in principle do not need be equal. This result corresponds to a topological subsector, which can also be directly obtained from \eqref{eq:bigOmega_expansion} by setting $z=\bar{z}=\omega$.

\subsection{Superblocks and Crossing Symmetry}
\label{subsec:superblocksandcrossingsymmetry}

In the presence of a defect, operators can acquire a non-vanishing expectation value, which explains why two-point functions (unlike the case of CFTs without a defect) can have a highly non-trivial dependence on the coordinates. Consider the OPE between the two external fields, the resulting expansion contains local operators whose expectation values in the presence of the defect is non-zero, the contribution of each local operator and its (super)conformal descendants can be neatly captured by a corresponding superconformal block:
\begin{equation}
\mathcal{F} (z,\bar{z},\omega) = \Omega^2 \sum_{\chi} \lambda_{\mathcal{O}\mathcal{O}\chi} a_{\chi} \mathcal{G}_{\chi} (z,\bar{z},\omega)\,.
\label{eq:bulkchannel}
\end{equation}
Here $\chi$ labels local operators which sit in irreps of the full $\mathfrak{psu}(2,2|4)$ superconformal algebra. As stated above, the relevant quantum numbers on this channel are $[\Delta, l, k ]$. The coefficients  $\lambda_{\mathcal{O}\mathcal{O}\chi}$ and $a_{\chi}$ correspond respectively to the three-point coupling and the one-point function associated to $\chi$. The superconformal block $\mathcal{G}_{\chi} (z,\bar{z},\omega)$ can be decomposed as
\begin{equation}
\mathcal{G}_\chi (z , \bar{z} , \omega) = \sum_{\Delta,k,l} c_{\Delta, k} (\chi) h_k (\omega) f_{\Delta, l} (z,\bar{z})\,,
\label{eq:bulksuperblock}
\end{equation}
which makes explicit the contribution of a full supermultiplet.

The $h_k (\omega)$ are R-symmetry blocks and $f_{\Delta, l} (z,\bar{z})$ are spacetime blocks. These functions are eigenfunctions of the Casimir operators of respectively $\mathfrak{so}(6)_\text{R}$ and $\mathfrak{so}(5,1)$. The coefficients $c_{\Delta, k} (\chi)$ are fixed kinematically and can be obtained, for example, by imposing that the superblock satisfies the Ward identities.

\bigskip

In addition to the standard OPE between two local operators, there is a new operator expansion specific to defect CFTs in which an individual local operator can be expanded by considering excitations constrained to the defect. We call these excitations  \textit{defect operators} and they allow us to write a second conformal block expansion which captures the contributions of superconformal families that live on the defect:
\begin{equation}
\mathcal{F}(z,\bar{z},\omega) = \sum_{\hat{\chi}} b_{\mathcal{O}\hat{\chi}}^2 \hat{\mathcal{G}}_{\hat{\chi}} (z,\bar{z},\omega)\,.
\label{eq:defectchannel}
\end{equation}
Here $\hat{\chi}$ corresponds to representations of $\mathfrak{osp}(4|4)$, which is the symmetry preserved by the defect. As mentioned before, the quantum numbers on this channel are $[\hat{\Delta},s,\hat{k}]$, while the coefficients $b_{\mathcal{O}\hat{\chi}}$ are bulk-to-defect couplings that characterize correlators between one bulk and one defect operators. These correlators are fixed by kinematics and therefore the $b_{\mathcal{O}\hat{\chi}}$ plays a similar role as three-point couplings in standard CFTs. The superconformal blocks $\hat{\mathcal{G}}_{\hat{\chi}} (z,\bar{z},\omega)$ can be decomposed as

\begin{equation}
\hat{\mathcal{G}}_{\hat{\chi}} (z , \bar{z} , \omega) = \sum_{\hat{\Delta} , \hat{k} , s} c_{\hat{\Delta} , \hat{k}} (\hat{\chi} ) \hat{h}_{\hat{k}} (\omega) \hat{f}_{\hat{\Delta},0,s} (z, \bar{z})\,.
\label{eq:defectsuperblock}
\end{equation}
The $\hat{h}_{\hat{k}} (\omega)$ are R-symmetry blocks, while the $\hat{f}_{\hat{\Delta},0,s} (z, \bar{z})$ are spacetime blocks, which capture the bosonic contributions of the defect algebra. Like before, these functions can be obtained by solving the Casimir equations, and the relative coefficients in \eqref{eq:defectsuperblock} can be fixed using the Ward identities.

\bigskip

\begin{figure}[t]
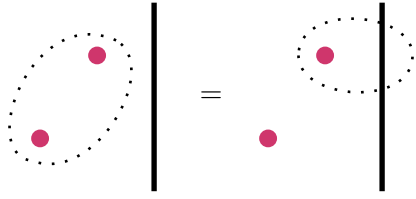

\centering
\crossingsymmetry
\caption{Pictorial representation of the crossing equation given in eq. \ref{eq:crossing}. The correlator can be expanded in two ways, either by performing a \textit{bulk} OPE (left-hand side) or a \textit{defect} OPE (right-hand side).}
\label{fig2}
\end{figure}

The two ways of expanding the correlator result in a ``crossing equation'':

\begin{equation}
\Omega^2 \sum_{\chi} \lambda_{\mathcal{O}\mathcal{O}\chi} a_{\chi} \mathcal{G}_{\chi} (z,\bar{z},\omega) \overset{!}{=} \sum_{\hat{\chi}} b_{\mathcal{O}\hat{\chi}}^2 \hat{\mathcal{G}}_{\hat{\chi}} (z,\bar{z},\omega)\,.
\label{eq:crossing}
\end{equation}
A pictorial representation of this relation is shown in fig. \ref{fig2}. This equation is the starting point of the defect bootstrap program, whose goal is to constrain that landscape of defect CFTs and hopefully solve individual models. In this work we will not follow the bootstrap approach, but rely instead on standard perturbation theory techniques. It will be nevertheless crucial to use defect CFT techniques in order to write our final result. 
Feynman integrals are often hard to solve in closed form and the superblock expansions will allow us to extract the CFT data, which in the end is the dynamical information that characterizes the correlator.

\bigskip

Before we jump to the perturbative calculation, we still need to specify the precise multiplets of both the $\mathfrak{psu}(2,2|4)$ and $\mathfrak{osp}(4|4)$ algebras that can appear on each channel. This analysis was done in \cite{Liendo:2016ymz} where the allowed multiplets were obtained by analytically continuing the results for a codimension-one defect (i.e. the boundary). 

\paragraph{Bulk channel selection rules.}The spectrum of allowed representations in the bulk OPE corresponds to the multiplets that can appear in the OPE of two $20'$ and in addition can have non-vanishing one-point functions. The list is as follows:
\begin{itemize}
\item the identity operator $\mathds{1}$;
\item $1/2$-BPS operators $\mathcal{B}_{[0,2k,0]}$ with $k=1,2$;
\item semishort blocks $\mathcal{C}_{[0,2,0],l}$ with $l \geq 0$ and even;
\item long blocks $\mathcal{A}^\Delta_{[0,0,0],l}$ with $\Delta \geq l+2$, $l \geq 0$ and even.
\end{itemize}
Here we are using the notation of \cite{Dolan:2002zh}; $l:=2j$ is the spin of the operator and $\Delta$ the scaling dimension. Notice that only $\mathfrak{so}(6)_\text{R}$ irreps of the type $[0,p,0]$ can contribute to this OPE. The superblocks are given explicitly in appendix \ref{subsec:bulkchannel}. One could also consider semishorts  $\mathcal{C}_{[0,0,0],l}$ multiplets, however these multiplets contain higher-spin conserved currents and are disallowed in interacting theories \cite{Maldacena:2011jn, Alba:2013yda}. We will see this type of multiplets at the leading order of our calculations, however we will interpret them as long multiplets $\mathcal{A}^\Delta_{[0,0,0],l}$ at the unitarity bound $\Delta = l + 2$, which will recombine when we add higher-order corrections.\footnote{When $\mathcal{A}^\Delta_{[0,0,0],l}$ hits the unitarity bound the decomposition is a bit more involved, there are four different multiplets appearing, not just $\mathcal{C}_{[0,0,0],l}$. However the other three multiplets do not contribute to our block expansion, see \cite{Dolan:2002zh} for the precise shortening/recombination rules.}

Summarizing, the bulk channel superconformal block expansion reads

\begin{equation}
\Omega^{-2} \mathcal{F}(z , \bar{z} , \omega) = A + B  \mathcal{G}_{\mathcal{B}_{[0,2,0]}} + C  \mathcal{G}_{\mathcal{B}_{[0,4,0]}} + \sum_{l} D_l \mathcal{G}_{\mathcal{C}_{[0,2,0],l}} + \sum_{\Delta,l} E_{\Delta,l} \mathcal{G}_{\mathcal{A}^\Delta_{[0,0,0],l}}\,,
\label{eq:bulkchannelexpansion}
\end{equation}
where we have suppressed the ($z,\bar{z}$) and $\omega$ dependence on the right-hand side for compactness. The dynamical information in this expansion is contained in the CFT data, which corresponds to the spectrum of operators and to the coefficients in front of the blocks. The latter are given by the product of bulk three-point function coefficients and defect one-point functions. In order to avoid cluttering, we have simplified the notation a bit:
\begin{equation*}
A := \lambda_{\mathcal{O} \mathcal{O} \mathds{1}} a_{\mathds{1}} = b_{\mathcal{O}\mathcal{O}}\,, \quad B := \lambda_{\mathcal{O}\mathcal{O}\mathcal{O}_{\mathcal{B}_{[0,2,0]}}} a_{\mathcal{O}_{\mathcal{B}_{[0,2,0]}}}\,, \quad ...
\end{equation*}
where the dots stand for similar definitions for the rest of coefficients.

\paragraph{Defect channel selection rules.} In the defect channel, the following representations have non-vanishing two-point functions in presence of the defect:

\begin{itemize}
\item the identity operator $\hat{\mathds{1}}$;
\item $1/2$-BPS operators: $(B,+)_{\hat{k}} := [2\hat{k},0,\hat{k}]$, with $\hat{k}=1,2$;
\item $1/4$-BPS operators: $(B,1)_{[\hat{k},s]} := [2\hat{k},s,\hat{k}]$, with $s \geq 0$ and $\hat{k}=0,1$;
\item long operators: $L^{\hat{\Delta}}_{[0,s]} := [\hat{\Delta},s,0]$, with $\hat{\Delta} \geq s+1$ and $s \geq 0$.
\end{itemize}
The blocks can be found in appendix \ref{subsec:defectchannel}. Note that at the unitarity bound $\hat{\Delta} = s+1$, the long operators shorten and the corresponding CFT data coefficients are indistinguishable from the $(B,1)_{[0,s]}$ ones, since

\begin{equation}
\hat{\mathcal{G}}_{L^{s+1}_{[0,s]}} = - \hat{\mathcal{G}}_{(B,1)_{[0,s]}}\,.
\end{equation}
This relation is classical and will be spoiled by anomalous dimensions at loop level, thus guaranteeing that the expansion in superblocks remains unique.\footnote{Like in the bulk case, multiplet shortening/recombination is a bit more involved, the precise relation can be easily calculated using superconformal characters (see for example \cite{Dolan:2008vc} for a closely related analysis).} Notice that unlike the bulk channel there is no theorem that guarantees the absence of these multiplets in the interacting theory, the result of \cite{Maldacena:2011jn, Alba:2013yda} assumes a local CFT, defect theories are generically non-local, especially in the one-dimensional case we are studying.

In summary, the expansion in the defect channel reads

\begin{align}
\mathcal{F}(z , \bar{z} , \omega) &= \hat{A} + \hat{B} \hat{\mathcal{G}}_{(B,+)_1} + \hat{C} \hat{\mathcal{G}}_{(B,+)_2} + \sum_{s=0}^\infty \hat{D}_s \hat{\mathcal{G}}_{(B,1)_{[0,s]}} 
+ \sum_{s=0}^\infty \hat{E}_s \hat{\mathcal{G}}_{(B,1)_{[1,s]}} + \sum_{\hat{\Delta},s} \hat{F}_{\hat{\Delta},s} \hat{\mathcal{G}}_{L^{\hat{\Delta}}_{[0,s]}}\,.
\label{eq:defectchannelexpansion}
\end{align}
The dynamical CFT data are the spectrum of defect excitations and the bulk-to-defect coefficients. Again we use a shorthand notation to avoid cluttering:

\begin{equation*}
\hat{A} := b^2_{\mathcal{O}\hat{\mathds{1}}} = a_{\mathcal{O}}^2\,, \quad \hat{B} := b^2_{\mathcal{O}\hat{\mathcal{O}}_{(B,+)_1}}\,, \quad ...
\end{equation*}
These are defect two-point function coefficients between the bulk external operator and a defect operator.

%!TEX root = ../2pt_function_wline.tex
%%%%%%%%%%%%%%%%%%%%%%%%%%%%%%%%%%%%%%%%%%%%%

\section{Perturbative Computation}
\label{sec:perturbative}

The goal of this section is to compute the correlator perturbatively up to next-to-leading order $\mathcal{O}(g^8)$. After introducing the Feynman rules, we list all the diagrams relevant for leading order $\mathcal{O}(g^6)$ and  next-to-leading order. At leading order the integrals involved are simple and it is not hard to write a closed-form expression for the correlator. At next-to-leading order the complexity increases dramatically, and of the three R-symmetry channels we managed to obtain one in closed form. Nevertheless, the symmetry analysis of the previous section will allow us to extract the CFT data individually. This is a challenging technical task that we perform in section \ref{sec:cft_data}.

\subsection{Action and Propagators}
\label{subsec:action}

The action of $\mathcal{N}=4$ SYM is given by (including ghosts and gauge fixing)
\begin{align}
S &= \frac{1}{g^2} \int d^4 x\ \text{Tr} \left\lbrace \frac{1}{2} \tensor{F}{_{\mu\nu}} \tensor{F}{^{\mu\nu}} + \tensor{D}{_\mu} \tensor{\phi}{_i} \tensor{D}{^\mu} \tensor{\phi}{^i} - \frac{1}{2} [ \tensor{\phi}{_i} , \tensor{\phi}{_j} ] [ \tensor{\phi}{^i} , \tensor{\phi}{^j} ] \right. \notag \\
& \left. \qquad \qquad \qquad \qquad \qquad \qquad + i \bar{\psi} \tensor{\gamma}{^\mu} \tensor{D}{_\mu} \psi + \bar{\psi} \tensor{\Gamma}{^i} [ \tensor{\phi}{_i} , \psi ] + \tensor{\partial}{_\mu} \bar{c} \tensor{D}{^\mu} c + \xi \left( \tensor{\partial}{_\mu} \tensor{A}{^\mu} \right)^2 \right\rbrace\;.
\label{eq:action}
\end{align}
Our conventions are gathered in appendix \ref{app:insertionrules}. The resulting propagators in Feynman gauge ($\xi = 1$) take the following form in configuration space:
\begin{subequations}
\begin{align}
\text{Scalars:} \qquad 
& \propagatorS = g^2 \tensor{\delta}{_{ij}} \tensor{\delta}{^{ab}} I_{12}\;, \label{subeq:propagatorS} \\
\text{Gluons:} \qquad 
& \propagatorG = g^2 \tensor{\delta}{_{\mu\nu}} \tensor{\delta}{^{ab}} I_{12}\;, \label{subeq:propagatorG} \\
\text{Gluinos:} \qquad 
& \propagatorF = i g^2 \tensor{\delta}{^{ab}} \slashed{\partial}_{\Delta} I_{12}\;, \label{subeq:propagatorF} \\
\text{Ghosts:} \qquad 
& \propagatorGh = g^2 \tensor{\delta}{^{ab}} I_{12}\;, \label{subeq:propagatorGh}
\end{align}
\label{eq:propagators}%
\end{subequations}
where we have defined for brevity
\begin{equation}
I_{12} := \frac{1}{(2\pi)^2 x_{12}^2}\;,
\label{eq:I12}
\end{equation}
with $x_{ij} := x_i - x_j$ and
\begin{equation*}
\slashed{\partial}_{\Delta} := \gamma \cdot \frac{\partial}{\partial \Delta}\;, \qquad \qquad \Delta := x_1 - x_2\;,
\end{equation*}
with $\gamma_\mu$ the Dirac matrices. The Feynman rules follow straightforwardly, and we have assembled the relevant ones into a set of insertion rules in appendix \ref{app:insertionrules}.

\subsection{Feynman Diagrams}
\label{subsec:feynmandiagrams}

We wish now to list the diagrams that are relevant for the computation up to order $\mathcal{O}(g^8)$. We start by noting that a large number of diagrams are irrelevant because of the color contractions:
\begin{equation}
\vanishtwoptSS\ ,\ \vanishtwoptSG\ ,\ \vanishtwoptGG\ ,\ \vanishBPS\ \propto f^{ada} = 0\;.
\label{subeq:vanishtwoptSS}
\end{equation}
Here the first three diagrams refer to insertions on the Wilson line, while the rightmost circle in the fourth one represents the trace of the operators defined in (\ref{eq:singletraceoperators}).

\bigskip

The lowest order at which the two-point function receives a contribution is the order $\mathcal{O}(g^4)$, which we will call \textit{identity order}. It corresponds to the two-point function of single-trace operators \textit{disconnected} from the line:
\begin{equation*}
\diagTLzero
\end{equation*}
As mentioned before, the disconnected two-point function is protected and does not receive radiative corrections at higher orders of the coupling. The R-symmetry factor of this diagram is $(u_1 \cdot u_2)^2$, and hence only the $0$-channel is non-zero at this order (see section \ref{subsec:twopointfunction}).

\bigskip

At leading order $\mathcal{O}(g^6)$, we find the first contribution to the two-point function that involves the defect:
\begin{equation*}
\diagTLone
\end{equation*}
This diagram has a R-symmetry factor $(u_1 \cdot u_2)(u_1 \cdot \theta)(u_2 \cdot \theta)$ and therefore only the $1$-channel is non-vanishing at leading order.

\bigskip

The number of diagrams increases significantly at next-to-leading order $\mathcal{O}(g^8)$, and in particular we find that all the three R-symmetry channels are represented. The planar contributions are gathered in table \ref{table:diagrams}.

\begin{table}
\centering
\caption{Feynman diagrams in the planar limit $N \to \infty$ for the computation of the two-point function with line defect at next-to-leading order (NLO). The configurations are classified in function of their R-symmetry channel. It should be understood that all (planar) path orderings on the Wilson line have to be considered for each diagram.}
\begin{tabular}{cl}
\\ \hline \\[-.5em]
$0$-Channel & \begin{tabular}{l} \resizebox{.12\textwidth}{!}{\diagXone} \quad \resizebox{.12\textwidth}{!}{\diagXdtwo} \quad \resizebox{.12\textwidth}{!}{\diagHd} \quad \resizebox{.15\textwidth}{!}{\diagYY} \end{tabular} \\[2em] \hline \\[-.5em]
$1$-Channel & \begin{tabular}{l} \resizebox{.12\textwidth}{!}{\diagSEone} \quad \resizebox{.12\textwidth}{!}{\diagXone} \quad \resizebox{.12\textwidth}{!}{\diagHhcone} \quad \resizebox{.12\textwidth}{!}{\diagIYIone} \\[2em] \resizebox{.12\textwidth}{!}{\diagSEtwo} \quad \resizebox{.12\textwidth}{!}{\diagXhctwo} \quad \resizebox{.12\textwidth}{!}{\diagHhctwo} \quad \resizebox{.12\textwidth}{!}{\diagIYItwo} \\[2em] \resizebox{.12\textwidth}{!}{\diagSEthree} \quad \resizebox{.12\textwidth}{!}{\diagXhcthree} \quad \resizebox{.12\textwidth}{!}{\diagHhcthree} \quad \resizebox{.12\textwidth}{!}{\diagIYIthree} \\[2em] \end{tabular} \\[5em] \hline \\[-.5em]
$2$-Channel & \multicolumn{1}{c}{\resizebox{.12\textwidth}{!}{\planarone}} \\[2em] \hline
\end{tabular}
\label{table:diagrams}
\end{table}

\subsection{Identity and Leading Orders}
\label{subsec:identityandleadingorders}

We proceed now with the computation of the identity and leading orders. We have seen that only one diagram contributes at $\mathcal{O}(g^4)$, and it is trivial to compute since there is no integral involved:
\begin{equation}
\diagTLzero\ = \frac{(u_1 \cdot \theta)^2 (u_2 \cdot \theta)^2}{x_1^2 x_2^2} \frac{g^4 N^2}{2^5 \pi^4} \Omega^2\;.
\label{eq:resultidentityorder}
\end{equation}
The corresponding $F_c$-functions defined in section \ref{subsec:twopointfunction} are easy to read:

\begin{subequations}
\begin{gather}
F_0 (z , \bar{z}) = \frac{g^4 N^2}{2^5 \pi^4}\,, \\
F_1 (z,\bar{z}) = F_2 (z,\bar{z}) = 0\,.
\end{gather}
\label{eq:gidentity}
\end{subequations}

The leading order is also simple to compute, since it contains only two elementary (and independent) one-dimensional integrals:
\begin{align}
\diagTLone\ &= (u_1 \cdot u_2)(u_1\cdot\theta)(u_2\cdot\theta) g^6 N I_{12} \int d\tau_3 \int d\tau_4\ I_{13} I_{24} \notag \\
&= \frac{(u_1 \cdot \theta)^2 (u_2 \cdot \theta)^2}{x_1^2 x_2^2} \frac{g^6 N}{2^6 \pi^4} \Omega\,.
\label{eq:resultleadingorder}
\end{align}
This translates into the following $F_c$-functions:
\begin{subequations}
\begin{gather}
F_1 (z , \bar{z}) = \frac{g^6 N}{2^6 \pi^4}\,, \\
\intertext{while the other channels vanish, i.e.}
F_0 (z , \bar{z}) = F_2 (z , \bar{z}) = 0\,.
\end{gather}
\label{eq:leadingorder}%
\end{subequations}
Notice that they are constant, i.e. both the identity and leading orders manifestly satisfy the reduced Ward identities given in eq. (\ref{eq:reducedWI}).

The full correlator up to order $\mathcal{O}(g^6)$ reads
\begin{equation}
\vvev{\mathcal{O}(x_1) \mathcal{O}(x_2)}_{\mathcal{O}(g^6)} = \frac{(u_1 \cdot \theta)^2 (u_2 \cdot \theta)^2}{x_1^2 x_2^2} \frac{g^4 N^2}{2^5 \pi^4} \left\lbrace \Omega^2 + \frac{g^2}{2 N} \Omega \right\rbrace\,.
\end{equation}

\subsection{Next-to-Leading Order}
\label{subsec:nexttoleadingorder}

We now turn our attention to the more challenging computation of the correlator at next-to-leading order. In order to have compact expressions, it is convenient to define the following R-symmetry factor:
\begin{equation}
\lambda_c := g^8 N^2 (u_1 \cdot u_2)^{2-c} (u_1 \cdot \theta)^c (u_2 \cdot \theta)^c\,,
\end{equation}
where $c=0,1,2$ refers to the R-symmetry channel just as it was the case in section \ref{subsec:twopointfunction}. The integrals used throughout this section are listed in appendix \ref{app:integrals}.

\subsubsection{2-Channel}
\label{subsubsec:2channel}

We start by considering the $2$-channel, which is the simplest one since it consists of only one diagram without any vertex. After discarding the non-planar diagrams and doing the Wick contractions, it reads
\begin{equation*}
\planarone\ = \frac{\lambda_2}{8} \int d\tau_3 \int d\tau_4 \int d\tau_5 \int d\tau_6\ \Theta(\tau_{3456})\ (I_{13}I_{25} + I_{15}I_{23})(I_{14}I_{26} + I_{16}I_{24})\,.
\end{equation*}
The structure of the integrand comes from the fact that all permutations of the points on the Wilson line had to be considered. $\Theta(\tau_{3456})$ is a path-ordering variable, defined explicitly in (\ref{eq:pathorderingsymbol2}). The $F$-function of the $2$-channel can be expressed as
\begin{equation*}
F_2 (z , \bar{z}) = \frac{1}{8} g^8 N^2\ I(1, x_2^2)\,,
\end{equation*}
with $I(x_1^2, x_2^2)$ the integral given above. The diagram depends manifestly only on $x_2^2$, i.e. the only variable is the distance between the operator and the line, while the distance between the two bulk operators is irrelevant for this channel.

\bigskip

The computation of the integral can be performed analytically, except for the last one-dimensional integral. We were able to find an exact expansion of the remaining integral using numerical data, which can be resummed into the following expression:
\begin{align}
F_2 (z , \bar{z}) &= \frac{g^8 N^2}{2^{12} \pi^6} \left\lbrace 3 \pi^2 - 4 i \pi \log 2 + 4 \tanh^{-1} \sqrt{z \bar{z}} \left( \log z \bar{z} + 4 \log 2 - 2 \tanh^{-1} \sqrt{z \bar{z}}  \right) \phantom{\frac{1}{2}} \right. \notag \\
& \qquad + 4 \log^2 \left( 1 - \sqrt{z\bar{z}} \right) + 2 \log \left( \sqrt{z\bar{z}} - 1 \right) \left[ -2 \log \left( 1 - \sqrt{z\bar{z}} \right) \right. \notag \\
& \qquad \qquad \left. + \log \left( \sqrt{z\bar{z}} - 1 \right) + 2 \log 2 \right] - 2 \log \left(  1 + \sqrt{z\bar{z}}\right) \log 4 \left( 1 + \sqrt{z\bar{z}} \right) \notag \\
& \qquad \qquad \qquad + 4 \text{Li}_2 \left( -\sqrt{z\bar{z}} \right) - 4 \text{Li}_2 \sqrt{z\bar{z}} - 4 \text{Li}_2 \frac{1}{2} \left( 1 - \sqrt{z\bar{z}} \right) \notag \\
& \qquad \qquad \qquad \qquad \left. + 4 \text{Li}_2 \frac{1}{2} \left( 1 + \sqrt{z\bar{z}} \right) \right\rbrace\,.
\label{eq:result2channel}
\end{align}
The procedure for finding this closed form is detailed in appendix \ref{subsubsec:2channelatNLO}, and is plotted in fig. \ref{fig:2and1channels} for the limit $z = \bar{z}$\footnote{Since the diagram only depends on one variable $x_2^2$, i.e. $z \bar{z}$, the plot shown in the figure actually represents the channel everywhere.}.

\subsubsection{1-Channel}
\label{subsubsec:1channel}

\begin{figure}
\centering
\begin{subfigure}{.5\textwidth}
  \centering
  \includegraphics[width=.95\linewidth]{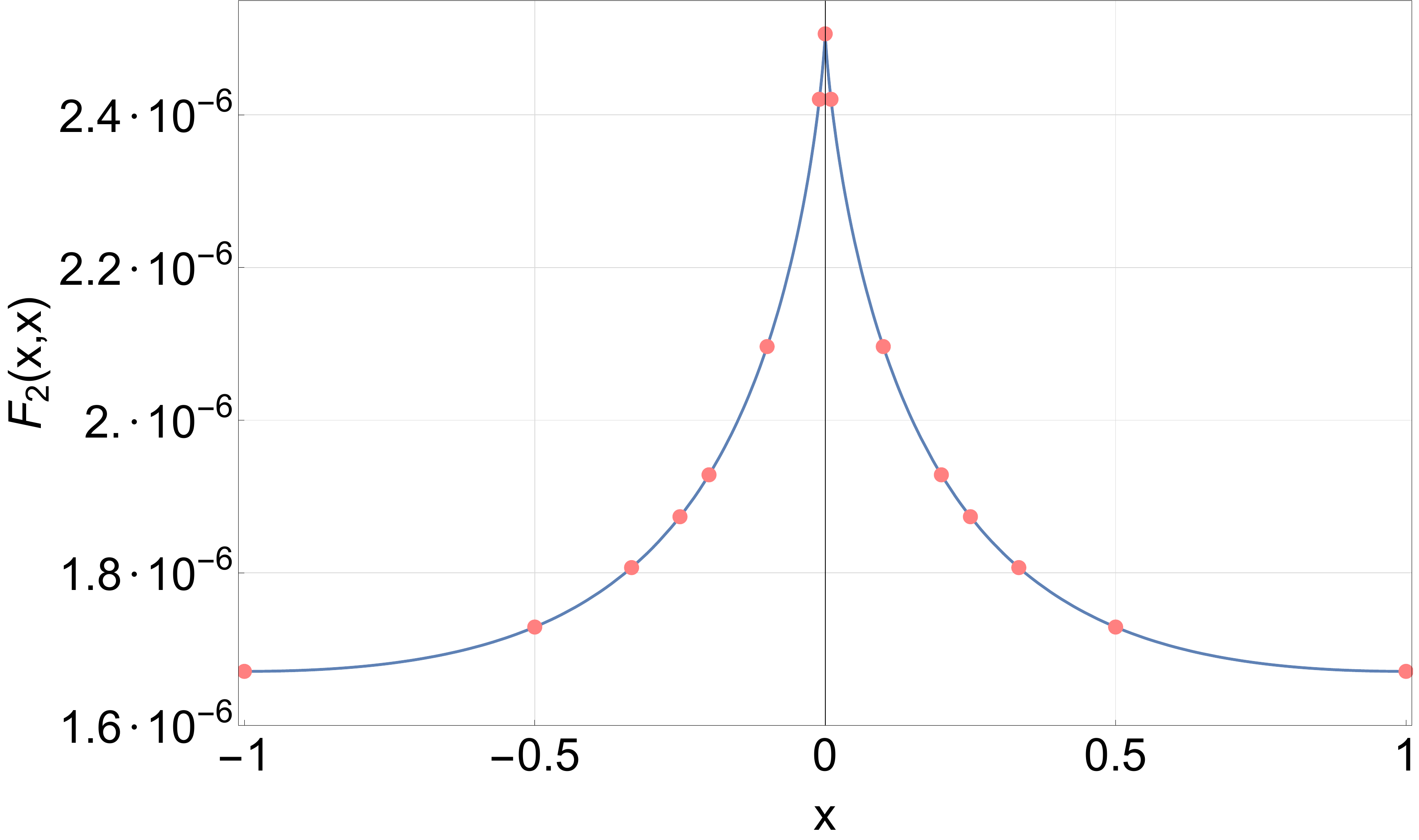}
\end{subfigure}%
\begin{subfigure}{.5\textwidth}
  \centering
  \includegraphics[width=.95\linewidth]{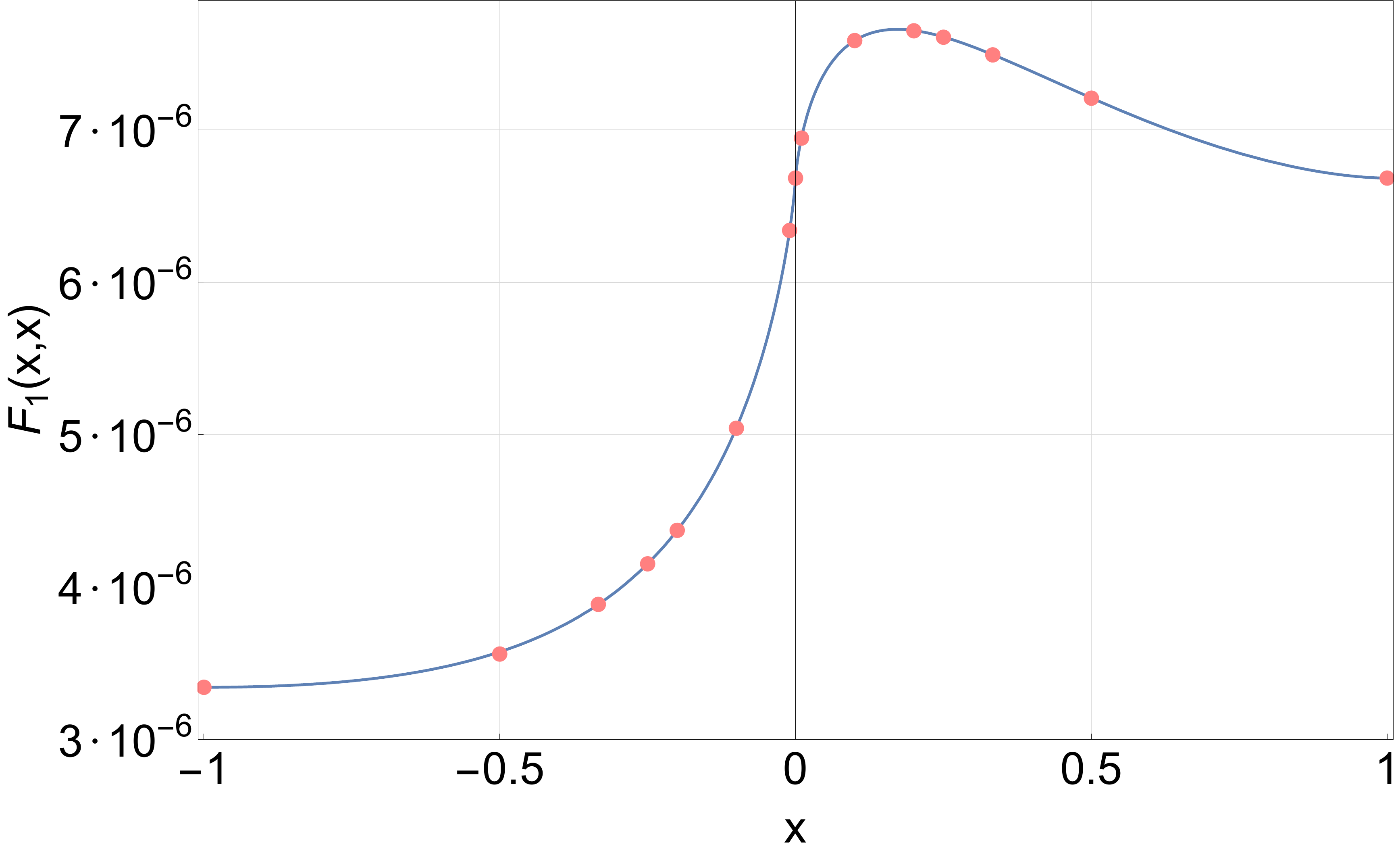}
\end{subfigure}
\caption{Plots of the $F$-functions for the $2$-channel (left) as well as for the $1$-channel (right). In both cases, the dots correspond to the data gathered by numerical integration, while the solid lines show the collinear limit of eq. (\ref{eq:result2channel}) for the $2$-channel (left) and the resummation of the $1$-channel as given in eq. (\ref{eq:1channelresummed}) (right). For both channels we observe a perfect agreement between numerical data and analytical expressions.}
\label{fig:2and1channels}
\end{figure}

The $1$-channel contains many more diagrams, which now include vertices and that we therefore expect to be significantly more difficult. We will not be able to solve the integrals analytically, however they can be computed numerically in the collinear limit $z = \bar{z}$. For this channel, we will first focus on the point $z = \bar{z} = 0$, and in section \ref{subsubsec:resummation} we will show that the channel can be resummed in the collinear limit.

\bigskip

The first category of diagrams that we study corresponds to the ones for which a divergence arises when we pinch the vertices towards one operator. We will refer to such diagrams as \textit{corner diagrams} (in which we also include the self-energy diagrams).

\bigskip

The X-diagram pinched at $x_1$ results in the following expression (after performing the contractions and taking the symmetry factors into account):
\begin{equation}
\diagXhctwo\ = - \lambda_1\ \int d\tau_3 \int d\tau_4\ I_{24}\ X_{1123}\,,
\label{eq:diagXhctwo}
\end{equation}
where we note that the pinching limit $X_{1123}$ is known analytically and is given by eq. (\ref{eq:X1123}). Because cancellations will occur, let us first collect the expressions associated to each diagram before considering the $\tau$-integrals. For now we simply note that this diagram is logarithmically divergent.

The H-diagram pinched at $x_1$ turns out to be
\begin{align}
\diagHhctwo\ &= - \lambda_1\ I_{12} \int d\tau_3 \int d\tau_4\ I_{13} I_{24} \left\lbrace - \frac{X_{1123}}{I_{12} I_{13}} + \frac{Y_{112}}{I_{12}} + \frac{Y_{113}}{I_{13}} \right. \notag \\
& \qquad \qquad \qquad \qquad \qquad \qquad \qquad \left. + \left(\frac{1}{I_{12}} + \frac{1}{I_{13}} - \frac{2}{I_{23}} \right) Y_{123} \right\rbrace\,,
\label{eq:diagHhctwo}
\end{align}
where we have made use of the pinched integral identity given in eq. (\ref{eq:F1314}). The three first terms are also logarithmically divergent.

The self-energy diagrams are straightforward to read using (\ref{eq:selfenergy}) and give
\begin{equation}
\diagSEone\ = 2\ \lambda_1 \int d\tau_3 \int d\tau_4\ I_{13} I_{24}\ Y_{112}\,,
\label{eq:diagSEone}
\end{equation}
as well as
\begin{equation}
\diagSEtwo\ = 2\ \lambda_1\ I_{12} \int d\tau_3 \int d\tau_4\ I_{24}\ Y_{113}\,.
\label{eq:diagSEtwo}
\end{equation}
It is well-known that these diagrams are also logarithmically divergent.

Comparing expressions in eq. (\ref{eq:diagXhctwo}-\ref{eq:diagSEtwo}) reveals that many terms occur several times with different signs. Hence it makes sense to group the diagrams in an \textit{upper-right corner} diagram as follows:
\begin{align}
\diagCone\ & :=\ \diagXhctwo\ +\ \diagHhctwo\ + \frac{1}{2} \left(\ \diagSEtwo\ +\ \diagSEone\ \right) \notag \\
& = - \lambda_1\ I_{12} \int d\tau_3 \int d\tau_4\ I_{13} I_{24} \left( \frac{1}{I_{12}} + \frac{1}{I_{13}} - \frac{2}{I_{23}} \right) Y_{123}\,.
\label{eq:diagCone}
\end{align}
Due to cancellations, the expression has simplified considerably and the integral is now finite. Note that this expression is consistent with the treatment of corner interactions performed in \cite{Drukker:2008pi}.

In a fully analogous way, we can treat the diagrams of the opposite corner, using the remaining half of the symmetric self-energy diagram:
\begin{align}
\diagCtwo\ & :=\ \diagXhcthree\ +\ \diagHhcthree\ + \frac{1}{2} \left(\ \diagSEthree\ +\ \diagSEone\ \right) \notag \\
& = - \lambda_1\ I_{12} \int d\tau_3 \int d\tau_4\ I_{13} I_{24} \left( \frac{1}{I_{12}} + \frac{1}{I_{24}} - \frac{2}{I_{14}} \right) Y_{124}\,.
\label{eq:diagCtwo}
\end{align}

The corner IYI-diagram at $x_1$ is defined as follows:
\begin{equation*}
\diagIYItwo\ = \lambda_1\ I_{12} \int d\tau_3 \int d\tau_4 \int d\tau_5\ \epsilon(\tau_3\ \tau_4\ \tau_5)\ I_{13} \left(\partial_{\tau_2} - \partial_{\tau_5} \right) Y_{245}\,,
\end{equation*}
where we recall that all permutations of the legs connected to the line are considered. The path-ordering symbol $\epsilon(\tau_3\ \tau_4\ \tau_5)$ is defined explicitly by eq. (\ref{eq:pathorderingsymbol}). This expression can be further simplified in the following way. Using integration by parts, one can rewrite $\left(\partial_{\tau_2} - \partial_{\tau_5} \right) Y_{245} \widehat{=} -\left(\partial_{\tau_4} + 2\partial_{\tau_5} \right) Y_{245}$. Since the derivatives now only act on the Wilson-line points, we can integrate by parts again with respect to $\tau_4$ and $\tau_5$, and use the fact that
\begin{equation*}
\partial_{\tau_4} \epsilon(\tau_3\ \tau_4\ \tau_5) = 2 \left( \delta(\tau_{45}) - \delta(\tau_{43}) \right)\,,
\end{equation*}
with $\tau_{ij} := \tau_i - \tau_j$ as usual. The $\delta$-functions kill one $\tau$-integral, and we are left with
\begin{equation}
\diagIYItwo\ = \lambda_1\ I_{12} \int d\tau_3 \int d\tau_4\ I_{13} \left( Y_{234} - Y_{244} \right)\,.
\label{eq:diagIYItwo}
\end{equation}

We can combine this diagram with the remaining half of the corresponding self-energy contribution to obtain the following \textit{upper-left corner} diagram:
\begin{align}
\diagCthree\ & :=\ \diagIYItwo\ + \frac{1}{2}\ \diagSEtwo \notag \\
&= \lambda_1\ I_{12} \int d\tau_3 \int d\tau_4\ I_{13}\ Y_{234}\,.
\label{eq:diagCthree}
\end{align}
Once again, we see that the pinched integrals cancel, thus making this expression finite. Similarly, we define the following \textit{lower-left corner} diagram:
\begin{align}
\diagCfour\ & :=\ \diagIYIthree\ + \frac{1}{2}\ \diagSEtwo \notag \\
&= \lambda_1\ I_{12} \int d\tau_3 \int d\tau_4\ I_{24}\ Y_{134}\;.
\label{eq:diagCfour}
\end{align}

Another class of diagrams consists of the symmetric diagrams, in the sense that the integrals are invariant under a permutation $x_1 \leftrightarrow x_2$. Using the $4$-point insertion rules from appendix \ref{subsec:bulkinsertions}, it is straightforward to write down the expressions corresponding to each diagram and to perform the Wick contractions. There is one X-diagram, which reads
\begin{equation}
\left. \diagXone\ \right|_1 = - \lambda_1\ I_{12} \int d\tau_3 \int d\tau_4\ X_{1234}\,,
\end{equation}
where the subscript $1$ means that we only consider the $1$-channel of the diagram\footnote{Indeed the diagram contributes in principle also to the $0$-channel, as we will see in the next subsection.}.

There is also one H-diagram, which gives:
\begin{equation}
\diagHhcone\ = - \lambda_1\ I_{12} \int d\tau_3 \int d\tau_4\ I_{13} I_{24}\ F_{13,24}\;,
\label{eq:Hdiagram}
\end{equation}
where the F- and X-integrals are defined in appendix \ref{subsec:standardintegrals}. Note that we have made use of the insertion rules (\ref{eq:Hinsertone}) and of the fact that
\begin{equation}
\int d\tau_3 \int d\tau_4\ \left\lbrace I_{13} I_{24}\ F_{13,24} + I_{14} I_{23}\ F_{14,23} \right\rbrace = 2 \int d\tau_3 \int d\tau_4\ I_{13} I_{24}\ F_{13,24}\,.
\end{equation}

In the same way, the symmetric IYI-diagram yields
\begin{equation*}
\diagIYIone\ = - \frac{\lambda_1}{2} \int d\tau_3 \int d\tau_4 \int d\tau_5\ \epsilon(\tau_3\ \tau_4\ \tau_5)\ I_{13} I_{25} \left( \partial_{\tau_1} - \partial_{\tau_2} \right) Y_{124}\,.
\end{equation*}
For this diagram, the $\tau_3$- and $\tau_5$-integrals can be performed as follows: insert the definition (\ref{eq:pathorderingsymbol}) of the path-ordering symbol $\epsilon(\tau_3\ \tau_4\ \tau_5)$, split e.g. the $\tau_5$-integral into pieces such that the signum functions involving $\tau_5$ can be eliminated, and integrate termwise. This results in the following expression:
\begin{equation*}
\int d\tau_5\ \epsilon(\tau_3\ \tau_4\ \tau_5)\ I_{25} = \frac{1}{(2\pi)^2 \left| x_2 \right|} \left\lbrace 2 \left( \tan^{-1} \frac{\tau_4}{\left| x_2 \right|} - \tan^{-1} \frac{\tau_3}{\left| x_2 \right|} \right) + \pi \sgn \tau_{34} \right\rbrace\,.
\end{equation*}
The $\tau_3$-integral is elementary, and the IYI-diagram turns out to be
\begin{equation}
\diagIYIone\ = - \frac{1}{2} \frac{\lambda_1}{(2\pi)^3 \left| x_1 \right| \left| x_2 \right|} \int d\tau_4\ \left( \tan^{-1} \frac{\tau_4}{\left| x_ 2 \right|} - \tan^{-1} \frac{\tau_4}{\left| x_ 1 \right|} \right) \left( \partial_{\tau_1} - \partial_{\tau_2} \right) Y_{124}\,.
\label{eq:diagIYIone}
\end{equation}
There remains only a one-dimensional integral to do for this diagram. All the symmetric diagrams are finite, and hence the $1$-channel is also finite on its own as expected.

\bigskip

Since the expressions for the Y-, X- and F-integrals are known analytically, we are left with one-dimensional and two-dimensional integrals. We can group them accordingly, and we define
the 1d integrals
\begin{equation}
\Omega F^{1\text{d}}_1 (z,\bar{z}) :=  \mathcal{F}_{Y_{123}} (z,\bar{z},\omega) + \mathcal{F}_{Y_{124}} (z,\bar{z},\omega) + \mathcal{F}_\text{IYI} (z,\bar{z},\omega)\;,
\label{eq:1dintegrals}
\end{equation}
with
\begin{align*}
\mathcal{F}_{Y_{123}} (z,\bar{z},\omega) & := \lambda_1\ I_{12} \int d\tau_3 \int d\tau_4\ I_{13} I_{24} \left( \frac{1}{I_{23}} - \frac{1}{I_{13}} \right) Y_{123} \notag \\
& = \frac{\lambda_1}{4 \pi |x_2|} I_{12} \int d\tau_3 \left( \frac{I_{13}}{I_{23}} - 1 \right) Y_{123}\,,
\end{align*}
where we have used the elementary integral given in eq. (\ref{eq:elementaryintegral1}), and
\begin{equation*}
\mathcal{F}_{Y_{124}} (z,\bar{z},\omega) := \frac{\lambda_1}{4 \pi |x_1|} I_{12} \int d\tau_4 \left( \frac{I_{24}}{I_{14}} - 1 \right) Y_{124}\,.
\end{equation*}
$\mathcal{F}_\text{IYI}$ is simply defined by the symmetric IYI-diagram of eq. (\ref{eq:diagIYIone}).

Similarly, grouping the $2$d integrals together results in
\begin{equation}
\Omega F^{2\text{d}}_1 (z,\bar{z}) := \mathcal{F}_{X} (z,\bar{z},\omega) + \mathcal{F}_{Y_{134}} (z,\bar{z},\omega) + \mathcal{F}_{Y_{234}} (z,\bar{z},\omega)\;,
\label{eq:2dintegrals}
\end{equation}
where we have defined
\begin{equation*}
\mathcal{F}_{X} (z,\bar{z},\omega) := \lambda_1\ I_{12} \int d\tau_3 \int d\tau_4\ I_{13} I_{24} \left( \frac{1}{I_{14} I_{23}} - \frac{1}{I_{13} I_{24}} - \frac{1}{I_{12} I_{34}} \right) X_{1234}\,,
\end{equation*}
as well as
\begin{equation*}
\mathcal{F}_{Y_{134}} (z,\bar{z},\omega) := \lambda_1\ I_{12} \int d\tau_3 \int d\tau_4\ I_{13} I_{24} \left( \frac{1}{I_{13}} - \frac{1}{I_{14}} + \frac{1}{I_{34}} \right) Y_{134}\,,
\end{equation*}
and
\begin{equation*}
\mathcal{F}_{Y_{234}} (z,\bar{z},\omega) := \lambda_1\ I_{12} \int d\tau_3 \int d\tau_4\ I_{13} I_{24} \left( \frac{1}{I_{24}} - \frac{1}{I_{23}} + \frac{1}{I_{34}} \right) Y_{234}\,.
\end{equation*}

We will not be able to solve these integrals analytically. However it is possible to numerically integrate both (\ref{eq:1dintegrals}) and (\ref{eq:2dintegrals}) on the line $z = \bar{z} =: x$. We found that the correlator corresponding to the $1$-channel takes the following form:
\begin{equation}
F_1 (x,x) = \frac{g^8 N^2}{3 \cdot 2^9 \pi^4} + \mathcal{O} (x \log x)\;.
\label{eq:1channelresult}
\end{equation}
The numerical integration could not be performed precisely enough in order to obtain the closed form of the coefficients at higher order, but the data will still be useful for checking the Ward identities. A plot of this channel can be found in fig. \ref{fig:2and1channels}. Eq. (\ref{eq:1channelresult}) gives the $1$-channel at the point $z \sim \bar{z} \sim 0$, which as we will see suffices for extracting the CFT data.

\bigskip

For completion we note some additional properties of the $0$-channel on the line $z = \bar{z}$:
\begin{subequations}
\begin{gather}
F_X(x,x) = 0 \quad \forall x \geq 0\,, \\
\intertext{as well as:}
F_1 (x,x) - F_X(x,x) = \text{const.} = \frac{g^8 N^2}{3 \cdot 2^9 \pi^4} \quad \forall x \leq 0\,.
\end{gather}
\end{subequations}
The first relation simply means that the X-integrals do not contribute to the correlator for $x \geq 0$, while the second one implies that \textit{only} the X-integrals are relevant for $x \leq 0$. We will not need to exploit these interesting properties in this work.

\subsubsection{0-Channel}
\label{subsubsec:0channel}

Similarly to the previous section, we now write down the integrals of the $0$-channel and show that the latter is finite on its own. In particular, we conclude that only one diagram contributes at this order and that its corresponding $F$-function is constant on the line $z = \bar{z}$ for $x \leq 0$. In section \ref{subsubsec:resummation} it will be shown that the $0$-channel can also be resummed in the full collinear limit, i.e. also for $x \geq 0$.

\bigskip

Let us start by reviewing the vanishing diagrams. There are two X-diagrams contributing to the $0$-channel, with either two scalar or two gluon fields contracted to the line. It is elementary to show that they cancel each other, i.e.
\begin{equation}
\left. \diagXone\ \right|_0 +\ \diagXdtwo\ = 0\,.
\label{eq:diagXdchannel}
\end{equation}
The reason why they have opposite sign is the presence of the $i$ in front of the gluon field in the definition of the Maldacena-Wilson line.

The next diagram will be referred to as the YY-diagram. The two Y-integrals are independent and hence we can factorize them and write
\begin{equation}
\diagYY = - \lambda_0\ \left( \int d\tau_3 \left(\partial_{\tau_1} - \partial_{\tau_2} \right) Y_{123} \right)^2\,.
\label{eq:diagYY}
\end{equation}
It is easy to see that the integral vanishes for any $x_2$, since the integrand is antisymmetric with respect to $\tau_3 \leftrightarrow - \tau_3$. This diagram therefore does not contribute to the $0$-channel and can also be discarded.

The last diagram that we must consider is an H-diagram, which is more intricate because of the complicated look of the $4$-point insertion given in eq. (\ref{eq:Hinserttwo}). However, noticing that
\begin{equation*}
\int d\tau_i\ \partial_{\tau_i} I_{ij} = 0
\end{equation*}
allows the diagram to be simplified to the following compact expression:
\begin{equation}
\diagHd\ = - 4 \lambda_0\ I_{12}\ \partial_{\tau_1} \partial_{\tau_2}  \int d\tau_3 \int d\tau_4\ H_{13,24}\,.
\label{eq:diagHd}
\end{equation}
This integral is the most difficult that we have encountered so far, since (i) the H-integral is not known analytically, and (ii) the fact that the derivatives are \textit{not} contracted as in the $1$-channel (see e.g. eq. (\ref{eq:Hdiagram})); this prevents us from reducing it to one-loop integrals using an identity such as the one given in eq. (\ref{eq:FXYidentity}). Note that the integral is finite, and hence the $0$-channel is finite on its own just like its siblings.

\bigskip

We will not be able to solve the ten-dimensional integral given in (\ref{eq:diagHd}) analytically. On the line $z = \bar{z} := x$, it can be reduced to a $4$d integral by using spherical coordinates, as explained in appendix \ref{subsubsec:0channelatNLO}. It is not easy to obtain an expansion of (\ref{eq:diagHd}) for $x \geq 0$, however we found numerically (see fig. \ref{fig:0channelandWI}) that
\begin{equation}
F_0 (x,x) = \text{const.} = - \frac{g^8 N^2}{3 \cdot 2^8 \pi^4} \quad \forall x \leq 0\,.
\label{eq:result0channel}
\end{equation}
This is a very important result, as it allows us to expand the correlator at $x \sim 0$ from below and to obtain infinitely many terms which can be compared to the superblock expansion given in eq. (\ref{eq:expsuperblocksbulk}) and (\ref{eq:expsuperblocksdefect}). We will use this in the next section in order to find closed-form expressions for the CFT data on the defect channel.

\subsubsection{Ward Identities in the Collinear Limit}
\label{subsubsec:wardidentities}

\begin{figure}
\centering
\begin{subfigure}{.5\textwidth}
  \centering
  \includegraphics[width=.95\linewidth]{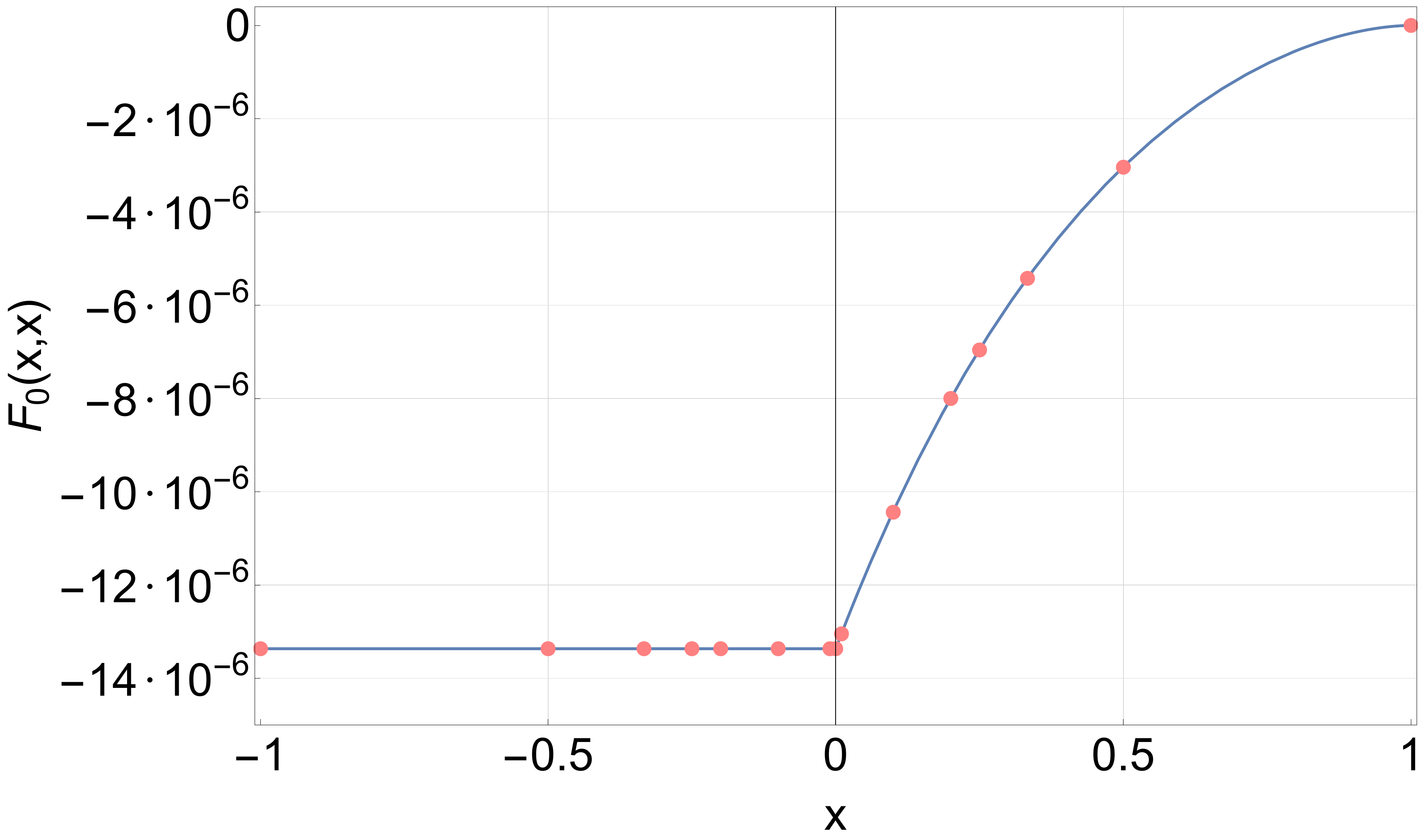}
\end{subfigure}%
\begin{subfigure}{.5\textwidth}
  \centering
  \includegraphics[width=.95\linewidth]{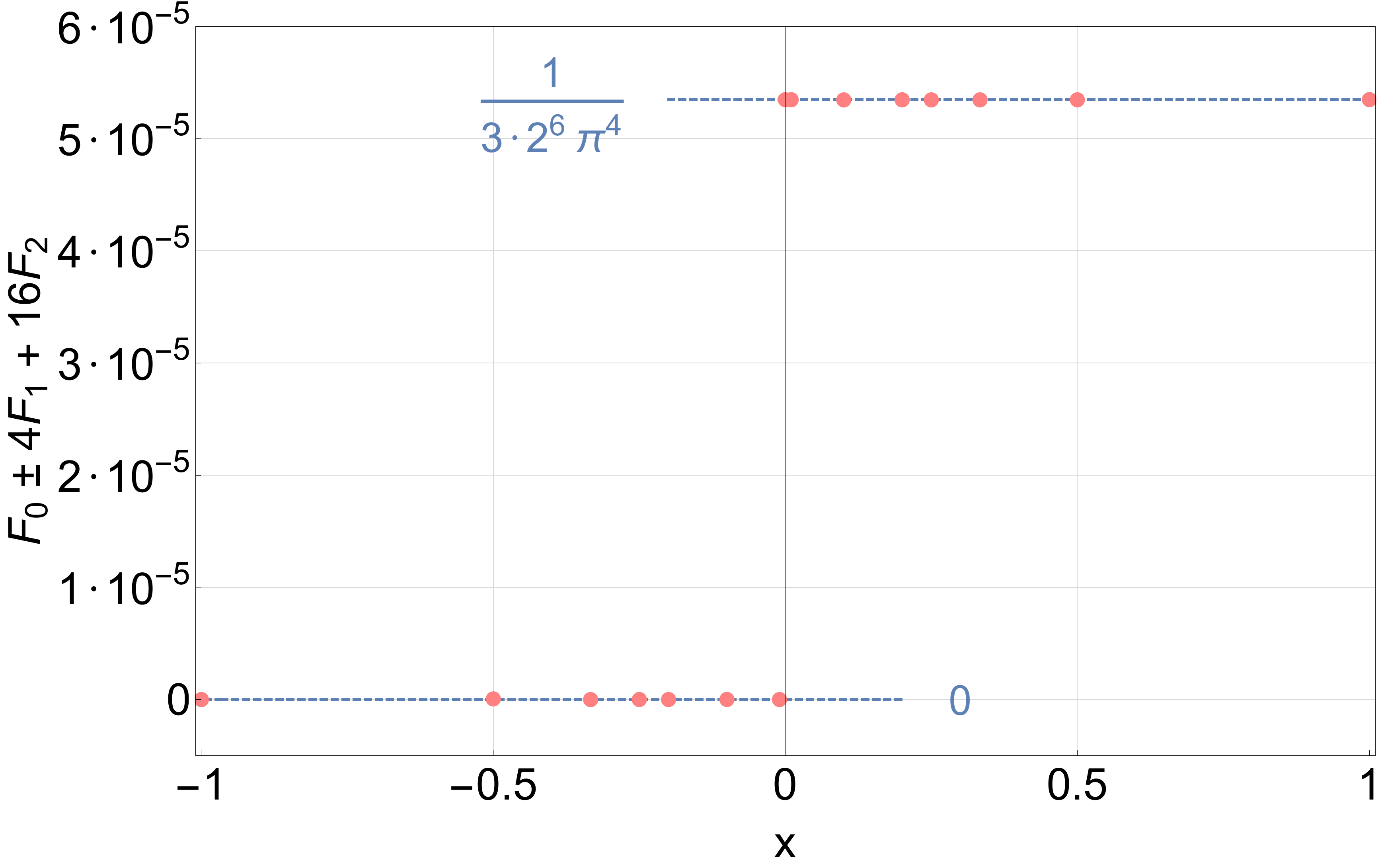}
\end{subfigure}
\caption{The left plot shows the $0$-channel after numerical integration (dots) and after resummation (solid line) as given by eq. (\ref{eq:0channelresummed}) in the collinear limit $z = \bar{z}$. We observe a perfect match, and note that $F_0$ is constant for negative $x$. The right plot shows the reduced Ward identities in the collinear limit for the $F_c$-functions computed numerically (dots). The constant behavior is exactly what is expected from eq. (\ref{eq:reducedWIcollinear}). The dashed lines are here to guide the eye.}
\label{fig:0channelandWI}
\end{figure}

We wish now to check whether the Ward identities that we presented in eq. (\ref{eq:reducedWIcollinear}) for the limiting case $z = \bar{z}$ are fulfilled by the expressions that we obtained in this section. The numerical data reveals that
\begin{subequations}
\begin{gather}
c_1 = \frac{g^8 N^2}{3 \cdot 2^6 \pi^4}\,, \\
c_2 = 0\,,
\end{gather}
\end{subequations}
as it can be seen in fig. \ref{fig:0channelandWI}. Although it is only performed in the collinear limit, the check of the Ward identities is a highly non-trivial confirmation of the validity of the expressions that were presented in the previous subsections. Note moreover that $c_2$ is in perfect agreement with the study of the correlator performed in \cite{Beccaria:2020ykg} at the point $z=\bar{z}=\omega=-1$.

\subsection{Summary}
\label{subsec:summary}

Let us summarize the results that were derived throughout this section. The identity and leading orders were easy to compute, and the corresponding $F_c$-functions turned out to be constant up to $\mathcal{O}(g^6)$:
\begin{subequations}
\begin{gather}
F_0 (z, \bar{z}) = \frac{g^4 N^2}{2^5 \pi^4}\,, \\
F_1 (z, \bar{z}) = \frac{g^6 N}{2^6 \pi^4}\,, \\
F_2 (z, \bar{z}) = 0\,.
\intertext{At NLO we were able to obtain a full analytical expression for the $2$-channel, which is given by eq. (\ref{eq:result2channel}). For the $0$-channel, we obtained an analytical result for the domain $z = \bar{z} := x$ with $x \leq 0$, while the $1$-channel could be solved only at the point $z = \bar{z} = 0$:}
F_0 (x, x) = - \frac{g^8 N^2}{3 \cdot 2^8 \pi^4} \quad \forall x \leq 0\,, \\
F_1 (0, 0) = \frac{g^8 N^2}{3 \cdot 2^9 \pi^4}\,.
\end{gather}
\end{subequations}
Each channel is finite as expected, and we also showed that the correlator satisfies the superconformal Ward identities order by order in the collinear limit.
%
%!TEX root = ../2pt_function_wline.tex
%%%%%%%%%%%%%%%%%%%%%%%%%%%%%%%%%%%%%%%%%%%%%

\section{CFT Data}
\label{sec:cft_data}

We are now ready to extract the CFT data by combining the perturbative computation of the previous section with the defect CFT analysis of section \ref{subsec:superblocksandcrossingsymmetry}.

\subsection{Expansion of Superblocks}
\label{subsec:expansionofsuperblocks}

\begin{figure}[t]
\centering
\includegraphics[width=.5\linewidth]{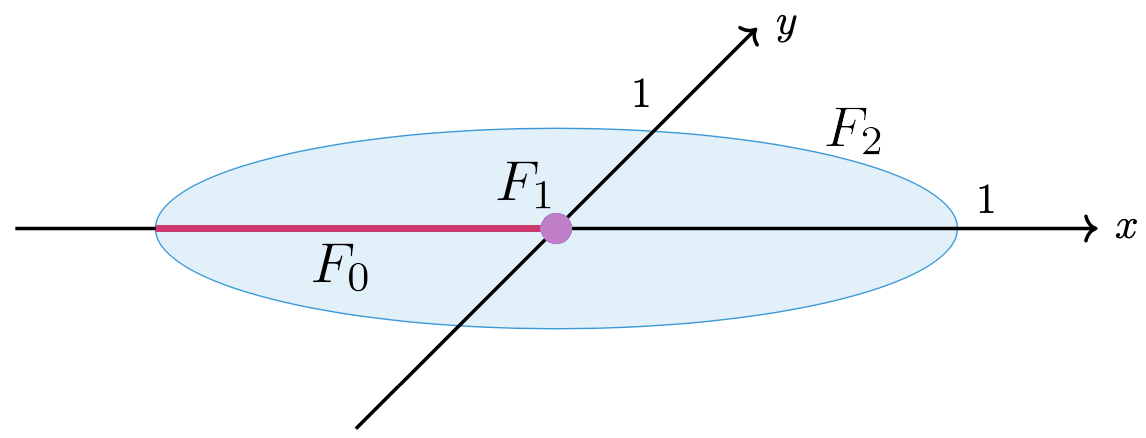}
\caption{Domains where the different R-symmetry channels are known analytically from the perturbative computation at next-to-leading order. Here $z:=x+i y$. The $2$-channel is known everywhere, the $0$-channel is known only on the line $z = \bar{z} =: x$ for $x \leq 0$, while the $1$-channel is known only at one point.}
\label{fig:domains}
\end{figure}

We need to expand the superblocks in the relevant OPE limits in order to compare the correlator obtained in (\ref{eq:bulkchannelexpansion}), (\ref{eq:defectchannelexpansion}) to the one computed perturbatively. 
We parameterize the anomalous corrections to
the scaling dimensions of
 bulk and defect operators as follows:
\begin{subequations}
	\begin{gather}
	\Delta_{\Delta, l} = \Delta^{(0)}_{\Delta, l} + \sum_{k=1}^\infty g^{2k} \gamma^{(k)}_{\Delta, l}\,, \\
	\hat{\Delta}_{\hat{\Delta}, s} = \hat{\Delta}^{(0)}_{\hat{\Delta}, s} + \sum_{k=1}^\infty g^{2k} \hat{\gamma}^{(k)}_{\hat{\Delta}, s}\,,
	\end{gather}
\end{subequations}
where the subscripts refer to the classical scaling dimensions and to the corresponding spin (either parallel or transverse).

\bigskip

For the bulk channel, the OPE limit is $z \sim \bar{z} \sim 1$ (i.e. the operator $\mathcal{O}(x_2)$ is close to $\mathcal{O}(x_1)$), in which the R-symmetry channels read
\begin{equation}
F_c (z,\bar{z}) = \left( \frac{(1-z)(1-\bar{z})}{\sqrt{z \bar{z}}} \right)^{2-c} \sum_{k,l,m} c^{(c)}_{k,l,m} \log^k \left[(1-z)(1-\bar{z})\right]\ (1-z)^{l/2} (1-\bar{z})^{m/2}\,.
\label{eq:expsuperblocksbulk}
\end{equation}
The $\log$ terms are a consequence of the anomalous dimensions, and arise naturally in loop computations in perturbation theory (see e.g. \eqref{eq:result2channel}). The coefficients $c_{k,l,m}^{(c)}$ only depend on $g$ and $N$, and are basically linear combinations of the CFT data. We spell out the first few relations in  (\ref{eq:bulkrelations0channel}-\ref{eq:bulkrelations2channel}). Eq. \eqref{eq:expsuperblocksbulk} is exact, but the expansion can be truncated accordingly for comparison with the perturbative computation. In principle, at any order in $g$, the relations given in (\ref{eq:bulkrelations0channel}-\ref{eq:bulkrelations2channel}) can be solved recursively and the solution is unique. The coefficients $c^{(c)}_{k,l,m}$ with $k \geq 1$ can only contain terms with anomalous dimensions $\gamma^{(n)}_{\Delta,l}$. Note that, as briefly mentioned before, it is \textit{not} necessary to know the full correlator in order to obtain the full CFT data; from the three R-symmetry channels, it suffices indeed to know one channel everywhere, one channel on the line (e.g. the limit $z = \bar{z}$) and one channel at one point, as it is the case at next-to-leading order from our perturbative computation (see fig. \ref{fig:domains}).

\bigskip

The OPE limit for the defect channel is at $z \sim \bar{z} \sim 0$ (i.e. the operator $\mathcal{O}(x_2)$ is close to the defect), and in that case the correlator can be expanded as:
\begin{equation}
F_c (z,\bar{z},\omega) = \left( \frac{(1-z)(1-\bar{z})}{\sqrt{z \bar{z}}} \right)^{2-c} \sum_{k,l,m} \hat{c}^{(c)}_{k,l,m} \log^k (z\bar{z})\ z^{l/2} \bar{z}^{m/2}\,.
\label{eq:expsuperblocksdefect}
\end{equation}
The previous considerations for the bulk channel apply analogously to the defect channel. The first few coefficients $\hat{c}_{k,l,m}$ are given by eq. (\ref{eq:defectrelations0channel}-\ref{eq:defectrelations2channel}) in appendix \ref{subsec:expdefectchannel}.

\bigskip

The defect channel is technically simpler than its bulk counterpart, which is evident by comparing the spacetime blocks in (\ref{eq:bulkstblock}) and (\ref{eq:defectstblock}): the defect blocks have a simple form in terms of products of two $_2F_1$ hypergeometric functions, while the bulk blocks consist of a less efficient infinite sum of generalized hypergeometric functions.  As a consequence, the bulk channel is severely limited by computing power, while the defect channel can be expanded to very high orders.

\subsection{Defect Channel}
\label{subsec:defectCFTdata}

In this section we present closed-form formulae for the defect CFT data up to next-to-leading order.

\subsubsection{Identity Order}
\label{subsubsec:defectCFTdataIO}

We have seen in section \ref{subsec:identityandleadingorders} that the $F$-functions are constant at identity order. It is easy to express the $0$-channel from (\ref{eq:gidentity}) such that it can be compared to (\ref{eq:expsuperblocksdefect}):
\begin{equation}
F_0 (z, \bar{z}) = \left( \frac{(1-z)(1-\bar{z})}{\sqrt{z \bar{z}}} \right)^2 \frac{g^4 N^2}{2^5 \pi^4} \sum_{l,m \geq 0} l m\, z^l \bar{z}^m\,.
\end{equation}
Using the relations given in (\ref{eq:defectrelations0channel}-\ref{eq:defectrelations2channel}) we immediately obtain
\begin{subequations}
\begin{gather}
\hat{C} = \frac{g^4 N^2}{2^9\pi^4}\,.
\intertext{The coefficients $\hat{E}_s$ are also easy to derive:}
\hat{E}_s = \frac{g^4 N^2}{2^8 \pi^4} \frac{(1+s)(2+s)}{1+2s}\,.
\intertext{It is convenient to characterize the longs by their \textit{twist} $\hat{\tau} := \hat{\Delta}-s$. The twist-two operators read}
\hat{F}_{s+2,s} = \frac{g^4 N^2}{2^7 \pi^4} \frac{(2+s)(3+s)}{(5+2s)}\,. \notag
\intertext{This relation can be generalized recursively for the longs with arbitrary \textit{even} twist:}
\left. \hat{F}_{\hat{\Delta},s} \right|_{\hat{\tau}\ \text{even}} = \frac{g^4 N^2}{2^9 \pi^4} \frac{\Gamma(\hat{\Delta} + 2) \Gamma(s + 3/2)}{\Gamma(\hat{\Delta} + 3/2) \Gamma(s + 1)} \frac{(\hat{\Delta} - s)(\hat{\Delta} + s + 1)}{(\hat{\Delta} + s)(\hat{\Delta} - s - 1)}\,,
\intertext{while the $\hat{F}_{\hat{\Delta},s}$ with \textit{odd} twist are zero. This is the supersymmetric equivalent of eq. (2.8) in \cite{Lemos:2017vnx}.}
\intertext{Since both the 1- and 2-channels vanish, we also have:}
\hat{A} = \hat{B} = \hat{D}_s - \hat{F}_{s+1,s} = 0\,.
\intertext{Note that it is not possible to distinguish the coefficients $\hat{D}_s$ and $\hat{F}_{s+1,s}$ just by looking at the identity order because of multiplet shortening, as explained in section \ref{subsec:superblocksandcrossingsymmetry}. However, looking at the $\log$ terms of LO and NLO allows them to be disentangled, and we find that they vanish individually:}
\hat{D}_s = \hat{F}_{s+1,s} = 0\,.
\end{gather}
\end{subequations}

\subsubsection{Leading Order}
\label{subsubsec:defectCFTdataLO}

At leading order we saw in section \ref{subsec:identityandleadingorders} that only the $1$-channel contributes. We can proceed in the exact same way as for the identity order, since the $F$-functions are also constant. The $1$-channel can be expressed as
\begin{equation}
F_1 (z,\bar{z}) = \frac{(1-z)(1-\bar{z})}{\sqrt{z \bar{z}}} \frac{g^6 N}{2^6 \pi^4} \sum_{l,m \geq 1} z^{l/2} \bar{z}^{m/2}\,.
\end{equation}
It is easy to solve for the CFT data by following the prescriptions presented in section \ref{subsec:expansionofsuperblocks}, and we obtain the following coefficients:
\begin{subequations}
\begin{gather}
\hat{A} = \hat{C} = 0\,, \qquad \hat{B} = \frac{g^6 N}{2^8 \pi^4}\,, \\
\hat{D}_s - \hat{F}_{s+1,s} = \frac{g^6 N}{2^7 \pi^4} \frac{1+s}{1+2s}\,, \qquad \hat{E}_s = 0\,, \qquad \left. \hat{F}_{\hat{\Delta},s} \right|_{\hat{\tau} \neq 1} = 0\,, \\
\left. \hat{\gamma}^{(1)}_{\hat{\Delta},s} \right|_{\hat{\tau}\ \text{even}} = 0\,.
\end{gather}
\end{subequations}
At this order, the coefficients $D_s$ and $F_{s+1,s}$ could not be disentangled\footnote{This would require to know at least the $\log$ terms at next-to-next-to-leading order (NNLO).}. We also note that the long operators with even twist $\hat{\tau}$ do not receive an anomalous contribution $\hat{\gamma}$.

\subsubsection{Next-to-Leading Order}
\label{subsubsec:defectCFTdataNLO}

At NLO the computation of the CFT data is more subtle, since we do not have the full correlator analytically. Still, we can solve order by order in the same fashion and guess the exact formula once a sufficient number of coefficients has been obtained.

\bigskip

It makes sense to start by the $2$-channel, since it is the only one that is known analytically everywhere. The expansion of (\ref{eq:result2channel}) in the defect OPE limit reads
\begin{align}
F_2(z,\bar{z}) &= \frac{g^8 N^2}{2^{12} \pi^4} - \frac{g^8 N^2}{2^{10} \pi^6} \sum_{l \geq 1} \left\lbrace \frac{2 \log 2 + H_{l-1/2}}{l} (z \bar{z})^l + \frac{1 + (1-2l) H_{l-1}}{(1-2l)^2} (z \bar{z})^{\frac{2l-1}{2}} \right\rbrace \notag \\
& \qquad \qquad \qquad \qquad \qquad + \frac{g^8 N^2}{2^{10} \pi^6} \log z\bar{z} \sum_{l \geq 1} \frac{1}{1-2l} (z \bar{z})^{\frac{2l-1}{2}}\,,
\label{eq:defectexp2channel}
\end{align}
where we introduced the harmonic numbers $H_n$:

\begin{equation*}
H_n := \sum_{k=1}^n \frac{1}{k}\,.
\end{equation*}

We also know the $0$-channel in the collinear limit $z = \bar{z}$ for $x \leq 0$, which turned out to be a constant function again (see eq. (\ref{eq:result0channel})), i.e.
\begin{equation}
\left. F_0 (x,x) \right|_{x \leq 0} = - \left( \frac{(1-z)(1-\bar{z})}{\sqrt{z \bar{z}}} \right)^2 \frac{g^8 N^2}{3 \cdot 2^8 \pi^4} \sum_{l,m} l m\, z^l \bar{z}^m\,.
\end{equation}
We can take the same limit in eq. (\ref{eq:expsuperblocksdefect}) in order to relate this expression to the CFT data. 
Moreover, the $1$-channel is known at one point, as given in eq. (\ref{eq:1channelresult}).

\bigskip

Comparing the $\log$ terms of (\ref{eq:defectexp2channel}) to (\ref{eq:expsuperblocksdefect}) and (\ref{eq:defectrelations2channel}), we find
\begin{subequations}
\begin{gather}
\left. g^2 \hat{F}_{s+1,s} \right|_{\mathcal{O}(g^6)} \hat{\gamma}^{(1)}_{s+1,s} = \frac{g^8 N^2}{2^8 \pi^6} \frac{1+s}{1+2s} H_{1+s}\,, \\
\left. \hat{\gamma}^{(2)}_{\hat{\Delta},s} \right|_{\hat{\tau}\ \text{even}} = 0\,.
\intertext{Curiously, the anomalous dimensions $\hat{\gamma}$ of long operators with even twist are seen to also vanish at this order. The power series terms result in the following formulae:}
\hat{A} = \frac{g^8 N^2}{2^{12} \pi^4}\,, \qquad \hat{B} = - \frac{g^8 N^2}{3 \cdot 2^{11} \pi^4}\,, \qquad \hat{C} = - \frac{g^8 N^2}{3 \cdot 2^{12} \pi^4}\,, \label{subeq:coeffA}\\
\resizebox{.85 \textwidth}{!} 
{
$\hat{D}_s - \hat{F}_{s+1,s} = \frac{g^8 N^2}{3 \cdot 2^{10} \pi^6} \frac{1+s}{1+2s} \left\lbrace \pi^2 + \left( \frac{1}{3} \frac{1}{(1+s)(1+2s)} - 12 \log 2 \right) H_{1+s} + \frac{1}{3} H^{(2)}_{1+s} \right\rbrace$
}\,, \\
\hat{E}_s = - \frac{g^8 N^2}{3 \cdot 2^9 \pi^6} \frac{(1+s)(2+s)}{1+2s} \left\lbrace \pi^2 + \frac{6}{2+s} H_{2+s} - 6 H^{(2)}_{2+s} \right\rbrace. \\
\intertext{As for the lower orders, it makes sense to distinguish the longs according to the parity of their twist. The CFT data for twist-even operators with $s=0$ (i.e. $\hat{\tau} = \hat{\Delta}$) is given by:}
\left. \hat{F}_{\hat{\Delta},0} \right|_{\hat{\tau}\ \text{even}} = - \frac{g^8 N^2}{3 \cdot 2^{12} \pi^4} \frac{\Gamma(3/2) \Gamma(\hat{\Delta}+2)}{\Gamma(\hat{\Delta}+3/2)} \frac{\hat{\Delta}+1}{\hat{\Delta}-1}, \notag \\
\intertext{while long operators with $s=0$ but odd twist exhibit a different structure:}
\resizebox{.9 \textwidth}{!} 
{
$\left. \hat{F}_{\hat{\Delta},0} \right|_{\hat{\tau}\ \text{odd}} = \frac{g^8 N^2}{3 \cdot 2^{12} \pi^6} \frac{\Gamma(3/2) \Gamma(\hat{\Delta}+2)}{\Gamma(\hat{\Delta}+3/2)} \frac{\hat{\Delta}+1}{\hat{\Delta}-1} \left\lbrace \pi^2 + 12 \left( \frac{1}{(\hat{\Delta}+3)^2} - \frac{1}{(\hat{\Delta}+2)^2} \right) + 3 \left( H^{(2)}_{\hat{\Delta}/2+1} - H^{(2)}_{\hat{\Delta}/2+3/2} \right) \right\rbrace$.
} \notag
%\intertext{For twist-two operators, we found:}
%\resizebox{.9 \textwidth}{!} 
%{
%$\hat{F}_{s+2,s} = \frac{g^8 N^2}{3 \cdot 2^9 \pi^6} \frac{\Gamma(2 \hat{\Delta})}{\Gamma(2\hat{\Delta} + 2)} \left\lbrace - \pi^2 \hat{\Delta}^2 (\hat{\Delta} + 1) + 6 (1 + 2 \hat{\Delta} (\hat{\Delta}+1)) - 6 \frac{\hat{\Delta} (3 \hat{\Delta}^2 - 1)}{\hat{\Delta} - 1} H_{\hat{\Delta}-1} - 6 \hat{\Delta}^2 (1 + \hat{\Delta}) H^{(2)}_{\hat{\Delta}} \right\rbrace$.
%} \notag
%\intertext{The CFT data for twist-three operators reads:}
%\resizebox{.9 \textwidth}{!} 
%{
%$\hat{F}_{s+3,s} = \frac{g^8 N^2}{2^7 \pi^6} \frac{\hat{\Delta}-1}{(1+2\hat{\Delta})(3-2\hat{\Delta})^2} \left\lbrace 1 - 2\hat{\Delta} (\hat{\Delta}-1) + 3(\hat{\Delta} (\hat{\Delta}-1)) + 3 (\hat{\Delta} (\hat{\Delta}-1) -1 ) H_{\hat{\Delta}-1} \right\rbrace$.
%} \notag
\intertext{The coefficients quickly grow in complexity as $s$ increases. Nevertheless, we managed to obtain an exact formula for even twist:}
\scalebox{.8}{$
\begin{aligned}[b] \left. \hat{F}_{\hat{\Delta},s} \right|_{\hat{\tau}\ \text{even}} & = \frac{1}{3 \cdot 2^{12} \pi^6} \frac{\Gamma(\hat{\Delta} + 2) \Gamma(\hat{\Delta} - \hat{\tau} + 3/2)}{\Gamma(\hat{\Delta} + 3/2) \Gamma(\hat{\Delta} - \hat{\tau} + 1)} \left\lbrace - \pi^2 \frac{(\hat{\Delta} -2)(\hat{\Delta}+s+1)}{(\hat{\Delta}+s)(\hat{\Delta}-s-1)} + 12 \frac{H_{\hat{\tau}}-H_{\hat{\tau}-1}-H_{\frac{\hat{\tau}-1}{2}}}{\hat{\Delta} - \hat{\tau}/2 + 1} \right. \\
& \qquad + 6 (H_{\hat{\Delta}+1} - H_{\hat{\Delta}-\hat{\tau}})(H_{\hat{\Delta}-\hat{\tau}/2} - H_{\hat{\tau}/2-1}) + 6 H^{(2)}_{\hat{\Delta}-\hat{\tau}/2} + \sum\limits_{j=1}^{\hat{\tau}/2} \frac{12}{(\hat{\Delta}-j+2)(\hat{\tau}-2(j+1))} \\
& \qquad \qquad \qquad \left. + 12 \sum\limits_{j=0}^{\hat{\tau}/2-1} \frac{h_{\hat{\tau},j+1}}{\hat{\Delta}-j+1} + 12 \sum\limits_{j=\hat{\tau}/2+1}^{\hat{\tau}} \frac{h_{\hat{\tau},\hat{\tau}-j+1}}{\hat{\Delta}-j+1} \right\rbrace,
\end{aligned}$}
\intertext{where we have defined the following help function:}
\resizebox{.9 \textwidth}{!} 
{
$h_{k,n} := H_{(k-1)/2} - H_{(k-2n+1)/2} - \sum\limits_{j=0}^n (1+j)\left( \frac{(k-2)^{j-1} \Gamma(k/2-n+1)}{2^j \Gamma(k/2-1)} + \frac{k^j \Gamma(k-n+2)}{\Gamma(k+1)} \right) S_{n-1}^{(j+1)}$.
} \notag
\intertext{We have introduced the Stirling numbers of the first kind $S_n^{(m)}$, defined to be the coefficients of the expansion of the Pochhammer symbols $(x)_n$ as a power series:}
(x)_n := \frac{\Gamma(x+n)}{\Gamma(n)} =: \sum_{m=0}^n S_n^{(m)} x^m\,.  \notag
\intertext{Similarly, it is possible to express the coefficients for odd twist in a closed form:}
\scalebox{.8}{$
\begin{aligned}[b] \left. F_{\hat{\Delta},s} \right|_{\hat{\tau}\ \text{odd}} &= \frac{g^8 N^2}{2^{11} \pi^6} \frac{\hat{\tau} (\hat{\tau} - 2 \hat{\Delta} -1)}{(\hat{\tau} - 1)(\hat{\tau} - 2 \hat{\Delta})} \frac{\Gamma(\hat{\Delta}+2) \Gamma(\hat{\Delta}-\hat{\tau}+3/2)}{\Gamma(\hat{\Delta}+3/2) \Gamma(\hat{\Delta}-\hat{\tau} +1)} \left\lbrace (H_{\hat{\Delta}+1} - H_{\hat{\Delta}-\hat{\tau}}) (2 H_{\hat{\tau}} - H_{(\hat{\tau}-1)/2}) \phantom{\sum_a^b} \right. \\
& \qquad + \frac{2}{\hat{\Delta}+2} \left( H_{\hat{\tau}-1} + H_{(\hat{\tau}+1)/2} - H_{\hat{\Delta}-\hat{\tau}} \right) + \frac{2}{\hat{\Delta}-\hat{\tau}} \left( H_{\hat{\tau}-1} + H_{(\hat{\tau}+1)/2} - H_{\hat{\Delta}+1} \right) \\
& \qquad\qquad + H_{\hat{\Delta}+(1-\hat{\tau})/2} \left( 2 ( H_{\hat{\Delta}+2} - H_{\hat{\Delta}-\hat{\tau}-1} ) + H_{\hat{\Delta}-\hat{\tau}} - H_{\hat{\Delta}+1} \right) \\
& \qquad\qquad\qquad \left. -2 \sum_{k=0}^{(\hat{\tau}-1)/2} \left( \frac{1}{1-k+\hat{\tau}} + \frac{1}{1+k-\hat{\tau} - \hat{\Delta}} \right) H_{1+\hat{\tau}-k} \right\rbrace.
\end{aligned}$}
\end{gather}
\end{subequations}

Even though these formulae look a bit messy, they do constitute a complete solution of our problem. We have obtained all of the relevant defect-channel coefficients up to NLO, and the correlator can now be expanded around the defect OPE limit $z \sim \bar{z} \sim 0$ to any desired order.

\subsubsection{Direct Computation of $\hat{A}$}
\label{subsubsec:coefficientA}

As a consistency check, let us compute the coefficient $\hat{A}$ independently for comparison with the result given in eq. (\ref{subeq:coeffA}). This is easy to do since its definition is just $\hat{A} := a_{\mathcal{O}}^2$, i.e. it is the square of the coefficient for the one-point function of a single-trace operator $\mathcal{O}(x)$ in the presence of the line defect. Hence we have to consider the following correlator:
\begin{equation}
\vvev{\mathcal{O}(x)} = \frac{(u \cdot \theta)^2}{x^2} F(x,\omega)\,,
\end{equation}
where $F$ is defined in a similar way to eq. (\ref{eq:superconformalinvariance}).

At leading order there is only one diagram:
\begin{align}
\diagoneptfunction\ &= \frac{1}{N} (u \cdot \theta)\ g^4\ \text{Tr}\ T^a T^b\ \text{Tr}\ T_a T_b \int d\tau_3 \int d\tau_4\ I_{13} I_{14} \notag \\
&= \frac{(u \cdot \theta)^2}{x^2} \frac{g^4 N}{2^6 \pi^2}\,,
\end{align}
which perfectly matches the coefficient $\hat{A}$ that we found at NLO.

\subsection{Bulk Channel}
\label{subsec:bulkCFTdata}

We move now our attention to the CFT data of the bulk channel. In this case, we face two issues that were not present in the defect channel: (i) the spacetime blocks are complicated functions (see (\ref{eq:bulkstblock})), hence it is difficult to expand them to a sufficiently high order; (ii) at NLO the perturbative computations of the $0$- and $1$-channels do not cover the bulk OPE limit $z \sim \bar{z} \sim 1$. The second problem will be solved by resumming the correlator in the collinear limit, while the first issue constitutes a limitation inherent to the bulk channel. We will be able to obtain closed forms for the CFT data up to leading order, while at NLO we give the first few coefficients for each representation and show that they reproduce known results.

\subsubsection{Resummation in the Collinear Limit}
\label{subsubsec:resummation}

We have derived in section \ref{sec:perturbative} analytical expressions for the $0$- and $1$-channel near $z \sim \bar{z} \sim 0$. Unfortunately this is not the case for the bulk OPE limit $z \sim \bar{z}$. We now resum these channels in the collinear limit $z = \bar{z} =: x$ in order to fill that gap. Note that, because of the Ward identities (see in particular eq. (\ref{eq:reducedWIcollinear})) and of the fact that the $2$-channel is already known analytically, we need only determine one R-symmetry channel in order to know the last one.

\bigskip

In principle, the expansion in superblocks given in (\ref{eq:defectchannelexpansion}) could be resummed with the use of the defect CFT data derived in the previous subsection. However this leads to a very challenging resummation, and instead of taking this road it is easier to consider the following procedure. The idea is to take the limit $z = \bar{z} =: x$ for the $2$-channel and to write an ansatz for the other channels based on this expression. It is then easy to use the expansion in superblocks in order to solve for the coefficients, provided the ansatz is complete, i.e. that there occurs no cancellation between the $0$- and $1$-channels. It happens to be the case here, and we find the following expression for the $1$-channel:
\begingroup
\allowdisplaybreaks
\begin{align}
F_1 (x,x) &= \frac{g^8 N^2}{3 \cdot 2^9 \pi^6} \left\lbrace - 2 \pi^2 + 12 \tanh^{-2} x + 6 i \pi \log 2 + 6 \log (1-|x|) \log (|x| - 1) \phantom{\frac{1}{1}} \right. \notag \\
& \qquad - 6 \log^2 (1-|x|) - 6 \tanh^{-1} |x| \log x^2 - 6 \log 16 \tanh^{-1} |x| \notag \\
& \qquad \quad + 3 \log (1+|x|) \log [4 (1+|x|)] - 3 \log (|x|-1) \log [4(|x|-1)] \notag \\
& \qquad \quad \quad + \Theta(x) \tanh^{-1} x \left[ 12 \log (1+x) - 6 \log 4 \right] + \Theta(-x) \left[ 6 (\Li_2 (-x) - \Li_2 (x)) \phantom{\frac{1}{1}} \right. \notag \\
& \qquad \quad \quad \quad \qquad \qquad \qquad \qquad \left. \left. -2\pi^2 - 6 \Li_2 \frac{1-x}{2} + 6 \Li_2 \frac{1+x}{2} \right] \right\rbrace\,.
\label{eq:1channelresummed}
\end{align}
\endgroup
It is worth checking that this expression indeed reproduces the numerical data obtained in section \ref{sec:perturbative}. Fig. \ref{fig:2and1channels} shows a plot of eq. (\ref{eq:1channelresummed}) with the numerical data, and we observe a perfect agreement.

As mentioned above, we immediately obtain the expression for the $0$-channel by using the reduced Ward identities in the collinear limit:
\begin{align}
F_0 (x,x) & = - \frac{g^8 N^2}{3 \cdot 2^8 \pi^4} + \frac{g^8 N^2}{2^6 \pi^6} \Theta(x) \left\lbrace 2 \log 2 \tanh^{-1} x \phantom{\frac{1}{1}} - \log^2(1+x) + \log(1-x) \log(1+x) \right. \notag \\
& \qquad \qquad \qquad \qquad \qquad \left. + \Li_2 (x) - \Li_2 (-x) + \Li_2 \frac{1-x}{2} - \Li_2 \frac{1+x}{2} \right\rbrace.
\label{eq:0channelresummed}
\end{align}
This result is shown against the numerical data in fig. \ref{fig:0channelandWI}, and a perfect match is also observed. We now have enough information about the correlator in order to extract the CFT data of the bulk channel.

\subsubsection{Identity Order}
\label{subsubsec:bulkCFTdataIO}

It is clear from the expansion in superblocks that this order is trivial, since the $\Omega^2$ of eq. (\ref{eq:resultidentityorder}) corresponds to the identity contribution in eq. (\ref{eq:bulkchannelexpansion}) (we recall that $A := b_{\mathcal{O}\mathcal{O}}$). The only non-vanishing coefficient is therefore given by:

\begin{equation}
A = \frac{g^4 N^2}{2^5 \pi^4}\,.
\end{equation}

\subsubsection{Leading Order}
\label{subsubsec:bulkCFTdataLO}

At leading order, we expand in the bulk OPE limit the result presented in eq. (\ref{eq:resultleadingorder}) in order to match the coefficients to the expansion given in (\ref{eq:expsuperblocksbulk}):
\begin{equation}
F_1 (z, \bar{z}) = \frac{(1-z)(1-\bar{z})}{\sqrt{z \bar{z}}} \frac{g^6 N}{2^6 \pi^4} \sum_{l,m \geq 0} \frac{(-1/2)_l (-1/2)_m}{(1)_l (1)_m} (1-z)^l (1-\bar{z})^m\,.
\end{equation}
In this case there are only two non-vanishing families of coefficients, corresponding to the $1/2$-BPS operators $\mathcal{B}_{[0,2,0]}$:
\begin{subequations}
\begin{gather}
B = \frac{g^6 N}{2^4 \pi^4}\,,
\intertext{and to the longs at the unitarity bound $\mathcal{A}^{l+2}_{[0,0,0],l}$:}
E_{l+2,l} = \frac{g^6 N}{\pi^4} \left( \frac{1}{2}\right)^{2(l+4)} (l+1) \frac{\Gamma(1/2)\Gamma(l+2)}{\Gamma(l+5/2)}\,.
\end{gather}
\end{subequations}

\subsubsection{Next-to-Leading Order}
\label{subsubsec:bulkCFTdataNLO}

At NLO the computation of the coefficients becomes a lot more involved, and computing power sets an upper limit to the amount of CFT data that can be extracted.

\bigskip

In this case it is best to expand (\ref{eq:result2channel}), (\ref{eq:1channelresummed}) and (\ref{eq:0channelresummed}) at the OPE limit $z \sim \bar{z} \sim 1$ for lines of the type $1-z := x + i k x$, $k \in \mathbb{N}$. Reading the $\log$ terms immediately gives the one-loop anomalous dimensions for the long operators at the unitarity bound:
\begin{subequations}
\begin{gather}
\gamma^{(1)}_{l+2,l} = \frac{N}{2\pi^2} H_{l+2}\,.
\intertext{This expression is well-known \cite{Beisert:2003jj}, and in particular the operator with ($\Delta, l$)=($2,0$) corresponds to the Konishi operator $\mathcal{K}_1$, for which $\gamma^{(1)}_{2,0}$ has already been computed in numerous ways \cite{Anselmi:1996dd, Bianchi:1999ge, Beisert:2003jj}. This provides another powerful check of our results.}
\intertext{The coefficients for the short operators read:}
A = 0\,, \qquad B = - C = - \frac{g^8 N^2}{3 \cdot 2^7 \pi^4}\,,
\intertext{while the long operators with odd $\Delta$ simply vanish:}
\left. E_{\Delta,l} \right|_{\Delta\ \text{odd}} = 0\,.
\end{gather}
\end{subequations}

We are left with the coefficients $D_l$ for the semishorts and with the $E_{\Delta,l}$ for the longs with $\Delta$ even. No closed form could be obtained, due to the large computing time inherent to the expansion of the superblocks. The first few values are gathered in table \ref{table:CFTdatabulk}, while values up to $l=16$ can be found in the ancillary \textsc{Mathematica} file of the {\tt arxiv.org} submission.

\bigskip

To conclude, we note that the coefficient $B$ can easily be checked against the literature, since it corresponds to the product of the one-point function coefficient $a_{\mathcal{O}}$ with the three-point function coefficient $\lambda_{\mathcal{O}\mathcal{O}\mathcal{O}}$. Since the latter is protected, only the one-point function is relevant. In the large $N$ limit, it is known exactly:

\begin{equation}
\vvev{\mathcal{W}_l \mathcal{O}(x)} = \frac{\vev{\mathcal{W}_c \mathcal{O}(x)}}{\vev{\mathcal{W}_c}} = \frac{\sqrt{\lambda}}{2\sqrt{2} N} \frac{I_2 (\sqrt{\lambda})}{I_1 (\sqrt{\lambda})}\,,
\end{equation}
written here in the convention of \cite{Okuyama:2006jc}. Expanding this expression for $\lambda := g^2 N \sim 0$, we find:

\begin{equation}
\vvev{\mathcal{W}_l \mathcal{O}(x)} = \frac{N}{16 \sqrt{2}} \left\lbrace \frac{2 g^2}{N} - \frac{g^4}{12} + \mathcal{O}(g^6) \right\rbrace\,,
\end{equation}
where the prefactor is convention-dependent and corresponds to the (protected) three-point function. The terms between the brackets perfectly match the coefficient $B$ up to NLO, when normalized with the two-point function:

\begin{equation}
\frac{B}{A} = \frac{2 g^2}{N} - \frac{g^4}{12} + \mathcal{O}(g^6)\,.
\end{equation}

\begin{table}
\centering
\caption{Bulk CFT data coefficients $D_l$ and $E_{\Delta,l}$ for low spin $l:=2j$ at NLO. Only the longs with even $\Delta$ are non-vanishing. The second line shows the CFT data for the longs at the unitarity bound $\Delta=l+2$. Note that all the values should be multiplied by $g^8 N^2$. Values up to $l=16$ can be found in the ancillary \textsc{Mathematica} file of the {\tt arxiv.org} submission.}
\begin{tabular}{c}
%\resizebox{\columnwidth}{!}{%
\begin{tabular}{cccc}
\\ \hline \\[-1em]
$D_0$ & $D_2$ & $D_4$ & $D_6$ \\
\hline \\[-.5em]
$-\frac{15 + 8\pi^2}{15360 \pi^6}$ & $-\frac{1365 + 512\pi^2}{4644864 \pi^6}$ & $-\frac{3185 + 1024\pi^2}{79073280 \pi^6}$ & $-\frac{5344955 + 1572864\pi^2}{1254629376000 \pi^6}$ \\[.75em]
\hline \\
\end{tabular} \\
\begin{tabular}{cccc}
\hline \\[-1em]
$E_{2,0}$ & $E_{4,2}$ & $E_{6,4}$ & $E_{8,6}$ \\
\hline \\[-.5em]
$- \frac{9 + \pi^2 + 18 \log 2}{4608 \pi^6}$ & $- \frac{85 + 14 \pi^2 + 350 \log 2}{501760 \pi^6}$ & $- \frac{2485 + 440 \pi^2 + 12936 \log 2}{187342848 \pi^6}$ & $- \frac{549207 + 100100 \pi^2 + 3264690 \log 2}{565373952000 \pi^6}$ \\[.75em]
\hline \\
\end{tabular} \\
\begin{tabular}{cccccc}
\hline \\[-1em]
$E_{4,0}$ & $E_{6,0}$ & $E_{6,2}$ & $E_{8,0}$ & $E_{8,2}$ & $E_{8,4}$ \\
\hline \\[-.5em]
$\frac{-3 + 2\pi^2}{73728 \pi^6}$ & $\frac{1}{147456 \pi^6}$ & $\frac{-15 + 64 \pi^2}{7962624 \pi^6}$ & $\frac{-15+8\pi^2}{165150720 \pi^6}$ & $\frac{263}{144179200 \pi^6}$ & $\frac{875 + 1536 \pi^2}{1518206976 \pi^6}$ \\[.75em]
\hline
\end{tabular}
\end{tabular}
\label{table:CFTdatabulk}
\end{table}
%
%!TEX root = ../2pt_function_wline.tex
%%%%%%%%%%%%%%%%%%%%%%%%%%%%%%%%%%%%%%%%%%%%%

\section{Conclusions}
\label{sec:conclusions}

Supersymmetric Wilson loops or lines have received a lot attention due to their role in AdS/CFT. The same can be said about correlators of 1/2-BPS operators, which have been a great source of non-trivial checks of the correspondence. Less work has been done on the interplay between local and non-local observables, and to bridge this gap was one of our main motivations. In this work we performed a detailed perturbative calculation of the two-point function of local chiral primaries in the presence of a supersymmetric line. We listed all the relevant diagrams up to next-to-leading order $\mathcal{O}(g^8)$, and by combining perturbative and defect CFT techniques we managed to obtain a full solution for the correlator as an expansion in (defect) conformal blocks. Roughly speaking, there are four infinite families of data that can be extracted from our results: bulk and defect OPE coefficients, bulk and defect anomalous dimensions. Of these four families, the bulk anomalous dimensions are the only ones that were known before, while the other three are brand new CFT data extracted from our calculations. For the bulk channel, we did not present an exact formula however the anomalous dimensions serve as highly non-trivial consistent checks for our results.

An interesting follow up to this work is to study the same two-point function but in the strong coupling limit $g^2N \to \infty$. Obviously perturbation theory cannot be used to study this regime, and even though holography can in principle provide us with the answer, explicit supergravity/string calculations are in general quite involved. A promising approach that has proven useful recently is to use analytic bootstrap techniques to constrain holographic correlators. The most systematic way of achieving this is by studying certain discontinuities of the correlator, which can be used to reconstruct the CFT data \cite{Caron-Huot:2017vep}. For planar $\mathcal{N}=4$ SYM at strong coupling this approach is particularly efficient \cite{Alday:2017vkk,Caron-Huot:2018kta}. In the supergravity approximation the spectrum of the theory is sparse and most operators appearing in the OPE are killed by the discontinuity of \cite{Caron-Huot:2017vep}. This means the full correlator can be reconstructed from a finite number of contributions. In defect CFT there exist similar formulas that reconstruct bulk and defect channels from a discontinuity \cite{Lemos:2017vnx,Liendo:2019jpu}, because the bulk channel OPE is not modified by the presence of a defect, it is likely that the two-point setup that we studied in this paper can also be reconstructed starting from a finite number of blocks. We plan to explore this in the near future.

A closely similar system that might benefit from the hybrid  perturbative/CFT approach used in this work is the four-point function of 1/2-BPS operators along the line. As discussed in the introduction, correlators of this type have been studied recently, but as far as we know, no one has studied the full four-point function perturbatively in the coupling yet.\footnote{Recent interesting work obtained some CFT data to high orders using integrability-based techniques \cite{Grabner:2020nis}.} 

Finally, there have been similar developments on line defects in ABJM \cite{Bianchi:2017ozk,Bianchi:2018scb,Bianchi:2020hsz} and $4d$ $\mathcal{N}=2$ theories \cite{Gimenez-Grau:2019hez}, and it would be interesting to study two-point functions of local operators in these systems as well, perhaps using some of the techniques presented in this work.

\acknowledgments

We are particularly grateful to T.~Klose for collaboration during several stages of this work.
We also thank A.~Gimenez-Grau, C.~Meneghelli, and M.~Preti for discussions and comments. PL is supported by the DFG through the Emmy Noether research group ``The Conformal Bootstrap Program'' project number 400570283. JB is funded by the Deutsche Forschungsgemeinschaft (DFG, German Research Foundation) -- Projektnummer 417533893/GRK2575 ``Rethinking Quantum Field 
Theory''.

\appendix

%!TEX root = ../2pt_function_wline.tex
%%%%%%%%%%%%%%%%%%%%%%%%%%%%%%%%%%%%%%%%%%%%%

\section{Superconformal Blocks}
\label{app:blocks}

In this appendix we list the superblocks that are used for writing down the correlator in section \ref{subsec:superblocksandcrossingsymmetry}.

\subsection{Bulk Channel}
\label{subsec:bulkchannel}

We start with the bulk channel. The spacetime blocks are eigenfunctions of the Casimir operator corresponding to the full conformal group $\mathfrak{so} (1,5)$. The solutions of the Casimir equation were originally studied in \cite{Billo:2016cpy}, in this paper however we will follow the conventions of \cite{Isachenkov:2018pef}:
\begin{align}
f_{\Delta,l} (z,\bar{z}) & = \sum_{m=0}^\infty \sum_{n=0}^\infty \frac{4^{m-n}}{m! n!} \frac{\left( - \frac{l}{2} \right)_m \left( \frac{l}{2} \right)_m \left( \frac{2 - l - \Delta}{2} \right)_m}{\left( - l \right)_m \left( \frac{3 - l - \Delta}{2} \right)_m} \frac{\left( \frac{\Delta - 1}{2} \right)_n^2 \left( \frac{\Delta + l }{2} \right)_n}{\left( \Delta - 1 \right)_n \left( \frac{\Delta + l + 1}{2} \right)_n} \frac{\left( \frac{\Delta + l}{2} \right)_{n-m}}{\left( \frac{\Delta + l - 1}{2} \right)_{n-m}} \notag \\
& \qquad \times\ _4F_3 \left( -n, -m , \frac{1}{2} , \frac{\Delta - l - 2}{2} , \frac{2 - \Delta + l - 2n}{2} , \frac{\Delta + l - 2m}{2} , \frac{\Delta - l - 1}{2} ; 1 \right) \notag \\
& \qquad \qquad \times\ _2F_1 \left( \frac{\Delta + l}{2} -m+n , \frac{\Delta + l}{2} -m+n , \Delta + l - 2(m-n) ; 1 - z\bar{z} \right) \notag \\
& \qquad \qquad \qquad \qquad \qquad \qquad \times \left[ (1-z)(1-\bar{z}) \right]^{\frac{\Delta - l}{2} + m + n} (1-z\bar{z})^{l - 2m}\,,
\label{eq:bulkstblock}
\end{align}
We note that there is no known closed form for the expansion of these blocks in the bulk OPE limit $z \sim \bar{z} \sim 1$. The R-symmetry blocks take a simple form:
\begin{equation}
h_{-k} (\omega) = \left( \frac{\omega}{(1-\omega)^2} \right)^{-k/2}\ _2F_1 \left( - \frac{k}{2} , - \frac{k}{2} , - k - 1 ; - \frac{(1-\omega)^2}{4 \omega} \right)\,.
\label{eq:bulkRsymmetry}
\end{equation}

In order to find the superblocks corresponding to each relevant representation of the full conformal group, an ansatz can be written based on the content of the supermultiplet. Applying the superconformal Ward identities given in eq. (\ref{eq:WI}) on this ansatz then fixes the coefficients in front of each term, up to an overall normalization constant. It is standard to set the coefficient of the term with the highest weight to $1$, and it is the convention that we use throughout this appendix and section \ref{sec:cft_data}. The blocks have been derived in \cite{Liendo:2016ymz} for the case of a codimension-one defect, from which we can extract the ones for the line defect by analytic continuation. Note that the superblock corresponding to the identity operator $\mathds{1}$ is just $1$.

\subsubsection{$\mathcal{B}_{[0,2k,0]}$ Superblocks}
\label{subsubsec:B02k0}

The superblocks corresponding to the $1/2$-BPS operators $\mathcal{B}_{[0,2k,0]}$ read
\begin{align}
\mathcal{G}_{\mathcal{B}_{[0,2k,0]}} & = f_{2k,0} h_{-2k} + \alpha_1 f_{2k+2,2} h_{-2k+2} + \alpha_2 f_{2k+4,0} h_{-2k+4}\,,
\end{align}
where we suppressed the dependence on $z, \bar{z}$ and $\omega$ on the RHS for compactness. As explained above, the coefficients $\alpha_i$ can be obtained by applying the Ward identities on the superblock. They are given explicitly in the ancillary \textsc{Mathematica} file attached as supporting material.

\subsubsection{$\mathcal{C}_{[0,2,0],l}$ Superblocks}
\label{subsubsec:C020}

The semishort operators $\mathcal{C}_{[0,2,0],l}$ lead to the following blocks:
\begin{align}
\mathcal{G}_{\mathcal{C}_{[0,2,0],l}} & = f_{l+4,l} h_{-2} + \beta_1 f_{l+6,l-2} h_0 + f_{l+6,l+2} ( \beta_{2,1} h_{-4} + \beta_{2,2} h_{-2} + \beta_{2,3} h_0 ) \notag \\
& \qquad + f_{l+8,l} ( \beta_{3,1} h_{-2} + \beta_{3,2} h_0 ) +\beta_4  f_{l+8,l+4} h_{-2} + \beta_5 f_{l+10,l+2} h_0\,.
\end{align}
As before, the coefficients $\beta_i$ are given explicitly in the ancillary \textsc{Mathematica} file.

\subsubsection{$\mathcal{A}^{\Delta}_{[0,0,0],l}$ Superblocks}
\label{subsubsec:A}

For the long operators $\mathcal{A}^{\Delta}_{[0,0,0],l}$, we obtain the following superblocks:
\begin{align}
\mathcal{G}_{\mathcal{A}^{\Delta}_{[0,0,0],l}} & = f_{\Delta,l} h_0 + f_{\Delta+2,l-2} ( \eta_{1,1} h_{-2} + \eta_{1,2} h_0 ) + f_{\Delta+2,l+2} ( \eta_{2,1} h_{-2} + \eta_{2,2} h_0 ) \notag \\
& \qquad + \eta_{3} f_{\Delta+4,l-4} h_0 + f_{\Delta+4,l} ( \eta_{4,1} h_{-4} + \eta_{4,2} h_{-2} + \eta_{4,3} h_0 ) + \eta_{5} f_{\Delta+4,l+4} h_0 \notag \\
& \qquad\qquad + f_{\Delta+6,l-2} ( \eta_{6,1} h_{-2} + \eta_{6,2} h_0 ) + f_{\Delta + 6,l+2} ( \eta_{7,1} h_{-2} + \eta_{7,2} h_0 ) \notag \\
& \qquad\qquad\qquad + \eta_8 f_{\Delta+8,l} h_0\,.
\end{align}
The coefficients $\eta_i$ can be found in the ancillary \textsc{Mathematica} file. These operators are not protected against corrections to their scaling dimension $\Delta$, and they are responsible for the $\log$ terms that appear in perturbation theory.

\subsection{Defect Channel}
\label{subsec:defectchannel}

We list now the superblocks present in eq. (\ref{eq:defectchannelexpansion}) for the expansion of the correlator in the defect channel. The spacetime blocks are in this case eigenfunctions of the Casimir operator of the symmetry group preserved by the defect, i.e. $\mathfrak{so}(1,2) \times \mathfrak{so}(3)$. As a consequence, they factorize and take an elegant form \cite{Billo:2016cpy}:
\begin{equation}
\hat{f}_{\hat{\Delta},0,s} (z,\bar{z}) = z^{\frac{\hat{\Delta} - s}{2}} \bar{z}^{\frac{\hat{\Delta} + s}{2}}\ _2F_1 \left( -s, \frac{1}{2}, \frac{1}{2} - s ; \frac{z}{\bar{z}} \right)\ _2F_1 \left( \hat{\Delta} , \frac{1}{2} , \frac{1}{2} + \hat{\Delta} ; z\bar{z} \right)\,.
\label{eq:defectstblock}
\end{equation}
The defect R-symmetry blocks are given by:
\begin{equation}
\hat{h}_{\hat{k}} (\omega) = \left( \frac{(1-\omega)^2}{\omega} \right)^{\hat{k}}\ _2F_1 \left( -\hat{k}-1 , -\hat{k} , -2(\hat{k}+1) ; - \frac{4 \omega}{(1-\omega)^2} \right)\,.
\label{eq:defectRsymmetry}
\end{equation}

In the same way as for the bulk channel, we write an ansatz for each representation of the (defect) $\mathfrak{osp}(4|4)$ algebra and solve for the coefficients of each term by applying the Ward identities. The blocks are also normalized such that the term with the highest weight has coefficient $1$. They can be found in \cite{Liendo:2016ymz} for the case of a codimension-one defect. As mentioned for the bulk channel, it is then possible to derive the blocks for the line defect simply by analytic continuation. Note that the superblock corresponding to the identity operator $\mathds{1}$ is just $1$.

\subsubsection{$(B,+)_{\hat{k}}$ Superblocks}
\label{subsubsec:B+}

For the $1/2$-BPS operators $(B,+)_{\hat{k}}$, we have the following blocks:
\begin{equation}
\hat{\mathcal{G}}_{(B,+)_{\hat{k}}} ( z , \bar{z} , \omega ) = \hat{f}_{\hat{k},0} \hat{h}_{\hat{k}}  + \hat{\alpha}_1 \hat{f}_{\hat{k}+1,1} \hat{h}_{\hat{k}-1} + \hat{\alpha}_2 \hat{f}_{\hat{k}+2,0} \hat{h}_{\hat{k}-2}\,,
\label{eq:blockB+}
\end{equation}
where as usual we suppressed the dependence on $z,\bar{z}$ and $\omega$. Again, the coefficients $\hat{\alpha}_i$ are obtained by applying the Ward identities on the superblock. They can be found in the additional \textsc{Mathematica} notebook.

\subsubsection{$(B,1)_{[\hat{k},s]}$ Superblocks}
\label{subsubsec:B1}

The blocks corresponding to the $1/4$-BPS representations $(B,1)_{[\hat{k},s]}$ read
\begin{align}
\hat{\mathcal{G}}_{(B,1)_{[\hat{k},s]}} ( z , \bar{z} , \omega ) &= \hat{f}_{1+\hat{k}+s , s} \hat{h}_{\hat{k}} + \hat{\beta}_{1} \hat{f}_{2+\hat{k}+s , s-1} \hat{h}_{\hat{k}-1} + \hat{f}_{2+\hat{k}+s , s+1} ( \hat{\beta}_{2,1} \hat{h}_{\hat{k}-1} + \hat{\beta}_{2,2} \hat{h}_{\hat{k}+1} ) \notag \\
& \qquad + \hat{f}_{3+\hat{k}+s , s} ( \hat{\beta}_{3,1} \hat{h}_{\hat{k}-2} + \hat{\beta}_{3,2} \hat{h}_{\hat{k}} ) + \hat{\beta}_4 \hat{f}_{3+\hat{k}+s , s+2} \hat{h}_{\hat{k}} \notag \\
& \qquad\qquad + \hat{\beta}_5 \hat{f}_{4+\hat{k}+s , s+1} \hat{h}_{\hat{k}-1}\,.
\label{eq:blockB1}
\end{align}
The coefficients $\hat{\beta}_i$ can be found in the \textsc{Mathematica} notebook.

\subsubsection{$L ^{\hat{\Delta}}_{[\hat{k},s]}$ Superblocks}
\label{subsubsec:Ldefect}

For the long operators $L^{\hat{\Delta}}_{[\hat{k},s]}$, the blocks take the following form:
\begin{align}
\hat{\mathcal{G}}_{L^{\hat{\Delta}}_{[\hat{k},s]}} ( z , \bar{z} , \omega ) &= \hat{f}_{\hat{\Delta} , s} \hat{h}_{\hat{k}} + \hat{f}_{\hat{\Delta}+1 , s-1} ( \hat{\eta}_{1,1} \hat{h}_{\hat{k}+1} + \hat{\eta}_{1,2} \hat{h}_{\hat{k}-1} ) + \hat{f}_{\hat{\Delta}+1 , s+1} ( \hat{\eta}_{2,1} \hat{h}_{\hat{k}+1} + \hat{\eta}_{2,2} \hat{h}_{\hat{k}-1} ) \notag \\
& \qquad + \hat{\eta}_3 \hat{f}_{\hat{\Delta}+2 , s-2} \hat{h}_{\hat{k}} + \hat{f}_{\hat{\Delta}+2 , s} (\hat{\eta}_{4,1} \hat{h}_{\hat{k}+2} + \hat{\eta}_{4,2} \hat{h}_{\hat{k}} + \hat{\eta}_{4,3} \hat{h}_{\hat{k}-2}) \notag \\
& \qquad\quad + \hat{\eta}_5 \hat{f}_{\hat{\Delta}+2 , s+2} \hat{h}_{\hat{k}} + \hat{f}_{\hat{\Delta}+3 , s-1} (\hat{\eta}_{6,1} \hat{h}_{\hat{k}+1} + \hat{\eta}_{6,2} \hat{h}_{\hat{k}-1}) \notag \\
& \qquad\quad\quad + \hat{f}_{\hat{\Delta}+3 , s+1} (\hat{\eta}_{7,1} \hat{h}_{\hat{k}+1} + \hat{\eta}_{7,2} \hat{h}_{\hat{k}-1}) + \hat{\eta}_8 \hat{f}_{\hat{\Delta}+4 , s} \hat{h}_{\hat{k}}\,.
\label{eq:blockL}
\end{align}
The coefficients $\hat{\eta}_i$ can be found in the \textsc{Mathematica} notebook. Like their bulk counterpart, scaling dimensions $\hat{\Delta}$ of (\ref{eq:blockL}) can receive anomalous corrections which give rise to $\log$ terms.

\subsection{Series Expansions and CFT Data}
\label{subsec:seriesexpansions}

In this section, we list a few relations between the CFT data coefficients (as defined in (\ref{eq:bulkchannelexpansion}), (\ref{eq:defectchannelexpansion})) and the expansions of the correlator in the OPE limits given in (\ref{eq:expsuperblocksbulk}), (\ref{eq:expsuperblocksdefect}).

\subsubsection{Bulk Channel}
\label{subsec:expbulkchannel}

We start by the bulk channel. We will give the first few relations for each R-symmetry channel as an illustration of their recursiveness. For the $0$-channel, they read
\begingroup
\allowdisplaybreaks
\begin{subequations}
\begin{gather}
\left. c^{(0)}_{0,-4,m} \right|_{m \neq \lbrace -4,-2 \rbrace} = 0\,, \qquad c^{(0)}_{0,-4,-4} = A\,, \qquad c^{(0)}_{0,-4,-2} = - A\,, \\
c^{(0)}_{0,-3,m} = 0\ \forall m\,, \\
\left. c^{(0)}_{0,-2,m} \right|_{m\ \text{odd}} = 0\,, \qquad c^{(0)}_{0,-2,-2} = A + \frac{B}{12} + E_{2,0}\,, \notag \\
 c^{(0)}_{0,-2,0} = - \frac{B}{24} - \frac{1}{2} E_{2,0} + \frac{1}{4} g^2 E_{2,0} \gamma^{(1)}_{2,0} + \frac{1}{4} g^4 E_{2,0} \gamma^{(2)}_{2,0}\,, \\
\left. c^{(0)}_{0,-1,m} \right|_{m\ \text{even}} = 0\,, \qquad ..., \\
\left. c^{(0)}_{1,-2,m} \right|_{m\ \text{odd}} = 0\,, \qquad c^{(0)}_{1,-2,-2} = \frac{1}{2} g^2 E_{2,0} \gamma^{(1)}_{2,0} + \frac{1}{2} g^4 E_{2,0} \gamma^{(2)}_{2,0}\,, \qquad ...
\end{gather}
\label{eq:bulkrelations0channel}%
\end{subequations}
Note that the $\log$ terms can only contain the CFT data related to the long representations. Moreover, the coefficient $A$ (corresponding to the identity operator) can only appear in this channel.

For the $1$-channel, we have
\begin{subequations}
\begin{gather}
\left. c^{(1)}_{0,-2,m} \right|_{m\ \text{odd}} = 0\,, \qquad c^{(1)}_{0,-2,-2} = \frac{B}{4}\,, \qquad c^{(1)}_{0,-2,0} = - \frac{B}{8}\,, \\
c^{(1)}_{0,-2,2} = - \frac{B}{24} - \frac{E_{2,0}}{8} - \frac{1}{32} g^2 E_{2,0} \gamma_{2,0}^{(1)} + \frac{1}{128} g^4 E_{2,0} \left\lbrace ( \gamma^{(1)}_{2,0} )^2 - 4 \gamma^{(2)}_{2,0} \right\rbrace\,, \\
c^{(1)}_{0,-1,3} = - \frac{3}{20} E_{3,0} - \frac{1}{50} g^2 E_{3,0} \gamma^{(1)}_{3,0} + \frac{1}{250} g^4 E_{3,0} \left\lbrace ( \gamma^{(1)}_{3,0} )^2 - 5 \gamma^{(2)}_{3,0} \right\rbrace\,, \\
\left. c^{(1)}_{0,0,m} \right|_{m\ \text{odd}} = 0\,, \qquad c^{(1)}_{0,0,0} = \frac{B}{16} + \frac{C}{20} + \frac{D_0}{4}\,, \qquad ... \\
c^{(1)}_{1,-2,2} = - \frac{1}{16} g^2 E_{2,0} \gamma^{(1)}_{2,0} - \frac{1}{64} g^4 E_{2,0} \left\lbrace ( \gamma^{(1)}_{2,0} )^2 + 4 \gamma^{(2)}_{2,0} \right\rbrace\,, \qquad ...
\end{gather}
\label{eq:bulkrelations1channel}
\end{subequations}

The first few relations involving the $2$-channel are
\begin{subequations}
\begin{gather}
\left. c^{(2)}_{0,0,m} \right|_{m\ \text{odd}} = c^{(2)}_{0,0,2} = 0\,, \qquad c^{(2)}_{0,0,0} = \frac{C}{16}\,, \qquad c^{(2)}_{0,0,4} = - \frac{C}{160} - \frac{D_0}{32}\,, \\
c^{(2)}_{0,1,m} = 0\ \forall m\,, \\
\left. c^{(2)}_{0,2,m} \right|_{m\ \text{odd}} = 0\,, \qquad c^{(2)}_{0,2,2} = \frac{1}{64} g^2 E_{2,0} \gamma^{(1)}_{2,0} + \frac{1}{256} g^4 E_{2,0} \left\lbrace 4 \gamma^{(2)}_{2,0} - (\gamma^{(1)}_{2,0} )^2 \right\rbrace\,, \\
\resizebox{.9 \textwidth}{!} 
{
$\left. c^{(2)}_{0,3,m} \right|_{m\ \text{even}} = 0\,, \qquad c^{(2)}_{0,3,3} = \frac{E_{3,0}}{80} + \frac{1}{100} g^2 E_{3,0} \gamma^{(1)}_{3,0} + \frac{1}{500} g^4 E_{3,0} \left\lbrace 5 \gamma^{(2)}_{3,0} - ( \gamma^{(1)}_{3,0} )^2 \right\rbrace\,, \qquad ...$
} \\
c^{(2)}_{1,2,2} = \frac{1}{128} g^4 E_{2,0} ( \gamma^{(1)}_{2,0} )^2\,, \qquad c^{(2)}_{1,2,4} = \frac{1}{256} g^4 E_{2,0} ( \gamma^{(1)}_{2,0} )^2\,, \qquad ...
\end{gather}
\label{eq:bulkrelations2channel}%
\end{subequations}
\endgroup
Not all superblocks contribute to the $2$-channel; indeed, only the $1/2$-BPS operators $\mathcal{B}_{[0,4,0]}$ and the longs $\mathcal{A}^{\Delta}_{[0,0,0],l}$ are relevant for this channel.

\bigskip

Note that all these expressions have been truncated at $\mathcal{O}(g^4)$ for the purpose of this work.

\subsubsection{Defect Channel}
\label{subsec:expdefectchannel}

We move now our attention to the defect channel. For the $0$-channel, the relations between CFT data and coefficients $\hat{c}^{(0)}_{k,l,m}$ read
\begingroup
\allowdisplaybreaks
\begin{subequations}
\begin{gather}
\hat{c}^{(0)}_{0,0,m} = \hat{c}^{(0)}_{0,1,m} = 0\,, \\
\left. \hat{c}^{(0)}_{0,2,m} \right|_{m\ \text{odd}} = 0\,, \qquad \hat{c}^{(0)}_{0,2,2} = 16 \hat{C}\,, \qquad \hat{c}^{(0)}_{0,2,4} = 8 \hat{E}_0, \qquad \hat{c}^{(0)}_{0,2,6} = 12 \hat{E}_1\,, \\
\left. \hat{c}^{(0)}_{0,3,m} \right|_{m\ \text{even}} = 0\,, \qquad \hat{c}^{(0)}_{0,3,3} = 8 g^2 \hat{F}_{1,0} \hat{\gamma}^{(1)}_{1,0} - 4 g^4 \hat{F}_{1,0} \left\lbrace ( \hat{\gamma}^{(1)}_{1,0} )^2 - 2 \hat{\gamma}^{(2)}_{1,0} \right\rbrace\,, \qquad ... \\
\hat{c}^{(0)}_{1,3,3} = 4 g^4 \hat{F}_{1,0} ( \hat{\gamma}^{(1)}_{1,0} )^2\,, \qquad \hat{c}^{(0)}_{1,3,5} = 6 g^4 \hat{F}_{2,1} ( \hat{\gamma}^{(1)}_{2,1} )^2\,, \qquad ...
\end{gather}
\label{eq:defectrelations0channel}%
\end{subequations}
Note that not all the defect superblocks contribute to the $0$-channel: only the coefficients $\hat{C}$, $\hat{E}_s$ and $F_{\hat{\Delta},s}$ are present.

For the $1$-channel, we have
\begin{subequations}
\begin{gather}
\hat{c}^{(1)}_{0,0,m} = 0\,, \\
\left. \hat{c}^{(1)}_{0,1,m} \right|_{m\ \text{even}} = 0\,, \qquad \hat{c}^{(1)}_{0,1,1} = 4 \hat{B}\,, \\
\left. \hat{c}^{(1)}_{0,2,m} \right|_{m\ \text{odd}} = 0\,, \qquad \hat{c}^{(1)}_{0,2,2} = 14 \hat{C} - 4 \hat{E}_0\,, \qquad ... \\
\left. \hat{c}^{(1)}_{1,1,m} \right|_{m\ \text{even}} = 0\,, \qquad \hat{c}^{(1)}_{1,1,3} = - g^2 \hat{F}_{1,0} \hat{\gamma}^{(1)}_{1,0} - \frac{1}{2} g^4 \hat{F}_{1,0} \left\lbrace ( \hat{F}_{1,0} )^2 + 2 \hat{\gamma}^{(2)}_{1,0} \right\rbrace\,, \qquad ...
\end{gather}
\label{eq:defectrelations1channel}
\end{subequations}

The relations for the $2$-channel are:
\begin{subequations}
\begin{gather}
\hat{c}^{(2)}_{0,0,0} = \hat{A}\,, \qquad \hat{c}^{(2)}_{0,0,m} = 0\,, \\
\left. \hat{c}^{(2)}_{0,1,m} \right|_{m\ \text{even}} = 0\,, \qquad \hat{c}^{(2)}_{0,1,1} = 2\hat{B} - (\hat{D}_0 - \hat{F}_{1,0})\,, \\
\resizebox{.85 \textwidth}{!} 
{
$\hat{c}^{(2)}_{0,1,3} = - \frac{2}{3} \hat{B} + (\hat{D}_0 - \hat{F}_{1,0}) - (\hat{D}_1 - \hat{F}_{2,1}) - \frac{1}{2} g^2 \hat{F}_{1,0} \hat{\gamma}^{(1)}_{1,0} + \frac{1}{4} g^4 \hat{F}_{1,0} \left\lbrace ( \hat{\gamma}^{(1)}_{1,0} )^2 - 2 \hat{\gamma}^{(2)}_{1,0} \right\rbrace\,,$
} \\
\left. \hat{c}^{(2)}_{0,2,m} \right|_{m\ \text{odd}} = 0\,, \qquad \hat{c}^{(2)}_{0,2,2} = \frac{16}{5} \hat{C} - 2 \hat{E}_0 + \hat{F}_{2,0}\,, \qquad ... \\
\hat{c}^{(2)}_{1,1,1} = \frac{1}{2} g^2 \hat{F}_{1,0} \hat{\gamma}^{(1)}_{1,0} + \frac{1}{2} g^4 \hat{F}_{1,0} \hat{\gamma}^{(2)}_{1,0}\,, \qquad ...
\end{gather}
\label{eq:defectrelations2channel}%
\end{subequations}
\endgroup
Note that the coefficient $\hat{A}$ is only present in this channel.
%
%!TEX root = ../2pt_function_wline.tex
%%%%%%%%%%%%%%%%%%%%%%%%%%%%%%%%%%%%%%%%%%%%%

\section{Insertion Rules}
\label{app:insertionrules}

We list in this section the insertion rules that are used for computing the Feynman diagrams in section \ref{sec:perturbative}. Those are derived from the Euclidean action of $\mathcal{N}=4$ SYM in $4$d Euclidean space, which is given in eq. (\ref{eq:action}). The propagators are given in (\ref{eq:propagators}). Note that the gauge group is $U(N)$ and that we work in the large $N$ limit. The generators obey the following commutation relation:
\begin{equation}
[ \tensor{T}{^a} , \tensor{T}{^b} ] = i \tensor{f}{^{ab}_c} T^c,
\end{equation}
in which $\tensor{f}{^{ab}_c}$ are the structure constants of the $\mathfrak{u}(N)$ Lie algebra. The generators are normalized as:
\begin{equation}
\text{Tr}\ \tensor{T}{^a} \tensor{T}{^b} = \frac{\tensor{\delta}{^{ab}}}{2}.
\end{equation}
Note that $\tensor{f}{^{ab0}} = 0$ and $\text{Tr}\ \tensor{T}{^a} = \sqrt{\frac{N}{2}} \tensor{\delta}{^{a0}}$. The (contracted) product of structure constants gives $\tensor{f}{^{abc}} \tensor{f}{_{abc}} = N (N^2 - 1) \sim N^3$, where the second equality holds in the large $N$ limit.

\subsection{Bulk Insertions}
\label{subsec:bulkinsertions}

We start by listing the bulk insertions rules needed for the computation of the Feynman diagrams, i.e. the insertions that follow from the $\mathcal{N}=4$ SYM action.

\bigskip

The only $2$-point insertion that is needed is the self-energy of the scalar propagator at one-loop, which is given by the following expression \cite{Erickson:2000af, Plefka:2001bu, Drukker:2008pi}:
\begin{align}
\propagatorSSEnotext\ &=\ \SSEone\ +\ \SSEtwo\ +\ \SSEthree\ +\ \SSEfour \notag \\
&= -2 g^4 N \tensor{\delta}{^{ab}} \tensor{\delta}{_{ij}} Y_{112}.
\label{eq:selfenergy}
\end{align}
The integral $Y_{112}$ is given in eq. (\ref{eq:Y112}) and presents a logarithmic divergence.

We also require only one $3$-point insertion, which is the vertex connecting two scalar fields and one gauge field. It is easy to obtain from the action (\ref{eq:action}) and it reads:
\begin{equation}
\vertexSSG\ = - g^4 \tensor{f}{^{abc}} \tensor{\delta}{^{ij}} \left( \partial_1 - \partial_2 \right)_\mu Y_{123}.
\label{eq:vertexSSG}
\end{equation}
The Y-integral is defined in eq. (\ref{subeq:Y123}) and its analytical expression can be found in (\ref{eq:Y123}).

Another relevant vertex is the $4$-scalars coupling. Similarly to the $3$-vertex, it is straightforward to read the corresponding Feynman rule from the action and perform the Wick contractions to get:
\begin{align}
\vertexSSSS &= -g^6 \left\lbrace f^{abe}f^{cde} \left( \delta_{ik}\delta_{jl} - \delta_{il}\delta_{jk} \right) + f^{ace}f^{bde} \left( \delta_{ij}\delta_{kl} - \delta_{il}\delta_{jk} \right) \right. \notag \\[-1.5em]
& \left. \qquad \qquad \qquad + f^{ade}f^{bce} \left( \delta_{ij}\delta_{kl} - \delta_{ik}\delta_{jl} \right) \right\rbrace X_{1234}.
\label{eq:vertexSSSS}
\end{align}
The X-integral can be found in eq. (\ref{eq:X1234}). We also use the $4$-coupling between $2$ scalars and $2$ gluons. This vertex reads:
\begin{equation}
\vertexSSGG =- g^6 \delta_{ij} \delta_{\mu\nu} (f^{ace}f^{bde} + f^{ade}f^{bce})\ X_{1234}.
\label{eq:vertexSSGG}
\end{equation}

There are two more sophisticated $4$-point insertions that we require. The first one reads:
\begin{equation}
\Hinsertone\ = g^6 \left\lbrace \tensor{\delta}{_{ik}} \tensor{\delta}{_{jl}} \tensor{f}{^{ace}} \tensor{f}{^{bde}} I_{13} I_{24} F_{13,24} + \tensor{\delta}{_{il}} \tensor{\delta}{_{jk}} \tensor{f}{^{ade}} \tensor{f}{^{bce}} I_{14} I_{23} F_{14,23} \right\rbrace,
\label{eq:Hinsertone}
\end{equation}
with $F_{13,24}$ as defined in (\ref{subeq:F1324}), while an analytical expression is given in (\ref{eq:FXYidentity}).

The last insertion rule needed is the following:
\begin{align}
\Hinserttwo &= g^6 \delta_{ij} f^{ace} f^{bde} \left[ 4 \partial_{1\mu} \partial_{2\nu} + 2 \left( \partial_{1\mu} \partial_{4\nu} + \partial_{3\mu} \partial_{2\nu} \right) + \partial_{3\mu} \partial_{4\nu} \right] H_{13,24} \notag \\
& \qquad \qquad \qquad \qquad + ( 1,\mu, a \leftrightarrow 2,\nu,b ).
\label{eq:Hinserttwo}
\end{align}
Note that, contrary to the integrals present in the other vertices, the H-integral, defined in eq. (\ref{subeq:H1324}), has no known closed form to the best of our knowledge.

\subsection{Defect Insertions}
\label{subsec:defectinsertions}

We derive here the defect insertion rules needed for the computation of the diagrams. We start by considering scalar and gluon insertions on the defect. The expression for the Maldacena-Wilson line given in eq. (\ref{eq:MaldacenaWilsonLoop}) can be expanded in the following way:
\begin{align*}
\mathcal{W}_l &= \frac{1}{N}\ \text{Tr} \left\lbrace \mathcal{P} \exp \int d\tau \left( i \tensor{\dot{x}}{^\mu} \tensor{A}{_\mu}(x) + \left| \dot{x} \right| \tensor{\theta}{_i} \tensor{\phi}{^i} \right) \right\rbrace \notag \\
& = 1 + \frac{1}{N}\ \text{Tr}\ \int d\tau \left( i \tensor{A}{_\mu} (x) \tensor{\dot{x}}{^\mu} + \tensor{\phi}{^i} (x) \left| \dot{x} \right| \tensor{\theta}{_i} \right) \notag \\
& \qquad + \frac{1}{2! N}\ \text{Tr}\ \mathcal{P} \int d\tau_1\ d\tau_2 \left( i \tensor{A}{_\mu} (x_1) \tensor*{\dot{x}}{_1^\mu} + \tensor{\phi}{^i} (x_1) \left| \dot{x}_1 \right| \tensor{\theta}{_i} \right) \left( i \tensor{A}{_\nu} (x_2) \tensor*{\dot{x}}{_2^\nu} + \tensor{\phi}{^j} (x_2) \left| \dot{x}_2 \right| \tensor{\theta}{_j} \right) \notag \\
& \qquad \qquad + ...\,
\end{align*}
in which each term refers to a certain number of points on the line.

\bigskip

It is obvious that the tree level insertion, which corresponds to Feynman diagrams where the operators are disconnected from the line, reads:
\begin{equation}
\wilsonrulezero\ = 1.
\label{eq:wilsonrulezero}
\end{equation}

At first order, the Wilson-line insertions are proportional to the trace of a single generator:
\begin{equation}
\wilsonruleS\ ,\ \wilsonruleG\ \propto\ \text{Tr}\ \tensor{T}{^a}.
\label{eq:wilsonruleS}
\end{equation}
This is zero when there is at least one vertex in the diagram, and we encounter only such cases in this work.

The first non-trivial contribution appears at second order. Using the cyclicity of the trace, it is possible to remove the path ordering in order to obtain:
\begin{equation}
\wilsonruleSS\ = \frac{1}{4N} \tensor{\theta}{_i} \tensor{\theta}{_j} \tensor{\delta}{^{ab}} \int d\tau_1 \int d\tau_2\ \bigl\langle \tensor*{\phi}{_{1,a}^i} \tensor*{\phi}{_{2,b}^j} ...\ \bigr\rangle.
\label{eq:wilsonruleSS}
\end{equation}
Note that all permutations (i.e. all possible path orderings) are considered in this expression, and the diagrammatic representation should be understood as such.

The second-order expression for two gluon emissions in also relevant, and it reads:
\begin{equation}
\wilsonruleGG\ = - \frac{1}{4N} \tensor{\delta}{^{ab}} \int d\tau_1 \int d\tau_2\ \tensor*{\dot{x}}{_{\smash{1}}^{\smash{\mu}}} \tensor*{\dot{x}}{_{\smash{2}}^{\smash{\nu}}}\ \bigl\langle \tensor*{A}{_{1,\mu,a}} \tensor*{A}{_{2,\nu,b}} ...\ \bigr\rangle.
\label{eq:wilsonruleGG}
\end{equation}

At third order, the only contribution that we have to consider is the one with two scalar and one gluon lines, which can be expressed as:
\begin{align}
\wilsonruleSSG\ & = - \frac{1}{8 N} \tensor{\theta}{_i} \tensor{\theta}{_j} \tensor{f}{^{abc}} \int d\tau_1 \int d\tau_2 \int d\tau_3\ \varepsilon \left( \tau_1 \tau_2 \tau_3 \right) \tensor*{\dot{x}}{_2^\mu}\ \bigl\langle \tensor*{\phi}{_{1,a}^i} \tensor*{A}{_{2,\mu,b}} \tensor*{\phi}{_{3,c}^j} ...\ \bigr\rangle \notag \\
& \qquad \qquad \qquad \qquad + \frac{i}{8 N} \tensor{\theta}{_i} \tensor{\theta}{_j} \tensor{d}{^{abc}} \int d\tau_1 \int d\tau_2 \int d\tau_3\ \tensor*{\dot{x}}{_2^\mu}\ \bigl\langle \tensor*{\phi}{_{1,a}^i} \tensor*{A}{_{2,\mu,b}} \tensor*{\phi}{_{3,c}^j} ...\ \bigr\rangle,
\label{eq:wilsonruleSSG}
\end{align}
where we defined the path-ordering symbols $\Theta \left( \tau_{ijk} \right)$ and $\varepsilon \left( \tau_i \tau_j \tau_k \right)$ as follows:
\begin{subequations}
\begin{gather}
\Theta \left( \tau_{ijk} \right) := \Theta \left( \tau_{ij} \right) \Theta \left( \tau_{jk} \right), \quad \text{with}\ \tau_{ij}:=\tau_{i}-\tau_{j}\\
\varepsilon \left( \tau_i \tau_j \tau_k \right) := \text{sgn} \left( \tau_{ij} \right) \text{sgn} \left( \tau_{ik} \right) \text{sgn} \left( \tau_{jk} \right).
\label{eq:pathorderingsymbol}
\end{gather}
\end{subequations}
The second definition is needed in order to account for the antisymmetry of $\tensor{f}{^{abc}}$. Note that the second term in (\ref{eq:wilsonruleSSG}) always vanishes in this work.

Finally, at order four we only need the expression with scalar lines:
\begin{align}
\wilsonruleSSSS\  &= \frac{1}{4! N} \tensor{\theta}{_i} \tensor{\theta}{_j} \tensor{\theta}{_k} \tensor{\theta}{_l} \int d\tau_1 \int d\tau_2 \int d\tau_3 \int d\tau_4 \left( \Theta \left( \tau_{1234} \right)\ \text{Tr}\ \tensor{T}{^a} \tensor{T}{^b} \tensor{T}{^c} \tensor{T}{^d} \right. \notag \\
& \qquad \qquad \qquad \qquad \left. + \Theta \left( \tau_{1243} \right)\ \text{Tr}\ \tensor{T}{^a} \tensor{T}{^b} \tensor{T}{^d} \tensor{T}{^c} + ... \right) \bigl\langle \tensor*{\phi}{_{1,a}^i} \tensor*{\phi}{_{2,b}^j} \tensor*{\phi}{_{3,c}^k} \tensor*{\phi}{_{4,d}^l} ...\ \bigr\rangle,
\label{eq:wilsonruleSSSS}
\end{align}
with:
\begin{equation}
\Theta \left( \tau_{ijkl} \right) \equiv \Theta \left( \tau_{ij} \right) \Theta \left( \tau_{jk} \right) \Theta \left( \tau_{kl} \right).
\label{eq:pathorderingsymbol2}
\end{equation}
We only encounter this expression in diagrams that do not contain vertices, and in that case it simply reduces to:
\begin{equation*}
\frac{1}{N} \tensor{\theta}{_i} \tensor{\theta}{_j} \tensor{\theta}{_k} \tensor{\theta}{_l} \text{Tr}\ \tensor{T}{^a} \tensor{T}{^b} \tensor{T}{^c} \tensor{T}{^d} \int d\tau_1 \int d\tau_2 \int d\tau_3 \int d\tau_4\ \Theta \left( \tau_{1234} \right) \langle \tensor*{\phi}{_{1,a}^i} \tensor*{\phi}{_{2,b}^j} \tensor*{\phi}{_{3,c}^k} \tensor*{\phi}{_{4,d}^l} ...\ \rangle.
\end{equation*}
%
%!TEX root = ../2pt_function_wline.tex
%%%%%%%%%%%%%%%%%%%%%%%%%%%%%%%%%%%%%%%%%%%%%

\section{Integrals}
\label{app:integrals}

In this appendix, we give some detail about the integrals used in this paper.

\subsection{Elementary Integrals}
\label{subsec:elementaryintegrals}

We list here the elementary integrals encountered in this work. We often run into:
\begin{subequations}
\begin{gather}
\int_{-\infty}^\infty \frac{d\tau}{x^2 + \tau^2} = \frac{\pi}{|x|}\,. \label{eq:elementaryintegral1} \\
\intertext{Because of the path-ordering the limits of integration are often not infinite. We sometimes come across the following expression:}
\int_{\tau_i}^\infty \frac{d\tau_j}{x^2 + \tau_j^2} = \frac{1}{|x|} \left( \frac{\pi}{2} - \tan^{-1} \frac{\tau_i}{|x|} \right)\,. \label{eq:elementaryintegral2}
\intertext{It also happens that both limits of integration are finite, and in this case the following expression holds:}
\int_{\tau_i}^{\tau_j} \frac{d\tau_k}{x^2 + \tau_k^2} = \frac{1}{|x|} \left( \tan^{-1} \frac{\tau_j}{|x|} - \tan^{-1} \frac{\tau_i}{|x|} \right)\,. \label{eq:elementaryintegral3}
\end{gather}
\end{subequations}

\subsection{Standard Integrals}
\label{subsec:standardintegrals}

In the computation of the Feynman diagrams at next-to-leading order, we encounter three-, four- and five-point massless Feynman integrals, which we define as follows:
\begin{subequations}
\begin{gather}
Y_{123} := \int d^4 x_4\ I_{14} I_{24} I_{34}\,, \label{subeq:Y123} \\
X_{1234} := \int d^4 x_5\ I_{15} I_{25} I_{35} I_{45}\,, \label{subeq:X1234} \\
H_{13,24} := \int d^4 x_{56}\ I_{15} I_{35} I_{26} I_{46} I_{56}\,, \label{subeq:H1324}
\intertext{with $I_{12}$ defined in eq. (\ref{eq:I12}). In the last expression we have defined $d^4 x_{56} := d^4 x_5\ d^4 x_6$ for brevity. The letter assigned to each integral makes sense when drawing the propagators. We also encounter the following expression:}
F_{13,24} := \frac{\left( \partial_1 - \partial_3 \right) \cdot \left(\partial_2 - \partial_4 \right)}{I_{13}I_{24}} H_{13,24}\,. \label{subeq:F1324}
\end{gather}
\end{subequations}
The notation presented above has already been used in e.g. \cite{Beisert:2002bb, Drukker:2008pi}. The $3$- and $4$-point massless integrals in Euclidean space are conformal and have been solved analytically (see e.g. \cite{tHooft:1978jhc, Usyukina:1992wz} and \cite{Drukker:2008pi} for the modern notation). The so-called X-integral is given by:
\begin{equation}
X_{1234} = \frac{1}{16 \pi^2} I_{13} I_{24}\ \Phi (r, s)\,,
\label{eq:X1234}
\end{equation}
where we have defined:
\begin{subequations}
\begin{gather}
\Phi(r,s) := \frac{1}{A} \text{Im} \left\lbrace \Li_2\ e^{i\varphi} \sqrt{\frac{r}{s}} + \log \sqrt{\frac{r}{s}} \log \left( 1 - e^{i\varphi} \sqrt{\frac{r}{s}} \right) \right\rbrace\,, \label{subeq:Phi} \\
e^{i\varphi} := i \sqrt{-\frac{1-r-s-4iA}{1-r-s+4iA}}\,, \qquad A := \frac{1}{4} \sqrt{4rs - (1 - r - s)^2}\,, \label{subeq:eivarphiandA} \\
r := \frac{I_{13}I_{24}}{I_{12}I_{34}}\,, \qquad \qquad \qquad \qquad s := \frac{I_{13}I_{24}}{I_{14}I_{23}}\,. \label{subeq:Xrands}
\end{gather}
\end{subequations}
The Y-integral can easily be obtained from this expression by taking the following limit:
\begin{equation}
Y_{123} = \lim_{x_4 \to \infty} (2\pi)^2 x_4^2\ X_{1234} = \frac{1}{16 \pi^2} I_{12}\ \Phi (r, s)\,, \label{eq:Y123}
\end{equation}
where in the last equality the conformal ratios are defined as:
\begin{equation}
r := \frac{I_{12}}{I_{13}}, \qquad \qquad s := \frac{I_{12}}{I_{23}}\,. \label{eq:Yrands}
\end{equation}
We note that both integrals are finite when the points are distinct. Furthermore, eq. (\ref{subeq:Phi}) implies that the function $\Phi$ vanishes in the limit $r \to \infty$ and $s \to \infty$, and that $\Phi(r,s) = \Phi(1/r,s/r)/r$ \cite{Beisert:2002bb}. The latter simply means that the conformal ratios can be defined arbitrarily, as long as consistency is respected.

The H-integral seems to have no known closed form so far, but (\ref{subeq:F1324}) can fortunately be reduced to a sum of Y- and X-integrals in the following way \cite{Beisert:2002bb}:
\begin{align}
F_{13,24} &= \frac{X_{1234}}{I_{12}I_{34}} - \frac{X_{1234}}{I_{14}I_{23}} + \left( \frac{1}{I_{14}} - \frac{1}{I_{12}} \right) Y_{124} + \left( \frac{1}{I_{23}} - \frac{1}{I_{34}} \right) Y_{234} \notag \\
& \qquad \qquad \qquad \qquad \qquad + \left( \frac{1}{I_{23}} - \frac{1}{I_{12}} \right) Y_{123} + \left( \frac{1}{I_{14}} - \frac{1}{I_{34}} \right) Y_{134}\,.
\label{eq:FXYidentity}
\end{align}

The integrals given above also appear in their respective pinching limits, i.e. when two external points are brought close to each other. The integrals simplify greatly in this limit, and they exhibit a logarithmic divergence which is tamed with the use of point-splitting regularization. For the Y-integral, we define:
\begin{equation*}
Y_{122} := \lim_{x_3 \to x_2} Y_{123}\,, \qquad \qquad \qquad \lim_{x_3 \to x_2} I_{23} := \frac{1}{(2\pi)^2 \epsilon^2}\,.
\end{equation*}
In this limit, the conformal ratios are now given by:
\begin{equation*}
r = 1\,, \qquad \qquad s = (2\pi)^2 \epsilon^2 I_{12}\,.
\end{equation*}
Inserting this in (\ref{eq:Y123}) and expanding up to order $\mathcal{O} (\log \epsilon^2)$, we obtain:
\begin{equation}
\diagYonetwotwo\ := Y_{112} = Y_{122} = - \frac{1}{16 \pi^2} I_{12} \left( \log \frac{\epsilon^2}{x_{12}^2} - 2 \right)\,.
\label{eq:Y112}
\end{equation}
This result coincides with the expression given in \cite{Drukker:2008pi}.

Similarly, the pinching limit of the X-integral reads:
\begin{equation}
\diagXoneonetwothree\ := X_{1123} = - \frac{1}{16 \pi^2} I_{12} I_{13} \left( \log \frac{\epsilon^2 x_{23}^2}{x_{12}^2 x_{13}^2} - 2 \right)\,,
\label{eq:X1123}
\end{equation}
which is again the same as in \cite{Drukker:2008pi}.

Finally, the pinching limit $x_2 \to x_1$ of the F-integral gives:
\begin{align}
\frac{(\partial_1 - \partial_3) \cdot (\partial_1 - \partial_4)}{I_{13} I_{24}}\ \diagFonethreeonefour\ & \equiv F_{13,14} = F_{14,13} = - F_{13,41} \notag \\
& = -\frac{X_{1134}}{I_{13} I_{14}} + \frac{Y_{113}}{I_{13}} + \frac{Y_{114}}{I_{14}} + \left(\frac{1}{I_{13}} + \frac{1}{I_{14}} - \frac{2}{I_{34}} \right) Y_{134}\,.
\label{eq:F1314}
\end{align}

\subsection{Numerical Integration}
\label{subsec:numerical integration}

This section is devoted to the numerical integrations of the $2$- and $0$-channels at NLO that are presented in section \ref{sec:perturbative}.

\subsubsection{$2$-Channel}
\label{subsubsec:2channelatNLO}

As discussed in section \ref{subsubsec:2channel}, the only integral that needs to be computed for the $2$-channel is the following:
\begin{align}
I(x_1^2 , x_2^2) & := x_1^2 x_2^2 \int d\tau_3 \int d\tau_4 \int d\tau_5 \int d\tau_6\ \Theta(\tau_{3456})\ \left( I_{13} I_{25} + I_{15} I_{23} \right) \notag \\
& \qquad \qquad \qquad \qquad \qquad \qquad \times \left( I_{14} I_{26} + I_{16} I_{24} \right)\,,
\label{eq:defI}
\end{align}
with $\Theta(\tau_{3456})$ defined in (\ref{eq:pathorderingsymbol2}). We first note that the integral does not depend on $x_{12}$, but only on the distances between the operators and the line defect. As a consequence, the integral is symmetric with respect to $x_2 \leftrightarrow -x_2$.

In order to do the integral, we first perform the $\tau_6$- and $\tau_4$-integrals and we get:
\begin{align*}
I(x_1^2, x_2^2) &= \frac{|x_1| |x_2|}{128 \pi^4} \int d\tau_3 \int d\tau_5\ \Theta(\tau_{35}) \left( I_{13} I_{25} + I_{15} I_{23} \right) \left\lbrace \pi \left( \tan^{-1} \frac{\tau_3}{|x_1|} - \tan^{-1} \frac{\tau_5}{|x_1|} \right) \right. \notag \\
& \qquad \qquad + \tan^{-1} \frac{\tau_3}{|x_2|} \left( 2 \tan^{-1} \frac{\tau_5}{|x_1|} + \pi \right) + \tan^{-1} \frac{\tau_5}{|x_2|} \left( 2 \tan^{-1} \frac{\tau_3}{|x_1|} - \pi \right) \\
& \qquad \qquad \qquad \qquad \qquad \qquad \left. -4 \tan^{-1} \frac{\tau_5}{|x_1|} \tan^{-1} \frac{\tau_5}{|x_2|} \right\rbrace\,. 
\end{align*}
We can perform one more integration analytically by treating it term by term. The first integral gives:
\begin{align*}
I_1 (x_1^2 , x_2^2) & = \frac{|x_1| |x_2|}{128\pi^3} \int d\tau_3 \int d\tau_5\ \Theta(\tau_{35}) \left( I_{13} I_{25} + I_{15} I_{23} \right) \left( \tan^{-1} \frac{\tau_3}{|x_1|} - \tan^{-1} \frac{\tau_3}{|x_1|} \right) \notag \\
&= \frac{1}{256 \pi^5} \int d\tau_3\ \tan^{-1} \frac{\tau_3}{|x_1|} \left\lbrace |x_1|\ I_{13}\ \tan^{-1} \frac{\tau_3}{|x_2|} + |x_2|\ I_{23}\ \tan^{-1} \frac{\tau_3}{|x_1|} \right\rbrace\,,
\end{align*}
where we have performed the $\tau_3$-integral in the second term and relabeled $\tau_5$ to $\tau_3$ in the second line. The second integral reads:
\begin{align*}
I_2 (x_1^2, x_2^2) &= \frac{|x_1| |x_2|}{128 \pi^4} \int d\tau_3 \int d\tau_5\ \Theta(\tau_{35}) \left( I_{13} I_{25} + I_{15} I_{23} \right) \left\lbrace \tan^{-1} \frac{\tau_3}{|x_2|} \left( 2 \tan^{-1} \frac{\tau_5}{|x_1|} + \pi \right) \right. \notag \\
& \qquad \qquad \qquad \qquad \qquad \qquad \qquad \qquad \qquad \left. + \tan^{-1} \frac{\tau_5}{|x_2|} \left( 2 \tan^{-1} \frac{\tau_3}{|x_1|} - \pi \right) \right\rbrace \notag \\
&= \frac{|x_1| |x_2|}{64 \pi^4} \int d\tau_5 \int d\tau_3\ \left( I_{13} I_{25} + I_{15} I_{23} \right)\ \tan^{-1} \frac{\tau_3}{|x_2|} \tan^{-1} \frac{\tau_5}{|x_1|} \notag \\
& \qquad \qquad \qquad + \frac{|x_1| |x_2|}{128 \pi^4} \left( \int d\tau_5 \int d\tau_3\ \Theta(\tau_{35}) - \int d\tau_5 \int d\tau_3\ \Theta(\tau_{53}) \right) \\
& \qquad \qquad \qquad \qquad \qquad \qquad \times \left( I_{13} I_{25} + I_{15} I_{23} \right)\ \tan^{-1} \frac{\tau_3}{|x_2|}\,.
\end{align*}
The first term vanishes and we are left with:
\begin{equation*}
I_2 (x_1^2, x_2^2) = \frac{1}{256 \pi^5} \int d\tau_3\ \tan^{-1} \frac{\tau_3}{|x_2|} \left\lbrace |x_1|\ I_{13}\ \tan^{-1} \frac{\tau_3}{|x_2|} + |x_2|\ I_{23}\ \tan^{-1} \frac{\tau_3}{|x_1|} \right\rbrace\,.
\end{equation*}
The remaining term can also be reduced to a one-dimensional integral:
\begin{align*}
I_3 (x_1^2 , x_2^2) &= - \frac{|x_1| |x_2|}{32 \pi^4} \int d\tau_3 \int d\tau_5\ \left( I_{13} I_{25} + I_{15} I_{23} \right)\ \tan^{-1} \frac{\tau_5}{|x_1|} \tan^{-1} \frac{\tau_5}{|x_2|} \notag \\
&= - \frac{1}{256 \pi^5} \int d\tau_3\ \tan^{-1} \frac{\tau_3}{|x_1|} \tan^{-1} \frac{\tau_3}{|x_2|} \left\lbrace |x_1|\ I_{13}\ \left( \pi - 2 \tan^{-1} \frac{\tau_3}{|x_2|} \right) \right. \notag \\
& \qquad \qquad \qquad \qquad \qquad \left. + |x_2|\ I_{23}\ \left( \pi - 2 \tan^{-1} \frac{\tau_3}{|x_1|} \right) \right\rbrace\,.
\end{align*}
Putting everything together, the integral becomes:
\begin{align*}
I(x_1^2 , x_2^2) &= \frac{1}{256 \pi^6} \int d\tau_3\ \left\lbrace |x_1|\ I_{13}\ \tan^{-2} \frac{\tau_3}{|x_2|} \left( 2 \tan^{-1} \frac{\tau_3}{|x_1|} + \pi \right) \right. \notag \\
& \qquad \qquad \qquad \qquad \qquad \qquad \left. + |x_2|\ I_{23}\ \tan^{-2} \frac{\tau_3}{|x_1|} \left( 2 \tan^{-1} \frac{\tau_3}{|x_2|} + \pi \right) \right\rbrace\,.
\end{align*}
The terms cubic in $\tan^{-1}$ vanish because of antisymmetry. The integral reduces therefore to the following compact expression:
\begin{equation}
I (x_1^2, x_2^2 ) = \frac{1}{256 \pi^5} \int d\tau_3\ \left\lbrace |x_1|\ I_{13}\ \tan^{-2} \frac{\tau_3}{|x_2|} + |x_2|\ I_{23}\ \tan^{-2} \frac{\tau_3}{|x_1|} \right\rbrace\,.
\label{eq:2channelintegral}
\end{equation}

\begin{table}
\centering
\caption{Coefficients for the expansion of $I(1,x_2^2)$ following the ansatz given in eq. (\ref{eq:ansatz}) and obtained numerically by computing (\ref{subeq:coefficients1}), (\ref{subeq:coefficients2}). A factor $g^8 N^2/2^9 \pi^6$ has been removed for all coefficients. Guessing the closed form leads to the expression given in (\ref{eq:result2channel}).}
\resizebox{\columnwidth}{!}{%
\begin{tabular}{cccccccccc}
\hline \\[-1em]
$a_0$ & $a_1$ & $a_2$ & $a_3$ & $a_4$ & $a_5$ & $a_6$ & $a_7$ & $a_8$ & $a_9$ \\
\hline \\[-.5em]
$\frac{1}{8}$ & $2\log 2 - 1$ & $- 1$ & $\frac{2 \log 2}{3} + \frac{2}{9}$ & $- \frac{2}{3}$ & $\frac{2\log 2}{5} + \frac{13}{50}$ & $-\frac{23}{45}$ & $\frac{2\log 2}{7} + \frac{71}{294}$ & $-\frac{44}{105}$ & $\frac{2\log 2}{9} + \frac{71}{324}$ \\[.5em]
\hline \\[-1em]
$a_{10}$ & $a_{11}$ & $a_{12}$ & $a_{13}$ & $a_{14}$ & $a_{15}$ & $a_{16}$ & $a_{17}$ & $a_{18}$ & $a_{19}$ \\
\hline \\[-.5em]
$-\frac{563}{1575}$ & $\frac{2\log 2}{11} + \frac{1447}{7260}$ & $- \frac{3254}{10395}$ & $\frac{2\log 2}{13} + \frac{617}{3380}$ & $- \frac{88069}{315315}$ & $\frac{2\log 2}{15} + \frac{1061}{6300}$ & $-\frac{11384}{45045}$ & $\frac{2\log 2}{17} + \frac{12657}{80920}$ & $-\frac{1593269}{6891885}$ & $\frac{2\log 2}{19} + \frac{132931}{909720}$ \\[.5em]
\hline \\[-1em]
$b_1$ & $b_2$ & $b_3$ & $b_4$ & $b_5$ & $b_6$ & $b_7$ & $b_8$ & $b_9$ & $b_{10}$ \\
\hline \\[-.5em]
$1$ & $0$ & $\frac{1}{3}$ & $0$ & $\frac{1}{5}$ & $0$ & $\frac{1}{7}$ & $0$ & $\frac{1}{9}$ & $0$ \\[.5em]
\hline
\end{tabular}%
}
\label{table:2channelcoefficients}
\end{table}

We were not able to solve this integral analytically, but with the help of numericals it is still possible to obtain the closed form. We start with the following ansatz, which is based on the expansion of the superblocks in the defect channel given in (\ref{eq:expsuperblocksdefect}):
\begin{equation}
I(1, x^2) = \sum_{k=0}^\infty a_k x^k + \log x \sum_{k=1}^\infty b_k x^k\,,
\label{eq:ansatz}
\end{equation}
where we have defined $|x_1| = 1$ and $|x_2| := x$ in order to lighten the notation. The coefficients obey the following relations:
\begin{subequations}
\begin{gather}
a_k = \frac{1}{k!} \lim\limits_{x \to 0} \partial_x^k \left\lbrace I(1,x^2) - \log x \sum_{l=1}^{k-1} b_l \frac{x^{2l-1}}{2l-1} \right\rbrace\,, \label{subeq:coefficients1} \\
b_k = \frac{1}{k!} \lim\limits_{x \to 0} \left\lbrace x\ \partial_x^{k+1} I(1,x^2) + \sum_{l=1}^{k-1} (-1)^{k-l+1} (k-l)! l!\ x^{l-k} \right\rbrace\,.
\label{subeq:coefficients2}
\end{gather}
\end{subequations}
Hence the coefficients can be computed numerically for decreasing $x$ until convergence. The convergence also confirms the validity of the ansatz given in (\ref{eq:ansatz}). The coefficients are given in table \ref{table:2channelcoefficients}. We managed to obtain accurate enough data to be able to guess the closed form for all the coefficients. Moreover, the resulting series are all identifiable and can be resummed in order to obtain the closed form of the full integral as given in eq. (\ref{eq:result2channel}).

\bigskip

It is worth checking that we got the closed form right. Fig. \ref{fig:2and1channels} shows a plot of the numerical data and of eq. (\ref{eq:result2channel}), which match perfectly. We have therefore managed to obtain an exact analytical expression for the $2$-channel.

\subsubsection{$0$-Channel}
\label{subsubsec:0channelatNLO}

We now discuss the case of the H-integral in the $0$-channel, which is by far the hardest that we have to consider in this work. It is a $10$-dimensional integral with two $\tau$-derivatives, acting on $x_1$ and $x_2$:
\begin{equation}
\diagHd\ = - 4 \lambda_0\ I_{12}\ \partial_{\tau_1} \partial_{\tau_2}  \int d\tau_3 \int d\tau_4\ H_{13,24}\,,
\label{eq:Hintegral1}
\end{equation}
with $H_{13,24}$ defined in (\ref{subeq:H1324}). Integration by parts can be used for removing the $\tau_2$-derivative:
\begin{align*}
\partial_{\tau_2} H_{13,24} &= \int d^4 x_{56}\ I_{15} I_{35} \partial_{\tau_2} I_{26} I_{46} I_{56} \\
&= - \int d^4 x_{56}\ I_{15} I_{35} I_{26} \left( \partial_{\tau_4} + \partial_{\tau_5} \right) I_{46} I_{56}\,.
\end{align*}
Since $\int d\tau_4\ \partial_{\tau_4} I_{46} = 0$, we can drop the first term in the last line. Using integration by parts with respect to the $x_5$-integral, we obtain:
\begin{align*}
\partial_{\tau_2} H_{13,24} \widehat{=} - \partial_{\tau_1} H_{13,24}\,,
\end{align*}
where the $\widehat{=}$ means that this equality is to be understood as valid in the context of eq. (\ref{eq:Hintegral1}) only. Here we have used again the fact that $\int d\tau_3\ \partial_{\tau_3} I_{35} = 0$.

We now have the following expression:
\begin{equation*}
\diagHd\ = + 4 \lambda_0 I_{12} \int d\tau_3 \int d\tau_4 \int d^4 x_5\ \partial_{\tau_1}^2 I_{15} I_{35}\ Y_{245}\,.
\end{equation*}
The Y-integral is known analytically, hence we find ourselves facing a $6$-dimensional integral. The derivatives give:
\begin{equation*}
\partial_{\tau_1}^2 I_{15} = 2\ (2\pi)^2 I_{15}^2 \left( 4\ (2\pi)^2 \tau_5^2\ I_{15} - 1 \right)\,,
\end{equation*}
and it is easy to do the $\tau_3$-integral using (\ref{eq:elementaryintegral1}):
\begin{equation*}
\diagHd\ = \frac{8}{(2\pi)^2} \lambda_0 I_{12} \int d^4 x_5 \frac{1}{\left(\vec{x}_5^2 \right)^{1/2} x_{15}^4} \left( \frac{4 \tau_5^2}{x_{15}^2} - 1 \right) \int d\tau_4\ Y_{245}\,,
\end{equation*}
where $\vec{x}$ means that the $\tau$-component is zero, i.e. $\vec{x} \equiv (x, y, z, 0)$.

\bigskip

This is as far as we can go analytically for a general $x_2$. Going to the limiting case $x_2 = (x_2, 0, 0, 0)$ (i.e. the line $z = \bar{z}$) encourages us to introduce $3$d spherical coordinates for $y_5, z_5, \tau_5$, because $y_5$ and $z_5$ now always appear in the combination $ y_5^2 + z_5^2$. The integration is independent of the azimuthal angle and we can kill one integral in exchange of a $2\pi$ factor. We obtain:
\begin{equation}
\diagHd\ = \frac{4}{(2\pi)^6} I_{12} \int_0^\infty dr \int_0^\pi d\theta \int_{-\infty}^\infty dx \int_{-\infty}^\infty d\tau_4\ \frac{r^2 \sin \theta}{R \left( d^2 \right)^2} \left( \frac{4 r^2 \cos^2 \theta}{d^2} - 1 \right) Y_{245}\,,
\label{eq:Hintegral2}
\end{equation}
where we have dropped the index $5$ in order to keep our expression compact. The functions $R$ and $d$ are defined as follows:
\begin{align*}
& R(x,r,\theta) \coloneqq \sqrt{x^2 + r^2 \sin^2 \theta}\,, \\
& d^2 (x,r) \coloneqq (1-x)^2 + r^2\,.
\end{align*}
We are thus left with a hard $4$-dimensional integral, and at that point we cannot go further analytically. It is however possible to compute this integral numerically for arbitrary value of $x_2$, and the result is shown in fig. \ref{fig:0channelandWI}, where we notice that it is constant for $x_2 \leq 0$. This remarkable feature is exploited in section \ref{subsec:defectCFTdata} for extracting the CFT data without having to know the full correlator analytically. We showed in section \ref{subsec:bulkCFTdata} that once the defect CFT data is known, it is possible to resum the resulting expression in the collinear limit. The corresponding expression is given by (\ref{subsec:bulkCFTdata}) and also plotted in fig. \ref{fig:0channelandWI} for comparison with the numerical data.

\bibliography{./auxi/2pt_function_wline.bib}
\bibliographystyle{./auxi/JHEP}

\end{document}